\documentclass[]{aa}  

\usepackage{graphicx}
\usepackage{natbib}
\bibpunct{(}{)}{;}{a}{}{,} 
\usepackage{txfonts}
\usepackage[colorlinks=true,allcolors=blue]{hyperref}

\usepackage{balance}
\usepackage{gensymb}
\usepackage{placeins}
\usepackage{comment}
\usepackage{longtable}
\usepackage{booktabs}

\begin{document}

\title{Multi-epoch VLBI observations of the blazar 3C\,66A: Spatial twisting and temporal oscillation of the parsec-scale jet}

   \author{
    Paloma~Thevenet \inst{\ref{kasi}, \ref{obspm}},
    Jeonguk~Kim  \inst{\ref{yonsei}, \ref{kasi}}~\thanks{The authors Paloma~Thevenet and Jeonguk~Kim contributed equally to this work.}, 
    Guang-Yao~Zhao \inst{\ref{mpifr}, \ref{kasi}}~\thanks{Corresponding author; \email{gyzhao@mpifr-bonn.mpg.de}},
    Bong~Won~Sohn \inst{\ref{kasi}, \ref{ust}},
    Suk-Jin~Yoon \inst{\ref{yonsei}}
    }
   \institute{
        {Korea Astronomy and Space Science Institute, Daejeon 34055, Republic of Korea}~\label{kasi}
        \and
        {LUX, Observatoire de Paris, Université PSL, CNRS, Sorbonne Université, 92190 Meudon, France}~\label{obspm}
        \and
        {Department of Astronomy \& Center for Galaxy Evolution Research, Yonsei University, Seoul 03722, Republic of Korea}~\label{yonsei}
        \and
        {Max-Planck Institut für Radioastronomie, Auf dem Hügel 69, Bonn 53121, Germany}~\label{mpifr}
        \and
        {Korea National University of Science and Technology, Daejeon 34113, Republic of Korea}~\label{ust}
        }

   \date{\today}

  \abstract
   {High-resolution very long baseline interferometry (VLBI) observations have revealed a growing number of active galactic nuclei (AGNs) that exhibit variations in their inner jet position angle (PA).
   Investigations of such jets can shed light on the understanding of precession mechanisms and instabilities occurring in the jet and the coupled accretion disk, since changes in the spatial orientation arise in the innermost region.
   }
   {
   {Previous VLBI kinematic studies of the blazar 3C\,66A have unveiled complex jet kinematic behaviors (e.g., inward/outward, sub- to super-luminal and nonradial motions).
   Using follow-up high-resolution VLBI observations and archival data, we investigate the morphology and the variations in orientation and core flux density of the 3C\,66A jet to gain a deeper insights into its kinematic behavior and physical origins.}
   }
   {We performed KVN and VERA array (KaVA) observations at 22/43\,GHz over three epochs in 2014 and collected 109 sets of Very Long Baseline Array (VLBA) archival data at 43\,GHz between 1996 -- 2025. 
   We imaged the parsec-scale jet and parameterized it using circular Gaussian fittings to the UV visibilities.
   Finally, we derived the inner jet PA and the core flux densities for the VLBA data.}
   {The jet presents a twisted morphology in the KaVA maps.
   The PA of the fitted Gaussian components is in the range between 170$\degree$ and 195$\degree$. 
   Our kinematic analysis using the VLBA data indicates that the PA oscillates with an amplitude of 7.77\,$\pm$\,0.79$\degree$ and a period of 10.94\,$\pm$\,0.22 years, presented for the first time in this work. 
   This oscillation is topped by a continuous clockwise shift of the PA by $-$0.83\,$\pm$\,0.07$\degree$/year.
   We also identified a strong core flux variability with possible periodicity and a 2$\sigma$ correlation between the core flux density and the inner jet PA change.
   We discuss possible physical models that could explain the observed features for this object; in particular, a supermassive black hole binary (SMBHB) system, Lense-Thirring (LT) effect, and jet or disk instabilities.
   }
   {The oscillation and continuous shift of the PA and the possible radio flux periodicity, together with the optical flux periodicity of $\sim$2 years that had previously been  confirmed in several independent studies, favor a jet precession scenario driven by orbital motion and disk-orbit misalignment in a SMBHB system. For the estimated central mass of $M=(1.42\pm0.19)\times10^8M_{\odot}$ from variability timescales, the separation between the putative black holes is $r=(1.65\pm0.08)\times10^{-2}$ pc.
   }


   \titlerunning{Spatial twisting and temporal oscillation in the jet of 3C\,66A}
   \authorrunning{P. Thevenet et al.}
   \maketitle

\section{Introduction} \label{introduction}

Blazars are a subclass of active galactic nuclei (AGNs) whose jet axes are closely aligned with the Earth's line of sight. The small viewing angle leads to a number of unique observational properties due to relativistic Doppler boosting.
These include superluminal motion  \citep[e.g.,][]{1966Natur.211..468R, 2019ApJ...874...43L} and violent flux variability across the electromagnetic spectrum \citep[e.g.,][]{2017Natur.552..374R, 2017MNRAS.465..161S}, as well as visible changes in the jet morphology and orientation \citep[e.g.,][]{2017A&A...602A..29B, 2018MNRAS.478..359K}.   
  
The periodic and erratic changes in the position angle (PA) of the inner jet have been detected in many blazars \citep{2013AJ....146..120L, Lister2021ApJ...923...30L, Kostrichkin2025MNRAS.537..978K}. 
Such PA variations could be associated with the jet launching mechanism because they occur near the jet base, yet there is no unique paradigm pertaining to their origin. Some proposed mechanisms are instabilities in the jet flow, disk precession, or jet precession. The  latter two can  arise for a number of reasons, such as the Lense-Thirring (LT) precession of the disk or the precession of the black hole (BH) spin axis \citep{Lense1918, Lu1992}, as well as the orbital motion in a supermassive black hole binary (SMBHB) in some cases \citep{Begelman1980Natur.287..307B}. One aspect of special interest is that inner jet PA changes often arise together with flux variability and  with the flux periodicity on occasion \citep{2009A&A...508.1205B, 2012Ap&SS.342..465L,2005A&A...431..831L, 2014MNRAS.445.1370K, 2018MNRAS.478.3199B, Jiang2023}.   
  
3C\,66A is a well-known BL Lac object. Very Large Array (VLA) observations indicate that the source has an extended jet with a PA of 170$\degree$ and weak radio lobes toward both the north and south \citep{1993ApJS...86..365P, 1996ApJS..107...37T}. Only the southern jet has been detected in very long baseline interferometry (VLBI) images. \citet{2007A&A...468..963C} find two bendings at $\sim$1.2 and $\sim$4\,mas from the core in 2.3, 8.4, and 22\,GHz VLBI images obtained between 2001 and 2002. The bent structure is also confirmed by \citet{2015AJ....149...46Z} using data from 2004--2005. 15\,GHz VLBI observations from 2003--2013 moreover show complex jet kinematics \citep{2019ApJ...874...43L}. The bright features identified present a wide range of apparent speeds, from apparently stationary to superluminal motions. In addition, some features are characterized by transverse motion, while some are in the direction of the core, as also reported in \citet{2015AJ....149...46Z}. 
Similar complex kinematics were reported at the mm-regime \citep{2001ApJS..134..181J, 2005AJ....130.1418J, 2017ApJ...846...98J}. Using 10 years of 43\,GHz data from the VLBA (Very Long Baseline Array) BU-BLAZAR monitoring program, \cite{Weaver2022} also reported a bent jet and found four stationary components, some of which present PA variability of more than 20$\degree$. Moreover, they determined six moving knots in the jet that reach apparent superluminal speeds.  Recently, \cite{Kostrichkin2025MNRAS.537..978K} used an automated algorithm to study jet orientation changes in hundreds of AGNs. From 15\,GHz and 43\,GHz data, these authors also found a variable PA in the jet of 3C\,66A.

In addition, 3C\,66A is known for its optical flux variability, found periodic on timescales between approximately 2 and 5 years in a significant number of independent studies \citep{Fan2002, Belokon2003, Kaur2017, Fan2018, OteroSantos2020, Cheng2022}. \cite{Cheng2022} argued that the two-year period is persistent in the historical light curve (LC) of 3C\,66A and is not a transient feature. Proposed explanations for the periodicity include instabilities in the accretion disk or orbital motion in a SMBHB. However, there is no consensus yet regarding the origin of this optical periodicity. On the other hand, \cite{OteroSantos2020} did not find any evidence of periodic behavior in the $\gamma$-ray LC of 3C\,66A.
  
In this paper, we confirm that the 3C\,66A parsec-scale jet is twisted (i.e., characterized by multiple bendings) using high-resolution KaVA (KVN and VERA array) images. Moreover, we report the detection of oscillation and continuous clockwise shift in the inner jet PA, based on long-term VLBA observations. The same PA periodicity is found in the downstream jet with a phase-shift and amplitude reduction. In addition, we report high radio flux variability, with three possible quasi-periods.
These features have motivated us to investigate plausible physical scenarios leading to the observed jet orientation and flux periodicities. In particular, we investigated jet precession that could be triggered by a SMBHB system or the LT effect. We also explore further instabilities and disk-related mechanisms that might be relevant to 3C\,66A. 
  
The paper is organized as follows. In Section \ref{obs}, we describe the datasets and the data reduction for the KaVa and VLBA data. In Section \ref{results}, we present our analysis and results concerning the jet structure and the core flux. In Section \ref{modeling_precession}, we model and constrain precession parameters for the jet of 3C\,66A. In Section \ref{modeling_SMBHB}, we investigate the SMBHB hypothesis and show that it can explain all the observables reported in this paper. In Section \ref{discussion}, we consider additional scenarios such as LT precession and instabilities to account for some of the properties of 3C\,66A. Our findings are summarized in Section \ref{conclusions}.
  
We adopted a standard $\Lambda$ cold dark matter ($\Lambda$CDM) cosmology with parameters $\Omega_\mathrm{M}=0.3$, $\Omega_\mathrm{\Lambda}=0.7$, and $H_0=68.0$ km s$^{-1}$ Mpc$^{-1}$, rounded from the latest Planck Collaboration results \citep{Planck2020A&A...641A...6P, Tristram2024A&A...682A..37T}. 3C\,66A has an uncertain spectroscopic redshift \citep[0.335 < z < 0.444;][]{2013ApJ...766...35F} due to the lack of spectral features. Here, we use the redshift of $z = 0.340$ that was more recently determined from a cluster the host galaxy of 3C\,66A belongs to \citep{2018MNRAS.474.3162T}. At this distance, 1.0\,mas corresponds to 5.0\,pc.

\section{Observations \& data reduction} \label{obs}
\subsection{KaVA observations} \label{KaVAobs}

  At the time of these observations, KaVA consisted of seven antennas at three stations (Yonsei, Ulsan, and Tamna) of the Korean VLBI network (KVN) and four (Mizusawa, Iriki, Ogasawara, and Ishigakijima) of the VLBI  Exploration of Radio Astrometry (VERA) array~\citep{2014PASJ...66..103N,2017PASJ...69...71H}. KaVA is the precursor of the East Asia VLBI Network (EAVN) \citep{An2018NatAs...2..118A, Akiyama2022Galax..10..113A}.
  In 2014, 3C\,66A was observed with KaVA during three epochs: April 16-17 (DOY = 106--107), September 30-October 1 (273--274), and December 15-16 (349--350).
  The observations for each epoch were conducted over two successive days, with the first day at 22\,GHz and the second at 43\,GHz. 
  On April 16-17, the on-source time for 3C\,66A was $\sim$6 hours at both frequencies, while it was $\sim$2 and $\sim$3 hours at 22 and 43\,GHz, respectively, for the latter two epochs.
  The KVN-Ulsan station failed to participate in the second epoch observations due to instrumental malfunctions, and during the observation on December 15, several stations, especially those providing long baselines, suffered from severe weather conditions.
  Single polarization (left-hand circular polarization) was recorded with a total bandwidth of 256 MHz at each frequency: 16 IFs (intermediate frequencies or sub-bands) and 16 MHz per IF. 
  The data were correlated at the Daejeon hardware correlator \citep{2015JKAS...48..125L}.

  A priori amplitude and phase calibration were conducted using the National Radio Astronomy Observatory (NRAO) AIPS software \citep{1983AJ.....88..688S}.
 We first used opacity-corrected system temperatures and antenna gain curves to perform amplitude calibration. 
  Additionally, we multiplied a factor of 1.35 to all visibility amplitudes in order to compensate for the known amplitude loss due to digital sampling and quantization \citep{2015JKAS...48..229L}. 
  The source 3C 84 or NRAO 150 was selected as the bandpass and manual fringe fitting calibrator, and KVN-Yonsei was used as the reference antenna.
  We corrected the parallactic angles for KVN antennas only. 
  The global fringe fitting was done with a solution interval of 1 and 0.5 minutes for the 22 and 43\,GHz data, respectively. 
  The output visibilities from AIPS were imported to the DIFMAP package \citep{1997ASPC..125...77S} for imaging and model fitting. 
  We performed an hybrid imaging procedure \citep{1981MNRAS.196.1067C}, with CLEAN \citep{1974A&AS...15..417H} and phase-only self-calibrations iteratively in the early stage and amplitude self-calibration at the end of the process. 
  We modeled the UV visibilities with several circular Gaussian components by $modelfit$. 
  The Gaussian model fitting determined the flux, separation from the core, PA, and full width at half maximum (FWHM) of each component.  
  The uncertainties of each parameter were calculated following the equations described in \citet{2008AJ....136..159L}.

\subsection{VLBA archival data} \label{VLBAobs}

  We used all available 3C\,66A VLBA data from observations performed at 43\,GHz between 1996 and 2025. 
  Out of the total 109 epochs, 68 were obtained within the VLBA-BU-BLAZAR program observed between October 2008 and June 2020.
  The detailed description of this program and the data reduction are presented in \citet{2017ApJ...846...98J} and \cite{Weaver2022}. The five epochs obtained in 2024-2025 are from the BEAM-ME project\footnote{\url{https://www.bu.edu/blazars/BEAM-ME.html}} by the same group. The observations were carried out with 9 to 10 antennas (36 or 45 out of 45 possible baselines) and the typical integration time is of 40 minutes.
  
  With the CLEAN algorithm, by carefully setting CLEAN windows, we imaged the self-calibrated UV data without further self-calibration.
  The 36 other epochs of the 43\,GHz VLBA archival data, spanning the period August 1996--November 2005, were taken from the NRAO Data Archive System. These observations were also carried out with 9 to 10 antennas and have a similar integration time. We performed the data reduction using AIPS and DIFMAP in a similar way to that described in Section~\ref{KaVAobs}. For the archival VLBA data, an opacity fitting was applied during the amplitude calibration. One of the south-west antennas ("FD", "PT", "LA," or "KP") was selected as the reference antenna and one of the bright, well-sampled sources (3C 279, 3C 345, 3C 454.3, or 0420$-$014) was selected as the calibrator source for bandpass calibration and manual pulse-calibration. The solution interval was set to 30 seconds. After AIPS calibration, imaging was performed in DIFMAP and we fitted circular Gaussian components to each self-calibrated VLBA dataset.

\section{Results} \label{results}
\subsection{Parsec-scale jet morphology} \label{morphology}

   \begin{figure}
   \centering
   \includegraphics[trim={0mm, 15mm, 00mm, 30mm},clip=true,width=\hsize]{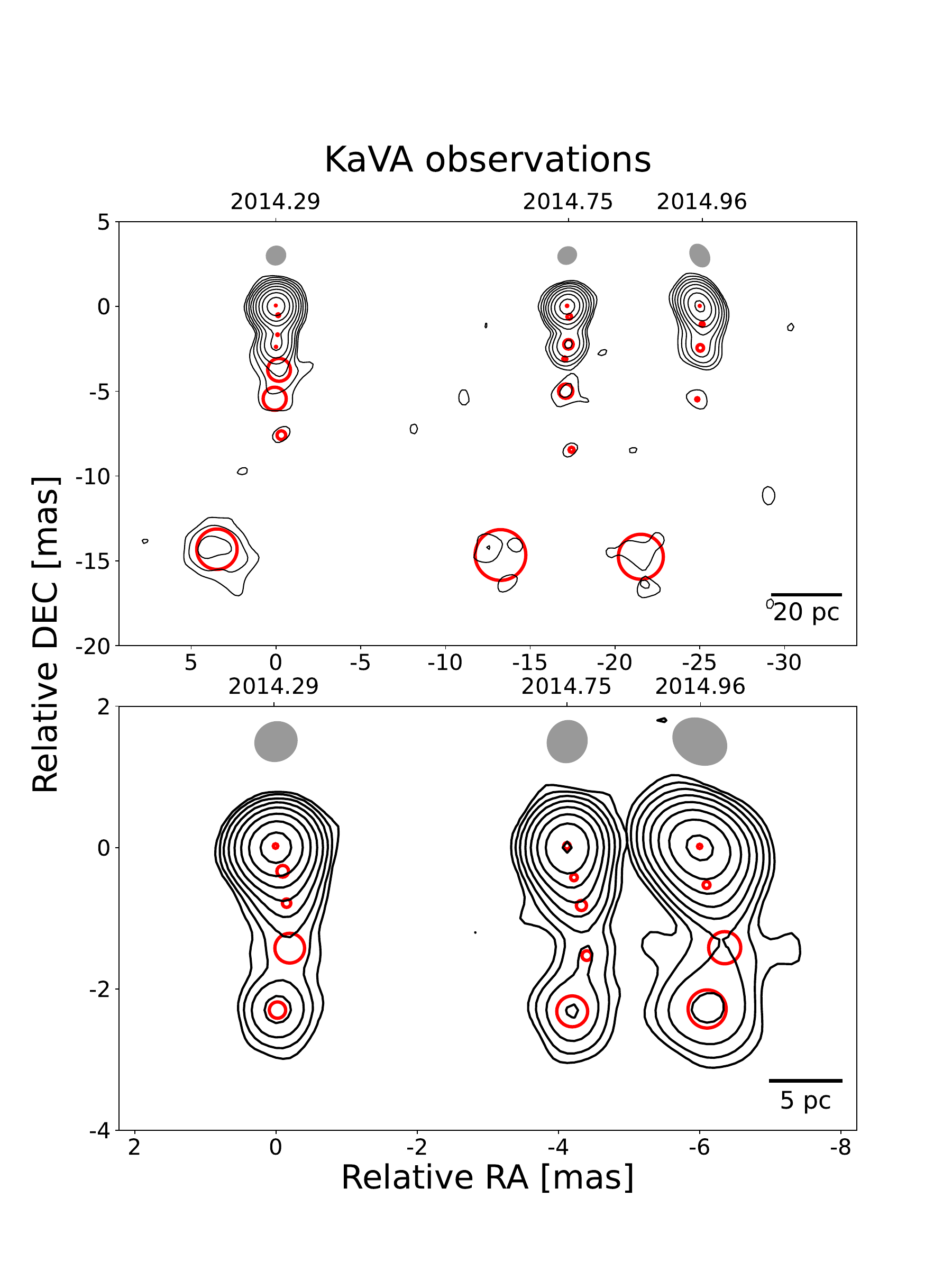}
      \caption{KaVA intensity maps of the 3C\,66A radio jet at 22\,GHz (top) and 43\,GHz (bottom) in 2014.
      The black contour levels start at three times the rms value, scaling twice. 
      The grey shaded ellipses above each contour show the restoring beam. 
      The detailed imaging parameters are summarized in Table \ref{tab1}. 
      The horizontal spacing is proportional to the observation time gap. 
      The exact observing time is in the unit of years. 
      The red circles are the model components obtained from $modelfit$. 
      The continuous curved jet extends downwards to $\sim$3\,mas on both panels. 
              }
         \label{fig1}
   \end{figure}

   \begin{figure}
   \centering
   \includegraphics[width=\hsize]{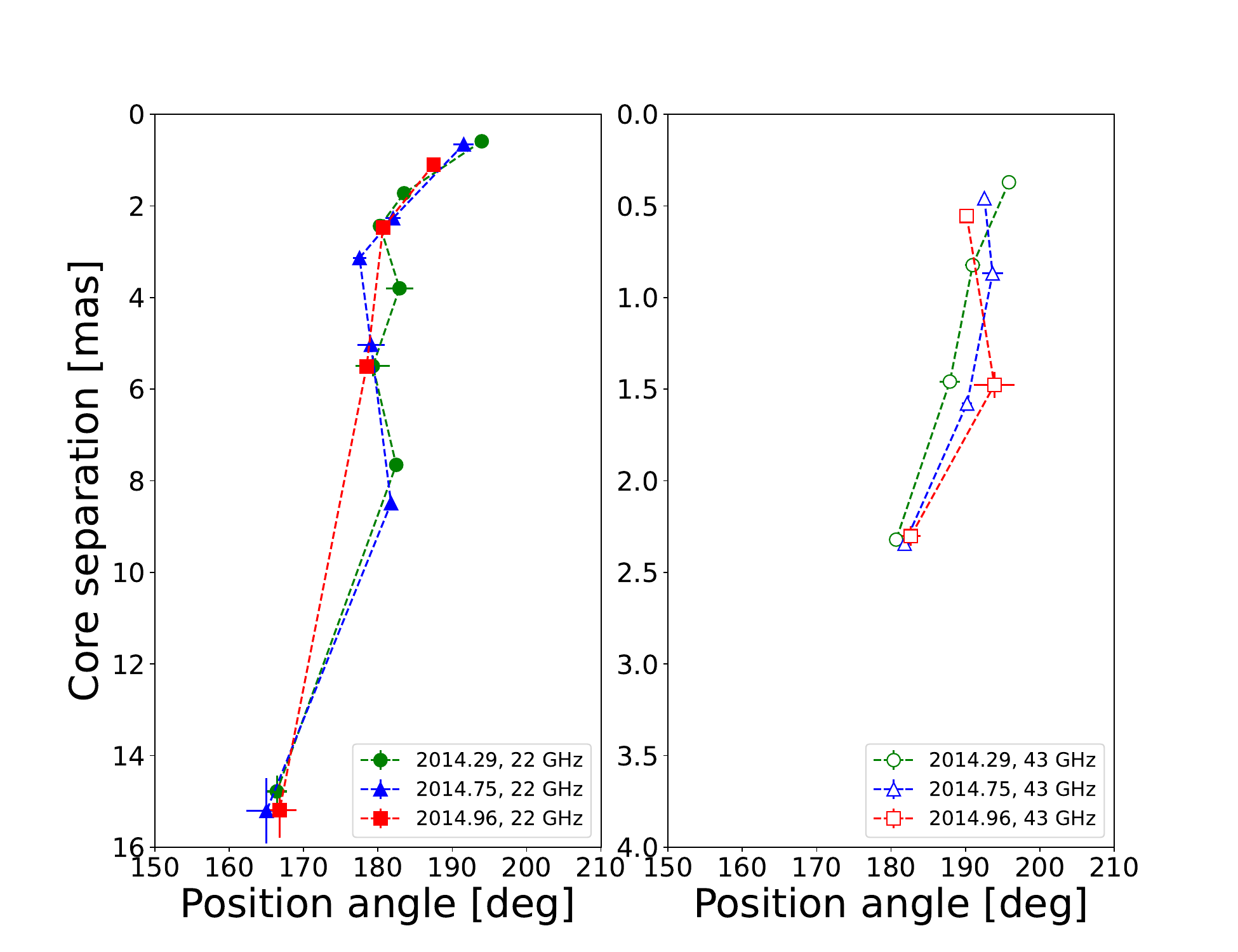}
      \caption{PA of the Gaussian components of the 3C\,66A jet as a function of the core separation obtained with the KaVA observations at 22 (left) and 43\,GHz (right) in 2014.  
      Each epoch is marked by green circles, blue triangles, and red squares. 
              }
         \label{fig2}
   \end{figure}

  Figure \ref{fig1} shows the total intensity contour maps obtained from the KaVA observations at 22 and 43\,GHz with the Gaussian jet components superimposed.  
  The mapping parameters are summarized in Table \ref{tab1}. 
  A core-jet structure is found in the north-south direction down to $\sim$16\,mas at 22\,GHz (upper panel of Figure \ref{fig1}) and down to $\sim$3\,mas at 43\,GHz (lower panel of Figure \ref{fig1}). 
  As the observing circumstances for each epoch are not the same (see Section \ref{KaVAobs}), we can see various dynamical ranges and restoring beam shapes (Table \ref{tab1}). 
  The parameters and errors of the Gaussian components for each epoch are listed in Table \ref{tab2}. 
  The brightness temperature of each component at the observing frequency $\nu$ in the source frame is calculated with the equation \citep[e.g.,][]{1982ApJ...252..102C}: 
  $ T_{\mathrm{b}}=1.22\times10^{12} S \nu^{-2} d^{-2} (1+z)$ K, 
  where $S$ is the flux density of the component in Jy, $z$ is the redshift of the source, and $d$ is the FWHM of the circular Gaussian component in mas. 
  Component $K$ is identified as the core, since it is located at the most upstream of the jet with the largest flux density, the highest brightness temperature, and the flattest spectrum \citep{2015AJ....149...46Z}.
  The core separations and PAs of the other components are relative to component $K$.
  
  We found evidence of multiple bendings in the VLBI jet of 3C\,66A.
   At $\sim$1.5\,mas downwards from the core in the 43\,GHz maps (bottom panel of Figure \ref{fig1}), the jet curves from the southwest to the southeast.
  In the 22\,GHz maps (top panel of Figure \ref{fig1}), the jet axis further bends to the south direction between 4 and 8\,mas.
  We found additional bending between 8 and 15\,mas based on the bright downstream knot visible at $\sim$15\,mas in the southeasterly direction.
  We define this structure as twisted, in the sense that there are several bendings with changing direction along the jet. This apparent morphology is confirmed by the PA distribution of the Gaussian components as a function of the core separation (Figure \ref{fig2}). The bendings found are consistent with the results of previous studies \citep{2007A&A...468..963C, 2015AJ....149...46Z} and are visible up to $\sim$20\,mas southwards in the lower frequency (2.3 and 8.4\,GHz) maps of \cite{2015AJ....149...46Z}. This twisted structure is also visible on smaller scales with the 43 GHz VLBA data, as illustrated by the ridgelines of six different epochs in Figure \ref{components}.

\subsection{Evolution of the jet orientation} \label{orientation}

   \begin{figure}
   \centering
   \includegraphics[trim={0mm, 10mm, 00mm, 10mm},clip=true,width=\linewidth]{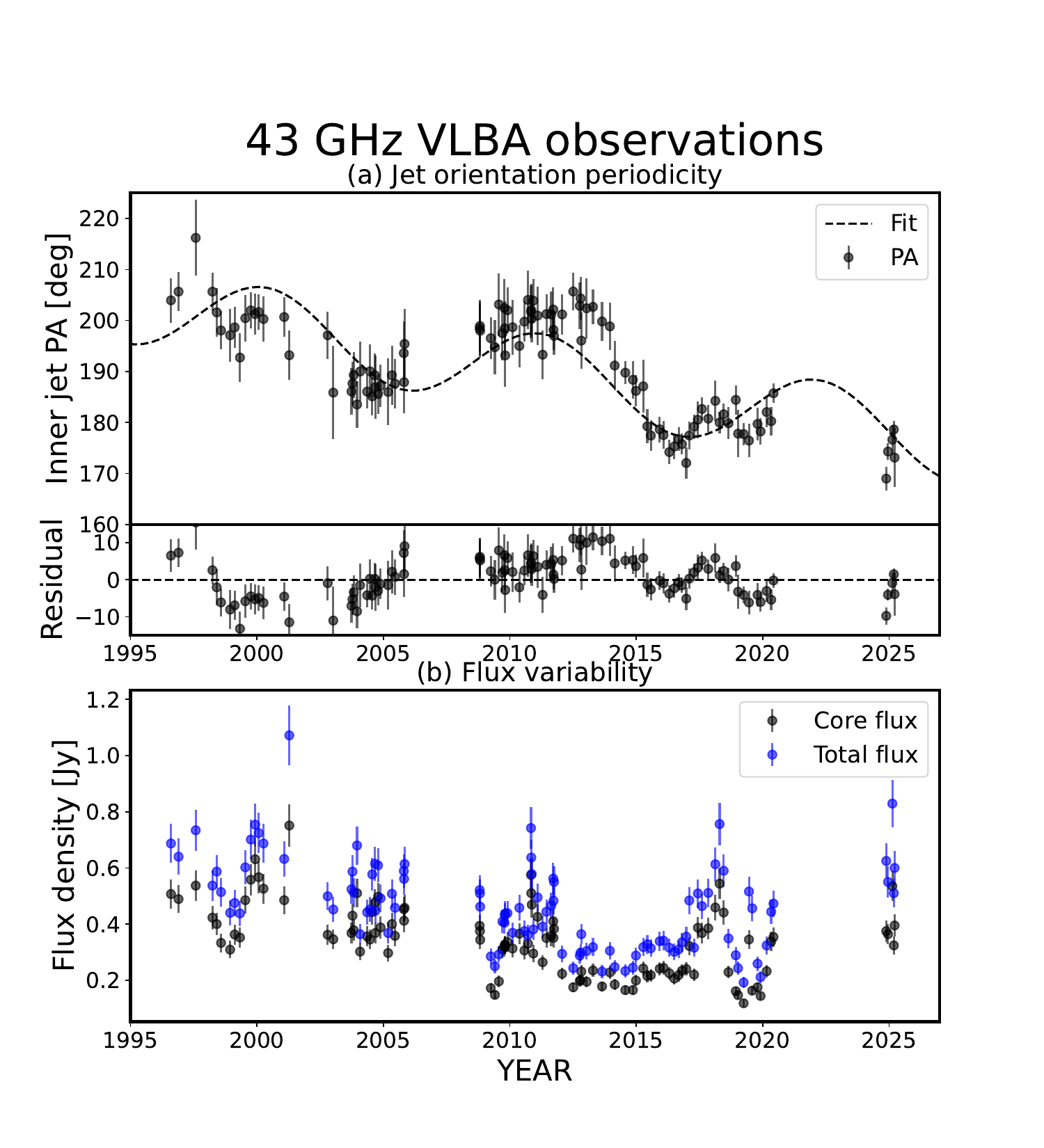}
      \caption{(a) Inner jet PA versus time at 43\,GHz. The dashed curve in the upper panel is the analytic sinusoidal+linear fit. The lower panel shows the residuals. The fitting parameters are summarized in Table \ref{tab3_PAfit}. (b) Time evolution of the total (blue) and core (black) flux density.}
         \label{fig3}
   \end{figure}

   \begin{table*}[ht]
      \caption{\centering Results of the analytic fit to the PA change.}
         \label{tab3_PAfit}
         \centering
         \begin{tabular}{cccccccc}
            \hline
            \noalign{\smallskip}
            Component & Method & A [deg] & P [yr] & S [deg/yr] & $t_0$ [yr] & $\theta_{0}$ [deg] & $\chi^2$ \\
            \noalign{\smallskip}
            \hline
            \noalign{\smallskip}
            Inner jet & Annular bins & 7.77 $\pm$ 0.79 & 10.94 $\pm$ 0.22 & $-$0.83 $\pm$ 0.07 & 2008.57 $\pm$ 0.22 & 191.87 $\pm$ 0.64 & 2.25 \\
            \noalign{\smallskip}
             & Circular Gaussians &  7.38 $\pm$ 0.65 & 11.14 $\pm$ 0.24 & $-$0.83 $\pm$ 0.07 & 2008.37 $\pm$ 0.25 & 193.31 $\pm$ 0.71 & 4.09\\
            \noalign{\smallskip}
            \hline
            \noalign{\smallskip}
            Downstream jet & Annular bins & 4.25 $\pm$ 0.60 & 10.50 $\pm$ 0.29 & $-$0.68 $\pm$ 0.06 & 2010.12 $\pm$ 0.23 & 188.53 $\pm$ 0.48 & 1.90 \\
            \noalign{\smallskip}
             & Circular Gaussians & 4.20 $\pm$ 0.58 & 10.03 $\pm$ 0.26 & $-$0.70 $\pm$ 0.06 & 2010.25 $\pm$ 0.23 & 189.04 $\pm$ 0.51 & 2.78\\
            \noalign{\smallskip}
            \hline
         \end{tabular}
   \end{table*}

  We studied the changes in the jet orientation for the 43\,GHz VLBA archival data. We used two methods to determine the PA of a region in the jet: estimating the PA of the circular Gaussian components obtained from \textit{modelfit} and estimating the PA from the clean components in annular bins. Since the clean components are the deconvolved solutions of the source structure, we were able to derive the PA of the jet by using the distribution of clean components \citep[e.g.,][]{2013AJ....146..120L,2014A&A...571L...2R}. We began by analyzing the inner jet PA evolution. Using Gaussian fitting, the inner jet region corresponds to the component J1 (see Figure \ref{components}). This component remains at a core separation of 0.2-0.6\,mas throughout all epochs. 
  For the clean components method, we set an annular bin (called annulus A) in the radius range from 0.15 to 0.45\,mas away from the (0, 0) central point.
  We adopted the lower limit of the annulus to prevent distortion of the determined PA by the core component.
  We defined the flux-weighted average of the PA from the clean components in the bin as the inner jet PA and estimate the statistical error of the PA using the bootstrap method. 
  We randomly resampled the clean components from the bin and calculate the flux-weighted mean PA from the bootstrapped sample.
  After repeating the calculation 10~000 times, we adopted the standard deviation as the statistical error. Moreover, we  considered systematic errors to the PA to be a fraction, $f_\mathrm{err}$, of the beam size. This fraction depends on the bandwidth $\Delta\nu$ (that ranges between 20 MHz and 486 MHz) of each observation: $f_\mathrm{err}=0.2(\Delta\nu_\mathrm{ref}/\Delta\nu)^{1/2}$, with $\Delta\nu_\mathrm{ref}=32$ MHz. We took the PA uncertainty to include both the statistical and systematic error contributions.
The inner jet PA evolution obtained from the Gaussian component J1 and and from the clean components in annulus A are almost identical (see Figure \ref{PA_three_components_Gaussian}).
 The upper panel of Figure \ref{fig3}(a) shows the result for the annular bin method, presenting drastic changes in the PA over time. From 1998 to 2004, the PA decreases gradually, and it increases afterwards until 2013.
  It decreases once again until 2017 and  increases until 2020, which is the last epoch observed prior to a four-year gap.
  
The variations observed in the inner jet PA yield evidence of a periodicity, which we confirmed using three periodicity search methods (details in Appendix \ref{app:periodicity_methods}). We applied these methods to the annulus A PA data. For all the methods, the standard deviation on the periodicity was obtained from the FWHM of the peak found in the period domain.
Firstly, we applied the well-established Jurkevich method \citep{Jurkevich1971}. We detected a clear minimum for a period of approximately 12 years, which reaches well below the threshold typically used to determine significance, hinting at a reliable periodicity signal. The drop in the total variance is very wide and its gaussian fit indicates a period of $P=14.44\pm5.07$ years. 
Secondly, we used the Lomb-Scargle (LS) periodogram \citep{1976Ap&SS..39..447L, 1982ApJ...263..835S} and obtained a periodicity at $P=10.78\pm2.34$ years, largely above both the $5\sigma$ significance level assuming red noise and the 4e-5 false alarm level (FAL); namely, the level needed to attain 0.004\% false alarm probability (FAP). 
Finally, we apply the weighted wavelet Z-transform (WWZ) method \citep{Foster1996, Grossman1984}.
We find a peak at $P=16.76\pm4.25$ years in the time-shift averaged WWZ power, once again with more than $5\sigma$ significance assuming red noise. We note that, while the WWZ method is characterized by time-frequency resolution and the ability to detect transient quasi-periods, it does not include data error when searching for periodicity. This could explain, in particular, the longer period found with respect to the LS periodogram.
The large standard deviations in all the periodicity search techniques are due to the "short" observational time span with respect to the detected period and to the wide data gaps in the historical PA curve. 
While the possibility of irregular variations cannot be fully ruled out, these independent search methods, which include red noise estimation, all confirm a $5\sigma$ periodicity signal.
Continuous monitoring of 3C\,66A will allow us to better constrain the exact value of the periodicity.

We fit the data to a harmonic function with a global linear trend to characterize the PA evolution \citep[e.g.,][]{MartiVidal2011A&A...533A.111M, 2014MNRAS.445.1370K},
expressed as  \begin{equation}
        \theta = A\sin{\left(\frac{2 \pi}{P}(t-t_{0})\right)}+S(t-t_{0})+\theta_{0}, \label{eq_fit}
  \end{equation} 
  where $A$ is the magnitude, $P$ the period, $t_{0}$ a reference time, $S$ the slope of the linear trend, and $\theta_{0}$ the PA at $t=t_{0}$. 
  We performed a nonlinear least square fit using the Levenberg-Marquart method
  that searches the parameter set where $\chi^2$ is minimized. 
  Table \ref{tab3_PAfit} gives the fitting parameters. 
  The result indicates that the PA changes sinusoidally with a magnitude of $7.77\pm0.79$ degrees and a period of $10.94\pm0.22$ years. 
  Additionally, the mean PA drifts linearly by $-0.83\pm0.07$ degrees per year. 
  As expected, the reduced chi-square $\chi_\mathrm{red}^2$ is $\sim$2.25.
  The residual of the fit is shown in the lower panel of Figure \ref{fig3}(a). Table \ref{tab1} also shows the fitting results for the inner jet PA obtained with the Gaussian component J1, which are in very good agreement with the values from the annular bin approach.
  We therefore extract two evolution timescales from the inner jet PA evolution: the $\sim$11 years sinusoidal motion and a clockwise
shift of $\sim-0.8$ degrees/year .

  The inner jet PA oscillation has not been noticed in previous studies (likely due to relatively short time spans) to the spatial resolution and to variability criteria.
  \citet{2001ApJS..134..181J, 2005AJ....130.1418J, 2017ApJ...846...98J} respectively covered 2 (1995--1997), 3 (1998--2001), and 5 years (2008--2013), making it difficult to detect the oscillation, although they used 43\,GHz VLBA data as in our study.
  3C\,66A was also monitored with the VLBA at 15\,GHz for 10 years (2003--2013) under the monitoring of jets in AGNs with VLBA experiments (MOJAVE) program \citep{2013AJ....146..120L}. 
  However, it remains difficult within this duration to detect the periodicity. While \cite{Lister2019} covered the period 1994--2016, observations at 15\,GHz have three times lower resolution and probe a relatively outer region compared to the 43\,GHz observations. \cite{Weaver2022} presented the results from the 43\,GHz VLBA data between 2007--2018 and estimated the PA evolution of eight knots in 3C\,66A, classifying the behavior as constant overall \citep[][Figure 6]{Weaver2022}. This is related to the criterion chosen for the definition of constant evolution and to the time span of $\sim$11 years. We note that the evolution of their knot A2, which corresponds to our inner jet component (J1 or annulus A), presents a PA increase peaking around 2013 and then a decrease until 2017, in agreement with part of the second period that we observe. The difference between component A2's minimum and maximum PA is slightly more than 20 degrees, which matches the amplitude of the PA sinusoidal fit that we obtain.
  Recently, \cite{Kostrichkin2025MNRAS.537..978K} analyzed the inner jet direction of 3C\,66A as one of the AGNs for which the PA change was studied with an automated algorithm. From 43\,GHz observations spanning slightly more than a period, the authors find a similar oscillation as the one reported here, although the data is fitted with a linear regression. The same is done for the 15\,GHz PA data, that is also analyzed over approximately 10 years and which reports less variability.  
  In comparison, our analysis covers approximately 30 years with high resolution (typical beam size of 0.4 $\times$ 0.2\,mas), so that we can detect nearly three periods of the PA swing in 3C\,66A.

  On top of the evolution of the PA for the inner jet, we study the PA evolution of two components further from the core (details in Appendix \ref{app:PA_components}). As described above, the PA of the additional components is also estimated using two methods and their resulting evolution is equivalent (the results are shown in Figures \ref{PA_three_components} and \ref{PA_three_components_Gaussian}). The second component is part of the core-jet structure that extends southwards over $\sim$1\,mas: Gaussian component J2 or annulus B. We call it downstream component, since it is found south of the inner jet one. The third component is the separated region found at $\sim$2.5\,mas from the core: Gaussian component J3 or annulus C. The position of these components can be seen clearly in Figure \ref{components}. 
   The downstream component presents a PA swing apparently similar to that of the inner jet one. We fit the PA evolution with Eq. \ref{eq_fit} and find a similar period of $10.50\pm0.29$\,yr topped with a slightly slower clockwise shift of $-0.68\pm0.06$\,degrees/year from the annular bin estimate. Moreover, the amplitude of the oscillations is significantly reduced to $4.35\pm0.56$\,degrees and the sine is phase-shifted by $\sim$1.5\,years compared to the inner jet oscillation. The fit to the PA evolution of Gaussian component J2 leads to a similar result as obtained for annulus B. In the case of the third component, there is no periodicity in the PA evolution, but it follows a decreasing linear trend. The annulus C PA yields a continuous shift of $-0.31\pm0.04$\,degrees/year, identical to the Gaussian J3 PA with a value of $-0.31\pm0.03$\,degrees/year. The details of the fits to the PA can be found in Table \ref{tab3_PAfit}. It appears that the components studied all undergo a long-term clockwise PA shift, which is strongest for the inner jet, weaker for the downstream jet, and less than half the rate for the furthest component.

\subsection{Flux density variability} \label{fluxvariability}

   \begin{figure}
     \begin{minipage}{\linewidth}
      \centering
      \includegraphics[width=\linewidth]{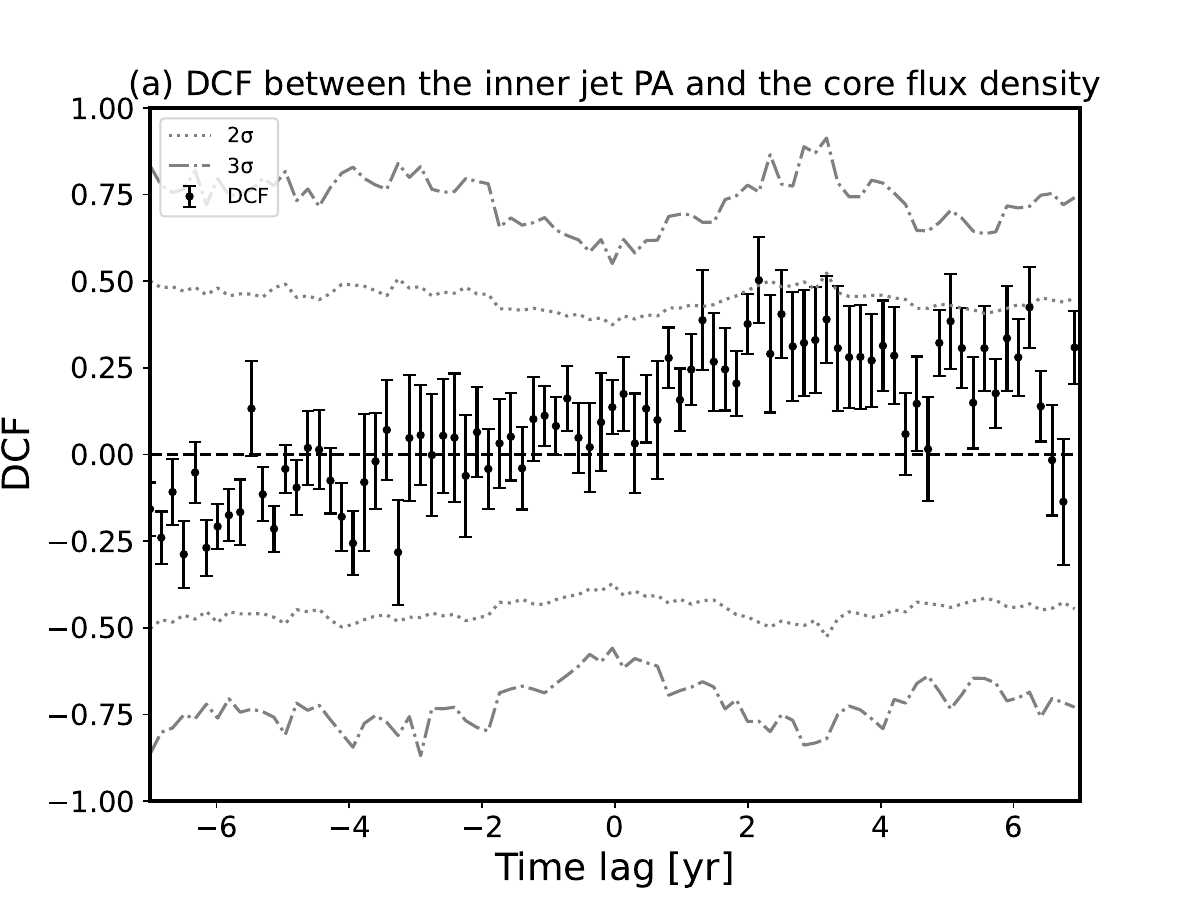}
     \end{minipage}
     
     \begin{minipage}{\linewidth}
      \centering
      \includegraphics[width=\linewidth]{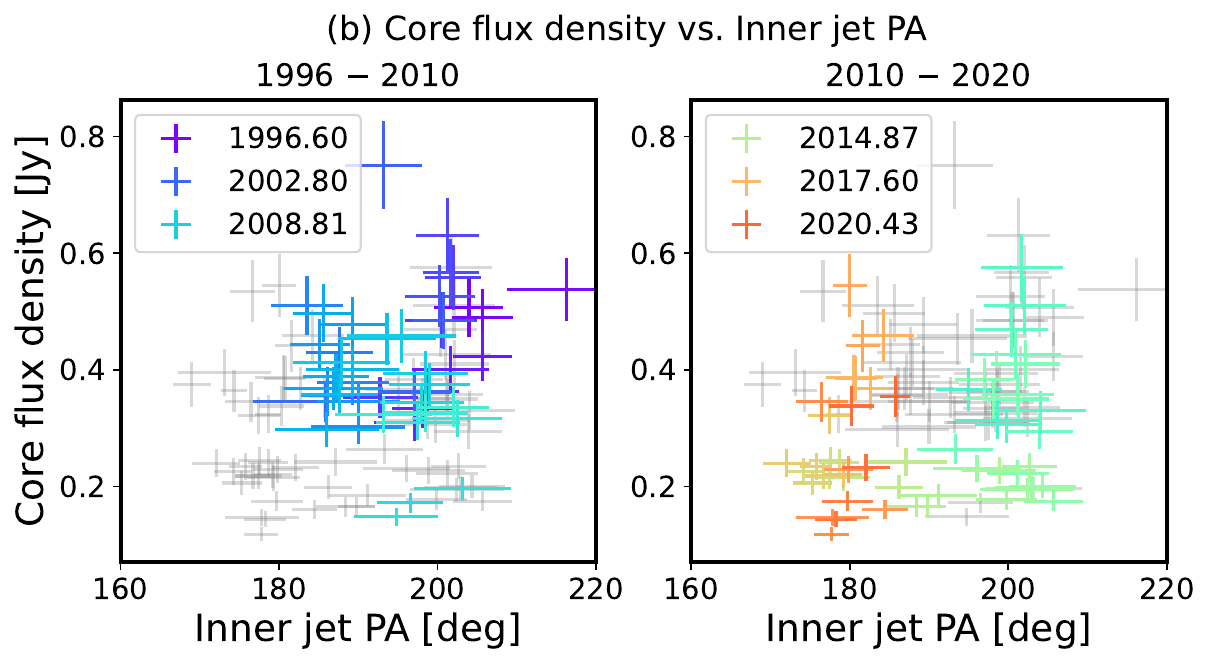}
     \end{minipage}
     \caption{(a)  DCF analysis between the PA and the core flux density. 
     The grey dotted and dash-dotted lines show the 2$\sigma$ and 3$\sigma$ significance levels, respectively. 
     (b) Core flux density vs. inner jet PA plot. Each panel displays the data between 1996 and 2010 (left) and between 2010 and 2025 (right). 
     We use changing colors to show the time sequence: grey denotes the values from all observing epochs, purple to dark blue to clear blue denotes the data observed between 1996, 2002, and 2010, while green to orange to red between 2010, 2017, and 2020.}
         \label{fig5}
   \end{figure}
 
  To compare the PA evolution with the radio flux density, we consider the sum of all clean components' flux density as the total flux density, and the sum of clean components within a radius of 0.20\,mas as the core flux density. The latter is in excellent agreement with the flux density obtained from the fitted circular Gaussian core, as well as with the peak flux density of the convolved map. We adopt one-tenth of the flux density as its error.
  
  Figure \ref{fig3}(b) shows the total and core flux density over time. The total flux density varies between 0.2 and 1.1\,Jy. Except for the overall higher flux, its evolution appears identical to the core flux density's, as it is core-dominated. The major features in the flux density variability therefore originate at the base of the jet.
  The outbursts in the flux can be associated with either the emergence of VLBI components or with geometric processes as a result of Doppler boosting.
  
  If the latter is the case, it should be manifested as the correlation between the core flux and the jet axis direction.
  To check the correlation between the core flux density and the inner jet PA, we apply the discrete correlation function \citep[DCF]{1988ApJ...333..646E}. We probe the time lags between $-7$ and $+7$ years. The time-lag bins are set as 62 days which is the median cadence of the observations. To test the significance of the correlation in the DCF analysis, we simulate 10~000 LCs having the same power spectral density (PSD) and probability distribution function (PDF) as those of the observations \citep{2013MNRAS.433..907E}.

  Figure \ref{fig5}(a) shows the DCF results.
  The grey dotted and dash-dotted lines mark the 2~$\sigma$ and 3~$\sigma$ significance levels, respectively. 
  The correlation coefficient between the inner jet PA and core flux density slightly exceeds 2~$\sigma$ at a positive time lag of $\sim$2 years, indicating a weak but possible correlation with PA changes leading the core flux density changes.
  The correlation is also hinted at by the behavior of the core flux density versus PA (Figure \ref{fig5}(b)). 
  The left and right panels of the plot show the results derived from the observations in 1996-2010, and 2010-2020, respectively (we leave the five 2024-2025 epochs in gray). Each panel shows crescent-like patterns for the two subdivided datasets that match the inner jet PA decrease-and-increase repeated twice, corresponding approximately to the two periods observed in the jet orientation evolution until 2020.
  This pattern could be the indication of a potential correlation, as argued by \cite{2014A&A...571L...2R} for the BL Lac object 0716+714. 

  Motivated by the optical flux periodicity detected in a number of studies on 3C\,66A \citep{Fan2002, Belokon2003, Kaur2017, Fan2018, OteroSantos2020, Cheng2022} and by the apparent high variability of the radio LC, we also perform a periodicity search for the core flux density using the three methods mentioned in Section \ref{orientation} (details in Appendix \ref{app:PA_components}). The analysis with the Jurkevic method yields a peak at $3.53\pm0.25$ years, at $6.92\pm0.21$ years and at $10.51\pm0.56$ years that reach the $F=0.5$ threshold, hinting at a strong periodic signals. The LS periodogram results indicate a peak at $6.22 \pm 0.85$ years which attains the 3$\sigma$ significance level considering red noise. Two other peaks, at $3.43\pm0.30$ years and $2.38\pm0.15$ years, reach the 2$\sigma$ significance level. Using the  WWZ method, we find two peaks in the time-shift averaged power that reach the $2\sigma$ significance level. The longest period is  $\sim$10 years and merges with the $\sim$7 years peak. We also find a stronger signal reaching the 3$\sigma$ significance level and which is a better-defined Gaussian peaking at $4.98\pm0.49$ years. Taken together, these results indicate three possible periodicities in the radio LC of 3C\,66A: at $\sim$10 years, $\sim$7 years, and $\sim$3-4 years. Each periodicity may not be exactly sinusoidal, despite the detection with methods such as the LS periodogram. For instance, the radio flux may be characterized by repetitive flares as has been observed in several sources \citep{Jiang2022arXiv220111633J, Britzen2023ApJ...951..106B}.
  
  In addition to the high variability of the core and total radio flux densities, we draw attention to their overall decrease, in particular between 1996 and 2020. The baseline is clearly decreasing over time, with a rate of $-9.40\pm1.43$\,mJy/year, obtained from a linear fit of the core flux density.

\section{Jet precession model} \label{modeling_precession}

\subsection{Case for two precession timescales}
\label{modeling_precession_timescales}

Radio-interferometric monitoring studies of AGNs find an increasing number of sources showing jet PA variations \citep[e.g.,][and reference therein]{2013AJ....146..120L, Lister2021ApJ...923...30L, Cui2021RAA....21...91C, Kostrichkin2025MNRAS.537..978K}, often accompanied by flux periodicity at one or more wavelengths \citep[e.g.,][]{2014MNRAS.445.1370K, 2018MNRAS.478.3199B, Dey2021MNRAS.503.4400D, Jiang2023}. While several possible scenarios to trigger this phenomenon have been suggested, jet precession is the most frequently considered model to explain periodic PA behavior \citep[e.g.,][]{MartiVidal2011A&A...533A.111M, 2015A&A...578A..86R, 2017ApJ...851L..39C, 2019A&A...621A..11Q, 2018MNRAS.478.3199B, Cui2023, Jiang2023}. 
The model predicts that newly-born components close to the core orient at a different PA from older components farther away from the core, resulting in a bent jet as witnessed in 3C\,66A. \cite{Fendt2022ApJ...933...71F} show with 2D (M)HD (magnetohydrodynamic) simulations that precessing nozzles lead to S-shaped jets, similar to the 3C\,66A twisted jet structure revealed by our KaVa observations. We believe that jet motion in 3C\,66A is governed by two precession timescales: the first associated to the $\sim$11 year PA oscillation and the second to its slow re-orientation. 

First, regarding the decade-long precession, the wide range of apparent speeds and the variability of the core flux density in 3C\,66A could be explained as a result of the variation in the Doppler factor depending on the viewing angle.
The possible correlation between the PA change and the core flux variability discussed in Section \ref{fluxvariability} is consistent with this argument. In addition, \cite{Mohana2021MNRAS.507.3653M} find a clear baseline flux decrease in the Fermi $\gamma$-ray LC of 3C\,66A occurring in mid-2011. Their results show a persistent high-baseline flux for 3 years beforehand and a persistent low-baseline one during the subsequent 8 years. They find evidence of a similar baseline flux drop in the optical band. To explain these baselines' variations the authors argue in favor of Doppler factor or jet inclination change. The flux drop date coincides with the second peak that we observe in the inner jet PA. Given the observational timescale analyzed by \cite{Mohana2021MNRAS.507.3653M}, the 3 years of high baseline match the period in which the PA changes westward to an average of 200$\degree$, while the low baseline matches the PA southward shift reaching 180$\degree$. We consider this coincident multi wavelength flux drop an additional argument in favor of jet precession for the decade-long PA oscillation. 
A similar relation between flux density and jet orientation change was reported, for example, by \cite{MartiVidal2011A&A...533A.111M} in the case of M81, as evidence of a precessing jet. More recent data continue to support the precession model for this source \citep{Jiang2023}. 

Second, the PA evolution in 3C\,66A presents a superposed long-term shift of approximately $-0.8\degree$ per year. This slow re-orientation could constitute a fraction of a longer periodic variation in precession models where two periods coexist. Indeed, this slow PA re-orientation, whether positive or negative (anti-clockwise or clockwise), has been observed in other sources and has been modeled as a precession occurring over periods of the order of magnitude of $10^3$ years \citep[e.g.,][]{2005A&A...431..831L, Jiang2023}. In our case, the continuous decrease of the radio core flux density also supports the precession origin for the long-term PA shift. Comparing the core baseline flux decrease to the inner jet PA evolution, we observe approximately a 10\,mJy flux decrease for a re-orientation of $-0.8\degree$. 
We chose not to model this long-term precession, given the much longer timescale on which it occurs compared to our observation time span. Indeed, we expect the period to be of approximately 1000 years based on comparison with \cite{2005A&A...431..831L}. We also performed a sinusoidal fitting of the inner jet PA with a long-term sine.  While a wide range of parameters allow to approximately recover the $-0.83\degree$/year continuous shift over the 29 years of our time span, the period must fall approximately between 500 and 1000 years.

\subsection{Model parameters and uncertainties}
\label{modeling_precession_MCMC}

\begin{table}
\caption{\centering Parameters of the precessing jet in 3C\,66A.}
\centering
\begin{tabular}{ p{4.4cm} p{0.3cm} p{2.2cm} p{0.5cm}  }
\noalign{\smallskip}
            Parameter, variable & & Value & Unit \\
            \noalign{\smallskip}
     \hline
     Half-opening angle of the cone, & $\Omega$ & 0.47 $\pm$ 0.17 & $\degree$\\
     Precession period, & $P_\mathrm{p}$ & 10.88 $\pm$ 0.24 & yr\\
     Reference time, & $t_0$ & 2016.80 $\pm$ 0.14 & yr\\
     Angle between the precession cone axis and the line of sight, & $\phi_0$ & 3.32 $\pm$ 1.18 & $\degree$\\
     Projected angle of the cone axis onto the sky plane & $\theta_0$ & 185.19 $\pm$ 0.51 & $\degree$\\
     \hline
\end{tabular}
\label{tab_MCMC_parameters}
\end{table}

\begin{figure}
   \centering
   \includegraphics[clip=true,width=\linewidth]{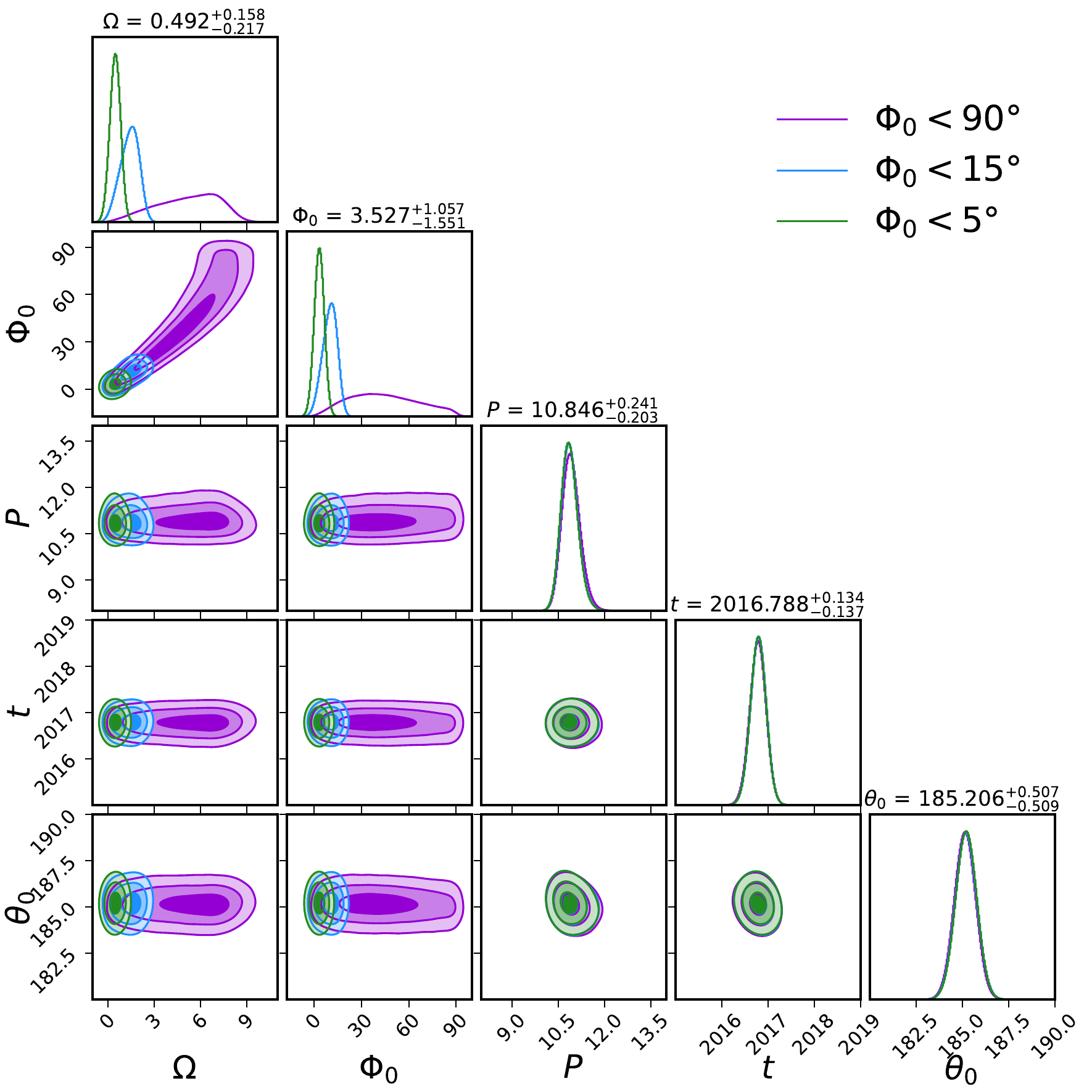}
      \caption{Corner plots of the likelihood analysis for the precession model given three ranges of $\phi_0$. The parameter values shown are the median and 1$\sigma$ credible intervals of the posterior samples.}
    \label{fig7_MCMC_corner_plots}
\end{figure}

We propose a simple precessing jet model to reproduce the approximate 11\,year period oscillation in the PA of 3C\,66A and use the value of $-0.83\degree$/year obtained from the sinusoidal plus linear fit of the PA (from the annular bin method) to account for its long-term decrease. With the reduced number of parameters, we obtain in this way a better constraint from the data.
The precession model consists of a solid cone of half-opening angle $\Omega$, precessing with period $P_\mathrm{p}$ (angular velocity $\omega=2\pi/P_\mathrm{p}$) about the precession cone axis directed along $z_\mathrm{p}$. We refer to Figure 11 in \cite{2018MNRAS.478.3199B} for an equivalent geometrical scheme. In order to express the PA $\theta$ in the observer reference frame (Cartesian coordinates $x$, $y$, $z$) as the projected jet axis from the jet reference frame, we must perform a series of rotations, as detailed for example in \cite{Cui2023}. The angle between the precession cone axis and the line of sight $z_0$ is denoted $\phi_0$ and the projected precession axis onto the sky plane makes an angle $\theta_0$. At any time $t$, the observed PA is:
\begin{align}
    \theta(t) &=\arctan\left(\frac{y(t)}{x(t)}\right) \label{eq_PA_precession},\\
    x(t) &= A(t)\cos\theta_0 - B(t)\sin\theta_0\text{, }\\
    y(t) &= A(t)\sin\theta_0 + B(t)\cos\theta_0,
\end{align}
where the quantities $A(t)$ and $B(t)$ are given by
\begin{align}
    A(t) &= \cos\Omega\sin\phi_0 + \sin\Omega\cos\phi_0\sin[\omega(t-t_0)],\\
    B(t) &= \sin\Omega\cos[\omega(t-t_0)],
\end{align}
and $t_0$ is the reference time. There are five unknown parameters in the model: $\Omega$, $P_\mathrm{p}$, $t_0$, $\phi_0$ and $\theta_0$. We perform a Bayesian analysis and use the Markov chain Monte Carlo (MCMC) Python module \textit{emcee}\footnote{\url{https://emcee.readthedocs.io/en/stable/user/install/}} \citep{emcee2013PASP..125..306F}. The details of the Bayesian analysis are given in Appendix \ref{app:precession_fitting}. The half-opening angle $\Omega$ of the precessing cone and the angle between the cone axis and the line of sight $\phi_0$ are degenerate, as visible in the purple corner plots shown in Figure \ref{fig7_MCMC_corner_plots}. These correspond to the case $\phi_0\in[0,90]\degree$, which is the expected parameter range when no additional information on the viewing angle $\Theta$ is known. The viewing angle relates to the precession model quantities as
\begin{equation}
    \Theta(t)=\arcsin\sqrt{x(t)^2+y(t)^2}. \label{eq_viewing_angle}
\end{equation}
A constraint on the viewing angle reduces the parameter range for $\phi_0$ and consequently for $\Omega$. Since 3C\,66A is a BL Lac object, we expect the viewing angle to be below $20\degree$. More restrictively, \cite{Weaver2022} find that the viewing angle of BL Lac objects peaks between $2-4\degree$. In the case of 3C\,66A, \cite{2017ApJ...846...98J} find an average viewing angle $\langle\Theta\rangle = 1.7 \pm 0.5\degree$ based on two moving knots. This value has been used for example in spectral energy distribution (SED) modeling by \cite{Mohana2021MNRAS.507.3653M}. \cite{Weaver2022} estimate, for 5 different jet knots in 3C\,66A, viewing angles ranging from $2.6\pm0.7\degree$ to $5.0\pm0.5\degree$. 
Given these constraints, we perform the likelihood analysis for two additional cases: the relaxed constraint case $\phi_0<15\degree$ which is equivalent to $\Theta<18\degree$ approximately (blue plots in Figure \ref{fig7_MCMC_corner_plots}), and the tight constraint case $\phi_0<5\degree$ corresponding to $\Theta<6\degree$ (green plots in Figure \ref{fig7_MCMC_corner_plots}). Such a joint constraint approach based on the viewing angle or inclination of the source has been employed by a number of authors \citep{2014MNRAS.445.1370K, Cui2023, Jiang2023}. With the additional viewing angle prior, we obtain a good fit to the data in both cases for an equivalent chi-square of 3.43. This is the same value as obtained when fitting the observed PA to a simple sinusoidal plus linear shift, showing that the precessing jet model fits the data as successfully as possible assuming the PA behavior is sinusoidal. We expect the deviation from the ideal value of 1 to be caused by fluctuations of the PA that occur on smaller scales than the precession period, typically between days and months, for example because of instabilities in the jet or component ejections \citep{2015AJ....149...46Z, Weaver2022}

We consider the tight constraint case $\phi_0<5\degree$ to be the most physically relevant one, due to the small reported viewing angles for 3C\,66A. We therefore detail the jet precession parameters for the prior $\phi_0<5\degree$ in Table \ref{tab_MCMC_parameters}. They are given as the mean and standard deviation of the MCMC samples.
The precessing cone has a very small half-opening angle and the precession axis is closely aligned to the line of sight, by a few degrees. The period obtained from the MCMC mean is of 10.88 years, which is in particularly good agreement with the periodicity detected with the LS periodogram,
peaking at 10.78 years, and also agrees with the Jurkevic search method within the uncertainties. We note that this period obtained from the MCMC samples does not depend on the prior chosen for the viewing angle.
In the following sections, we take the precession period, $P_\mathrm{p}=10.88\pm0.24$ year, to describe the inner jet PA short-term oscillation.

\subsection{Possible precession-driven Alfvén waves}
\label{modeling_precession_MHDwaves}

Before exploring the origins of the observed precession, we include some considerations on the short-term PA oscillation of the downstream jet. Indeed, a $\sim$10-year period was also measured for the annulus of $0.45<r<1.00$\,mas using the annular binning method, as well as for the second component using the Gaussian fitting method (see Appendix \ref{app:PA_components}). In both cases, the periodic PA pattern shows a reduced amplitude and a phase shift. At the large redshift of 3C\,66A, 1.0\,mas corresponds to approximately 5.0\,pc. If we consider the deprojected distance, taking the approximate viewing angle $\Theta=5\degree$ \citep{Weaver2022}, 1.0\,mas corresponds to $\sim$60\,pc, or more in the case of a smaller viewing angle. Studying the PA change over small annuli along the jet, we determine that the $\sim$10 year PA oscillation is observed up to 1.3\,mas from the core, equivalent to $\sim$80\,pc in deprojected distance. However, reducing the annular size implies that we have less data to measure the PA at each epoch and we cannot fit the periodic behavior reliably. We therefore have two confident measures of the oscillation pattern only: at 0.3\,mas and 0.7\,mas from the core on average. This makes it challenging to interpret the change in this pattern with distance from the core. 

Nonetheless, the reduced PA oscillation amplitude with $r$ implies that the pattern is not ballistic. 
Regarding the phase-shift, we consider two cases. First, it may be caused by a complex geometrical structure of the jet, in which case the PA oscillation in the downstream jet directly translates the precession motion of that region. 
Second, we can interpret the phase-shift as a signature of propagation of the periodic PA pattern. If we assume the latter, we can take the shift to be $\Delta\tau_\mathrm{shift}\approx1.5$ years between the inner and downstream jet regions ($0.3$ and $0.7$\,mas from the core, respectively). The pattern appears to be a downstream-propagating wave-like disturbance with a rough speed of $0.3$\,mas/year, which corresponds to the apparent speed $\beta_\mathrm{wave}^\mathrm{app}\approx5$. Here, we note the similarity with the transverse patterns propagating superluminally on the jet ridge line of BL Lacertae, observed by \cite{Cohen2015ApJ...803....3C} from 15\,GHz MOJAVE data. The authors interpret this motion as Alfvén waves that propagate downstream, assuming a well-ordered helical magnetic field. They moreover find a correlation between the waves' timing and direction and the swinging of a recollimation shock, located at approximately 0.34 pc (projected). This leads the authors to argue that the swinging of the shock component at the jet base excites the MHD waves, which propagate in a way resembling "waves on a whip" along the jet. The approximate propagation speed estimated for the periodic PA pattern in 3C\,66A is similar to that of the transverse waves in BL Lacertae. In addition, \cite{Cohen2015ApJ...803....3C} considered the transverse jet motion over a distance of approximately 13 pc deprojected, which is of the same order of magnitude as the one probed by our inner and downstream jet measurements. The periodic PA pattern observed in the downstream jet of 3C\,66A may therefore be the signature of an Alfvén wave excited by the precession of the jet nozzle and propagating on the longitudinal component of the magnetic field. 

\section{SMBHB system} \label{modeling_SMBHB}

Jet precession can originate from mechanisms such as the misalignment of the central engine spin axis with surrounding accretion disks by spin-induced disk precession \citep{1975ApJ...195L..65B,2004ApJ...616L..99C}, from disk instabilities \citep{Pringle1997MNRAS.292..136P,Lai2003ApJ...591L.119L}, or from a SMBH companion \citep{Begelman1980Natur.287..307B}. The SMBHB scenario, through the phenomenon of jet precession, has been proposed as an interpretation of the PA periodicity for a large number of sources \citep{Caproni2004ApJ...602..625C, 2018MNRAS.478.3199B, Abraham2018, Dey2021MNRAS.503.4400D, Kun2023MNRAS.526.4698K, OteroSantos2023MNRAS.518.5788O}. The presence of a SMBHB has also widely been brought forward to explain flux periodicity found from radio to high energies, whether accompanied by PA changes or not \citep{Li2022RAA....22e5017L, Kun2022ApJ...940..163K, Mao2024MNRAS.531.3927M}.
Regarding 3C\,66A, the two distinct precession timescales found, as well as the known optical flux periodicity, lead us to argue in favor of the SMBHB model. While it is clear that a SMBHB system can lead to periodic behavior, several mechanisms can explain the observed PA, optical and possible radio periodicities of 3C\,66A when assuming a SMBHB system. We show in Section \ref{modeling_SMBHB_orbital_motion} that the short-term PA period can be associated with orbital motion in a SMBHB system. We then consider two scenarios to explain the long-term PA period: spin-orbit precession in Section \ref{modeling_SMBHB_SO}; and disk-orbit precession in Section \ref{modeling_SMBHB_DO}. 
We discuss the origin of 3C\,66A's flux variability in Section \ref{modeling_SMBHB_flux}. We show that orbital motion and disk-orbit precession in a SMBHB can naturally account for all the observed properties of 3C\,66A. We consider the implications regarding other detection methods for this SMBHB system in Section \ref{modeling_SMBHB_detection}.

\subsection{Orbital motion}
\label{modeling_SMBHB_orbital_motion}

In the case of the PA, the periodicity is associated with a precessing jet both for short-term ($\sim$11 years) and long-term ($\sim$$10^3$ years) periods. 
The $\sim$11 years PA oscillation is only detectable up to approximately 80 pc from the core.
On the other hand, a long-term PA shift is also measured for the furthest component (J3 or annulus C) at approximately 2.5\,mas from the core, so at the deprojected distance of $\sim$150\,pc. It is therefore clear the the decade-scale jet precession affects only a part of the jet close to the core, while the long-term precession appears to act on the full jet. These findings strongly support the fact that the short-term and long-term periodicities have different physical origins. 

We argue that the short-term precession can be associated with the orbital motion of the primary BH in the binary system, while the long-term precession cannot. We can estimate the expected central SMBHB mass $M_8$, in units of $10^8M_{\odot}$, given an orbital period $P$ \citep{Begelman1980Natur.287..307B}: $M_8\approx 18^{-8/5}P^{8/5}q^{-3/5}$, and compare it to the expected mass. Here $q=M_2/M_1$ is the mass ratio of the system, with $M_1$ and $M_2$ the masses of the primary and the secondary BH, respectively. For 3C\,66A, the primary BH mass $M_1$ can be computed from optical variability timescale measurements \citep[e.g.,][]{Kaur2017, Fan2018} that give $M_1=(1.20\pm0.05)\times10^8M_{\odot}$ (see Appendix \ref{app:mass_estimate}). We consider the typical range of mass ratios 30:1$-$3:1 derived by \cite{Gergely2009ApJ...697.1621G} for SMBH mergers. Taking $q=0.18\pm0.15$, we obtain the central BH mass $M=(1.42\pm0.19)\times10^8M_{\odot}$. If the long-term period of at least 500 years, corrected for redshift, corresponds to the binary's orbital period, then $M\approx 2\times10^{10}-4\times10^{11}M_{\odot}$ for $q=0.99-0.01$. These values are two to three orders of magnitude above the expected mass of the central BH in 3C\,66A. They also reach the maximum observable SMBHB mass estimated to be $\sim5\times10^{10}M_{\odot}$ \citep{King2016MNRAS.456L.109K}. It therefore appears highly improbable that the observed long-term precession in 3C\,66A translates orbital motion in the SMBHB system. 
On the other hand, we can assume that the decade-scale precession reveals the orbital motion of the system. The inherent short-term precession periodicity is $P^\mathrm{PA}=P_\mathrm{obs}^\mathrm{PA}/(1+z)=8.12\pm0.18$ years. The estimated SMBHB mass, following \cite{Begelman1980Natur.287..307B}, then lies in the range $M\approx3\times10^{7}-4\times10^8M_{\odot}$ for $q=0.99-0.01$, which is in agreement with expectations for 3C\,66A. 

\begin{table}
\caption{\centering Parameters of the putative SMBHB system in 3C\,66A, with orbital motion and disk-orbit precession.}
\centering
\begin{tabular}{ p{2.7cm} p{0.9cm} p{2.4cm} p{1.3cm}  }
\noalign{\smallskip}
            Parameter, variable & & Value & Unit \\
            \noalign{\smallskip}
     \hline
     Total mass, & $M$ & 1.42 $\pm$ 0.19 & $10^8$ $M_{\odot}$\\
     Mass ratio, & $q$ & 0.18 $\pm$ 0.15 & \\
     Observed period, & $P_\mathrm{obs}$ & 10.88 $\pm$ 0.24 & yr\\
     Orbital period, & $P^\mathrm{orb}$ & 16.72 $\pm$ 0.42 & yr\\
     Separation, & $r_\mathrm{sep}$ & 1.65 $\pm$ 0.08 & $10^{-2}$ pc\\
     PN parameter, & $\epsilon$ & 4.10 $\pm$ 0.58 & $10^{-4}$\\
     Merger timescale, & $T_\mathrm{merge}$ & $\approx$ 934 & Myr\\
     Strain amplitude, & $h$ & $\approx$ 3.93 & $10^{-19}$\\
     \hline
\end{tabular}
\label{SMBHB_parameters}
\end{table}

\subsection{Spin--orbit precession}
\label{modeling_SMBHB_SO}

In this section, we interpret the short- and long-term periods as precession due to orbital motion versus precession induced by a misalignment between the primary BH spin and the orbital plane axis. This is done in analogy to the quasar S5\,1928+738, which presents a similar periodic and monotonic trend in its inclination and PA evolution \citep{2014MNRAS.445.1370K, Kun2023MNRAS.526.4698K}. We take the period of the precession model fit to the inner jet PA to be the observed orbital period, $P_\mathrm{obs}$. The intrinsic binary period is therefore assumed to be $P^\mathrm{PA}=8.12\pm0.18$ yr. 
From the equation of motion of a system with total mass $M=M_1+M_2$, the binary separation can be estimated to Newtonian order \citep{Kidder1995PhRvD..52..821K}:
\begin{equation}
    r_\mathrm{sep} = \left(\frac{P}{2\pi}\right)^{2/3}(GM)^{1/3} \label{SMBHB_radius}
\end{equation}
with $G$ the gravitational constant. Given the central BH mass $M=(1.42\pm0.19)\times10^8M_{\odot}$, the binary separation between the two putative BHs is $r_\mathrm{sep}=(1.02\pm0.05)\times10^{-2}$\,pc $=(2.10\pm0.10)\times10^{3}$\,AU. In gravitational radii $R_\mathrm{g}=GM_1/c^2=5.7\times10^{-6}$\,pc, relative to the primary BH, the separation is $r_\mathrm{sep}=(1.78\pm0.08)\times10^3R_\mathrm{g}$.

The post-Newtonian (PN) parameter, quantifying the validity of the Newtonian approximation, is given by
\begin{equation}
    \epsilon = \frac{GM}{r_\mathrm{sep}c^2}. \label{SMBHB_PN}
\end{equation}
We have $\epsilon=(6.64\pm0.94)\times10^{-4}$ which lies just outside the expected inspiral phase range $0.001<\epsilon<0.1$ \citep{Gergely2009ApJ...697.1621G}.
The coalescence of such a SMBHB is  predicted to occur within the gravitational (merger) timescale following
\begin{equation}
    T_\mathrm{merge}=\frac{5GM}{32c^3\epsilon^{4}}\frac{(1+q)^2}{q}.
    \label{SMBHB_Tmerge}
\end{equation}
For the assumed mass ratio of 3C\,66A, $T_\mathrm{merge}\approx136$\,Myr. 

From the computed PN parameter, we can estimate the spin-orbit precession timescale of the system and establish the validity of this interpretation for the long-term re-orientation of the PA in 3C\,66A. We consider that the precession is induced by the misalignment between the primary's spin $\boldsymbol{S_1}$ and the orbital motion angular momentum $\boldsymbol{L}$. In the expected mass ratio range for merging SMBHs, the spin $\boldsymbol{S_2}$ of the secondary BH can be neglected \citep{Gergely2009ApJ...697.1621G}. Ignoring 2PN effects, of second order $\mathcal{O}(\epsilon^2)$, the primary BH precesses around the total angular momentum vector $\boldsymbol{J}=\boldsymbol{S_1}+\boldsymbol{L}$ with period \citep{2014MNRAS.445.1370K}:
\begin{equation}
    P_\mathrm{SO}= 2\pi\Omega_\mathrm{SO}^{-1}=2\pi\frac{GM}{2c^3}\epsilon^{-5/2}\frac{(1+q)^2}{q}.
    \label{SMBHB_Pso}
\end{equation}
We obtained a spin-orbit precession period $P_\mathrm{SO}\approx47000$ years, which is largely above the expected period obtained from the fit of the slow PA re-orientation. Even considering the extreme case $q=0.99$, the spin-orbit precession period remains an order of magnitude above our expectation. We therefore disfavor spin-orbit precession to account for the long-term jet precession in 3C\,66A.

\subsection{Disk-orbit precession}
\label{modeling_SMBHB_DO}

We  show in Sections \ref{modeling_SMBHB_orbital_motion} and \ref{modeling_SMBHB_SO} that orbital motion can account for the decade-scale jet precession, but that spin-orbit precession is unlikely to drive the long-term PA change in 3C\,66A. Ignoring the LT effect, discussed in Section \ref{discussion_LT_disk}, the only satisfying explanation for the long-term jet precession is that the SMBHB orbital plane is misaligned with the primary BH's accretion disk. In this case, the $\sim$11 year period in the PA of 3C\,66A is expected to result from a combination of orbital motion and disk-orbit precession \citep{Katz1982ApJ...260..780K, Bate2000MNRAS.317..773B}. Indeed, in a binary system the gravitational torque from the secondary BH affects the disk of the primary and induces an additional oscillation in the accretion disk's orientation, that depends both on the precession frequency from the disk-orbit misalignment and on the orbital frequency of the system.
We therefore interpret our short-term precession as an effect of orbital motion and disk-orbit misalignment (also called "wobbling" or "nutation" in the literature), while the long-term precession results from disk-orbit misalignment alone. A similar case is argued for the double periodicities found in the PA evolution of M81 \citep{Jiang2023} or of the famous SMBHB candidate OJ 287 \citep{Katz1997ApJ...478..527K, Britzen2023ApJ...951..106B}.

\cite{Bate2000MNRAS.317..773B} found that the short-term precession, or "wobbling", corresponds to approximately half of the binary period, while the long-term precession should be on the scale of 20 periods. This implies that our observed binary period would be of approximately 20 years and the long-term period of approximately 400 years, which is in agreement with the fit to the PA re-orientation of 3C\,66A. We first check the possibility of disk-orbit precession in the case of 3C\,66A more rigorously, by determining the expected rate of disk precession from a misalignment with the orbital plane \citep{Katz1997ApJ...478..527K}:
\begin{equation}
    \Omega_\mathrm{DO}=-\frac{3}{4}\frac{GM_2}{r_\mathrm{sep}}\left(\frac{r_\mathrm{d}}{r_\mathrm{sep}}\right)^2\frac{\cos(\theta_\mathrm{d})}{(GM_1r_\mathrm{d})^{1/2}}, \label{SMBHB_rate_DO}
\end{equation}
where $r_\mathrm{d}$ is the radius of the accretion disk and $\theta_\mathrm{d}$ the angle between the disk and the orbital plane. \cite{Katz1997ApJ...478..527K} stresses that this corresponds to evaluated torques acting onto the disk and is not expected to match exactly the precession period observed. Nonetheless, the order of magnitude between both should be in agreement. For instance, in the case of the X-ray binary system SS 433, for which this model was applied before SMBHB systems, the derived timescale $P_\mathrm{DO}$ is approximately half of the observed precession timescale \citep{Katz1980ApJ...236L.127K}. We assume that the accretion disk around the primary BH precesses as a rigid body, implying that $r_\mathrm{d}<r_\mathrm{sep}$, and take $r_\mathrm{d}=0.5\times10^{-2}$ pc as roughly half the binary separation. For the disk inclination to the orbital plane, we choose $\theta_\mathrm{d}=20\degree$ in analogy to SS 433. From the separation $r_\mathrm{sep}=(1.02\pm0.05)\times10^{-2}$ pc, obtained by assuming that the observed short-term period translates directly the orbital period, we get $P_\mathrm{DO}\approx200$ years, which is of the right order of magnitude. We therefore consider disk-orbit precession to be a valid mechanism to explain the long-term period observed in 3C\,66A. 

In a system undergoing disk-orbit precession, we expected the orbital period observed to be modulated by the disk-orbit precession period, so that we must refine our SMBHB parameter estimation with respect to Section \ref{modeling_SMBHB_SO}.
We applied the results of \cite{Katz1982ApJ...260..780K} to obtain a more accurate value of the observed orbital period of the system \citep{Caproni2013MNRAS.428..280C, Britzen2023ApJ...951..106B}:
\begin{equation}
    P_\mathrm{obs}^\mathrm{orb}=P_\mathrm{obs}^\mathrm{prec}\left(\frac{P_\mathrm{obs}^\mathrm{prec}}{2P_\mathrm{obs}}-1\right)^{-1}, \label{SMBHB_orbital_period_DO}
\end{equation}
where $P_\mathrm{obs}^\mathrm{prec}$ is the long-term precession period that we obtain from the slow PA re-orientation. We take $P_\mathrm{obs}^\mathrm{prec}= 750\pm250$ years, with 750 the central estimate and 250 the uncertainty reflecting the plausible range $[500-1000]$ years, obtained from the long-term period fit of the PA. Then, Eq. \ref{SMBHB_orbital_period_DO}  yields $P_\mathrm{obs}^\mathrm{orb}=22.41\pm0.56$ years, equivalent to $P^\mathrm{orb}=16.72\pm0.42$ years in the source rest frame. 
Using Eqs. \ref{SMBHB_radius} and \ref{SMBHB_PN}, we determined, in this case, a larger binary separation of $r_\mathrm{sep}=(1.65\pm0.08)\times10^{-2}$\,pc $= (3.4\pm0.2)\times10^{3}$\,AU$=(2.89\pm0.14)\times10^3R_\mathrm{g}$ and a smaller PN parameter of $\epsilon=(4.10\pm0.58)\times10^{-4}$. The latter lies again outside the expected inspiral phase range; at the current stage, the bound system is dominated by dynamical friction. From Eq. \ref{SMBHB_Tmerge}, we also estimate the expected merger timescale, $T_\mathrm{merge}\approx934$\,Myr. Given the new binary separation from the updated orbital period, Eq. \ref{SMBHB_rate_DO} finally gives the disk-orbit precession timescale for the system: $P_\mathrm{DO}\approx846$\,years, which falls perfectly in the range obtained from the fitting of the long-term period.

Since disk-orbit precession implies the misalignment between the disk and orbital motion, we expect that either the primary BH spin is also misaligned with the orbital motion, or that it is misaligned with the accretion disk. The first would result in spin-orbit precession, which in this case would be negligible because occurring on a timescale of $10^5$ years. The second could lead to LT precession, discussed in Section \ref{discussion_LT_disk}. 

The above estimates strongly support orbital motion and disk-orbit precession as the driving mechanisms for the observed short- and long-term precession timescales in 3C\,66A. In the rest of Section \ref{modeling_SMBHB}, we therefore refer to the orbital parameters computed for this model as those of the putative SMBHB system. We summarize these parameters in Table \ref{SMBHB_parameters}.

\subsection{Flux variability}
\label{modeling_SMBHB_flux}

The variability in the flux of 3C\,66A, whether optical or radio, can be caused by geometrical effects in the precessing jet or by flaring from the accretion disk because of perturbed accretion rates due to the secondary BH's influence on the primary's disk. This is, for example, the most supported case to explain the flares observed in OJ 287 \citep{Lehto1996ApJ...460..207L, Dey2018ApJ...866...11D}.

For 3C\,66A, our favored scenario regarding radio and optical flux variability is Doppler boosted emission from the precessing jet. This is indeed one of our motivations to fit the PA change with a precession model in Section \ref{modeling_precession}. The possible correlation between the radio core flux density and the inner jet PA supports this theory, as well as the periodicity timescales themselves \citep{Rieger2004ApJ...615L...5R}. The optical period $P_\mathrm{obs}^\mathrm{optical}=2.38\pm0.19$\,years \citep{Cheng2022} of 3C\,66A also motivates the jet-boosted emission scenario.
Indeed, if we assume that the optical flux variability is dominated by the jet rather then the accretion disk and is the result of Doppler boosting, the intrinsic period is $P^\mathrm{optical}=\delta P_\mathrm{obs}^\mathrm{optical}/(1+z)=7.92\pm3.31$\,years, given the Doppler factor $\delta=4.46\pm0.19$ obtained from observations (see Appendix \ref{app:mass_estimate}). We find it interesting that the value of $P^\mathrm{optical}$ matches the intrinsic precession period $P^\mathrm{PA}$ resulting from the short-term PA oscillation. 

The flux variability may also occur due to modulation of the disk's accretion rate, leading to flaring events \citep{Gold2014PhRvD..89f4060G}. In this case, the periodicity may or may not need to be corrected for Doppler boosting: if the flaring comes from an inner disk region subjected to relativistic effects, some boosting may occur, but if it comes from a standard Shakura-Sunyaev accretion disk region it is negligible \citep{Shakura1973A&A....24..337S}. In either case, the mass accretion rate can increase due to a SMBHB once, twice or more times per orbit \citep{Hayasaki2013PASJ...65...86H}. Given our knowledge of the source, it is difficult to link the mass accretion variability to orbital parameters if we assume this origin for the optical flux periodicity. In the context of disk-orbit precession, considering that the optical flux originates from a standard, isotropic disk, the $\sim$2 year period would have to correspond to enhanced mass accretion some ten times per binary orbit, which appears unlikely. 

We have shown that Doppler boosted emission from the jet, which precesses due to orbital motion, can explain the observed optical periodicity of 3C\,66A. The correlation between the radio core flux density and inner jet PA change supports a similar origin for the radio variability. While accretion disk modulation may contribute to the flux variability in 3C\,66A, we do not believe it is a dominant effect.

\subsection{Implications regarding other detection methods}
\label{modeling_SMBHB_detection}

Inspiraling SMBHBs lose energy through GW emission that may be detected through techniques such as pulsar timing array (PTA) or the Laser Interferometer Space Antenna (LISA) in the upcoming decade \citep[e.g.,][]{Bi2023SCPMA..6620402B, EPTA2024A&A...685A..94E, Agazie2024, Sah2025ApJ...993..118S, AmaroSeoane2023LRR....26....2A, deBruijn2020ApJ...905L..13D}. Based on the SMBHB parameters obtained for 3C\,66A, we can estimate the strain amplitude for this system. The computed central mass, $M$, and assumed ratio, $q$, yield the chirp mass, $M_\mathrm{chirp}=M^{9/5}\left[\frac{q}{(1+q)^2}\right]^{3/5}=(4.18\pm1.52)\times10^7M_{\odot}$.
The strain amplitude is then \citep{Taylor2016, Feng2019}:
\begin{equation}
    h=\frac{2(GM_\mathrm{chirp}^{5/3}(2\pi f)^{2/3}}{c^4D_\mathrm{L}},
\end{equation}
where $f=1/P$ is the orbital frequency in the source rest frame and $D_\mathrm{L}=1847.9$ Mpc is the luminosity distance. This distance is computed with the \textit{CosmoCalc}\footnote{\url{https://www.astro.ucla.edu/~wright/CosmoCalc.html}} tool \citep{Wright2006}. The resulting strain amplitude is of the order of $10^{-19}$.
Given the frequency and strain estimated for the SMBHB in 3C\,66A, no current PTA collaboration has the required sensibility to detect GW from this system \citep[see Fig. 11 in][]{DOrazio2023SMBHBinaries}. Nonetheless, we encourage PTA investigation of 3C\,66A in order to place stringent upper limits on the chirp mass of the putative SMBHB system. 

In addition to GWs, detecting the orbital motion of SMBHBs is a very promising method to directly test the presence of these systems at the center of galaxies. The continuous increase in sensitivity and resolution of VLBI arrays and subsequent data analysis prone us to believe that firm detections of SMBHB motions will soon be possible \citep{An2018RaSc...53.1211A, Zhao2024ApJ...961...20Z, Gurvits2025A&A...700A.168G}. 3C\,66A is a bright source and the orbital period of its putative SMBHB is of the order of decades, making it a serious candidate for orbital motion detection with VLBI.
Interestingly, 3C\,66A is very close in the sky to the radio galaxy 3C\,66B ($z = 0.0213$), which is a famous candidate SMBHB due to a 1-year core orbital motion found by \cite{Sudou2003}. This motion was obtained using 3C\,66A as the stationary position reference after observing both sources simultaneously in each beam at 2.3 and 8.4\,GHz. We note that there is a non-negligible possibility that the motion attributed to the core of 3C\,66B is at least partially the result of core motion in 3C\,66A.
In particular, at lower frequencies the observed core is most likely blended with the inner jet, which was found at higher frequencies in this work.
Although \cite{Iguchi2010} further support the SMBHB hypothesis for 3C\,66B with a periodicity found at mm wavelengths, no other evidence of a SMBHB has been found. Moreover, PTA measurements refuted the presence of a SMBHB in 3C\,66B for the parameter range derived by \cite{Sudou2003} \citep{Jenet2004}. It should be noted, however, that later PTA studies put an upper limit on the chirp mass of the system that is consistent with the one refined by \cite{Iguchi2010} \citep{Feng2019, Arzoumanian2020, Arzoumanian2023, Agazie2024, Tian2025arXiv250814742T}. If the detected orbital motion corresponds to the core of 3C\,66B, this near-merger SMBHB would have an extremely short merger timescale (of the order of years to thousands of years) compared to the typical life-time of these systems, meaning that we caught it at a very rare moment. We believe it possible that 3C\,66B hosts a SMBHB, but the strong evidence found for the candidate 3C\,66A motivates further core motion observations of each source using different references. 

\section{Discussion} \label{discussion}

We have interpreted the twisted jet, PA change, and flux periodicity in 3C\,66A with jet precession caused by a SMBHB, since this scenario can easily explain all of the observed features. However, other scenarios could explain some of the properties listed in Section \ref{results}. Here, we review additional mechanisms that could account for at least part of the features observed in 3C\,66A.

\subsection{Lense-Thirring disk precession}
\label{discussion_LT_disk}

The LT effect corresponds to frame dragging inducing the precession of particles orbiting in misalignment with respect to the equatorial plane of a Kerr BH \citep{Lense1918}. Equivalently, in a single Kerr BH system where the viscous accretion disk is misaligned with the spin axis of the BH, frame-dragging effects combined with the internal viscosity lead to precession of the accretion disk (also known as Bardeen-Petterson effect) that can be modeled as a rigid body 
\citep{Bardeen1975, Nelson2000}. This precession can be extended to the entire disk-jet system, as shown by 3D GRMHD (general relativistic MHD) simulations \citep{Liska2018MNRAS.474L..81L}. Through the Bardeen-Petterson effect, the accretion disk axis and the BH spin are progressively forced to align.
The LT effect is claimed to be responsible for the jet precession observed in a number sources, such as 3C 454.3 \citep{Qian2014RAA....14..249Q} or M87 \citep{Cui2023}. We investigate whether the LT effect leading to the precession of the disk may cause the observed short and or long-term jet precession in 3C\,66A.

We consider a BH with dimensionless parameter $-1\leq a \leq 1$ such that its angular momentum is $|\boldsymbol{J}|=GM^2|a|/c$. In this part of our study, we take the single BH mass $M$ to be the primary BH mass of $M_1=(1.20\pm0.05)\times10^8M_{\odot}$ derived in Section \ref{modeling_SMBHB} from variability timescales.
Following the approach by \cite{Caproni2004_LT}, the inner radius of the accretion disk is assumed to be the radius of the marginally stable orbit, expressed as $R_\mathrm{ms}=\xi_\mathrm{ms}R_\mathrm{g}$, where $R_\mathrm{g}=GM/c^2$ is the gravitational radius and the dimensionless variable, $\xi_\mathrm{ms}$, is
\begin{align}
    \xi_\mathrm{ms} &= 3+A_2\mp[(3-A_1)(3+A_1+2A_2)]^{1/2},\text{ where}\\
    A_1 &= 1+(1-a^2)^{1/3}[(1+a)^{1/3}+(1-a)^{1/3}] \text{ and}\\
    A_2 &= (3a^2+A_1^2)^{1/2}.
\end{align}
The $\mp$ sign in the $\xi_\mathrm{ms}$ definition is $-$ for prograde motion and $+$ for retrograde motion. The outer radius of the precessing disk is dimensionlessly given by $\xi_\mathrm{out}=R_\mathrm{out}/R_\mathrm{g}$, in which case some choice on the value of this ratio must be made. The viscosity in the disk is modeled through the disk surface density, that we take to be either a power-law $\Sigma_\mathrm{d}(\xi)=\Sigma_0\xi^\mathrm{S}$ \citep{Nelson2000} or an exponential $\Sigma_\mathrm{d}(\xi)=\Sigma_0e^\mathrm{\sigma\xi}$ \citep{DAngelo2003ApJ...599..548D}, where $\Sigma_0$, $s$ and $\sigma$ are constants.

\begin{figure}
   \centering
   \includegraphics[width=\linewidth]{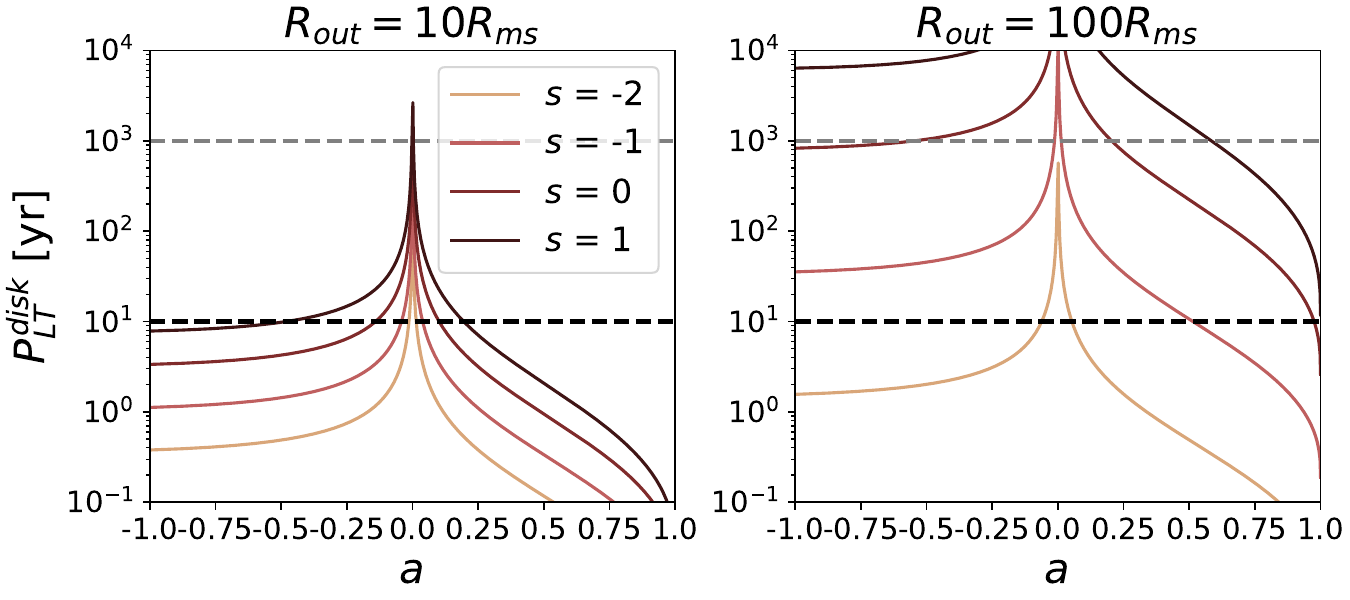}
   \includegraphics[width=\linewidth]{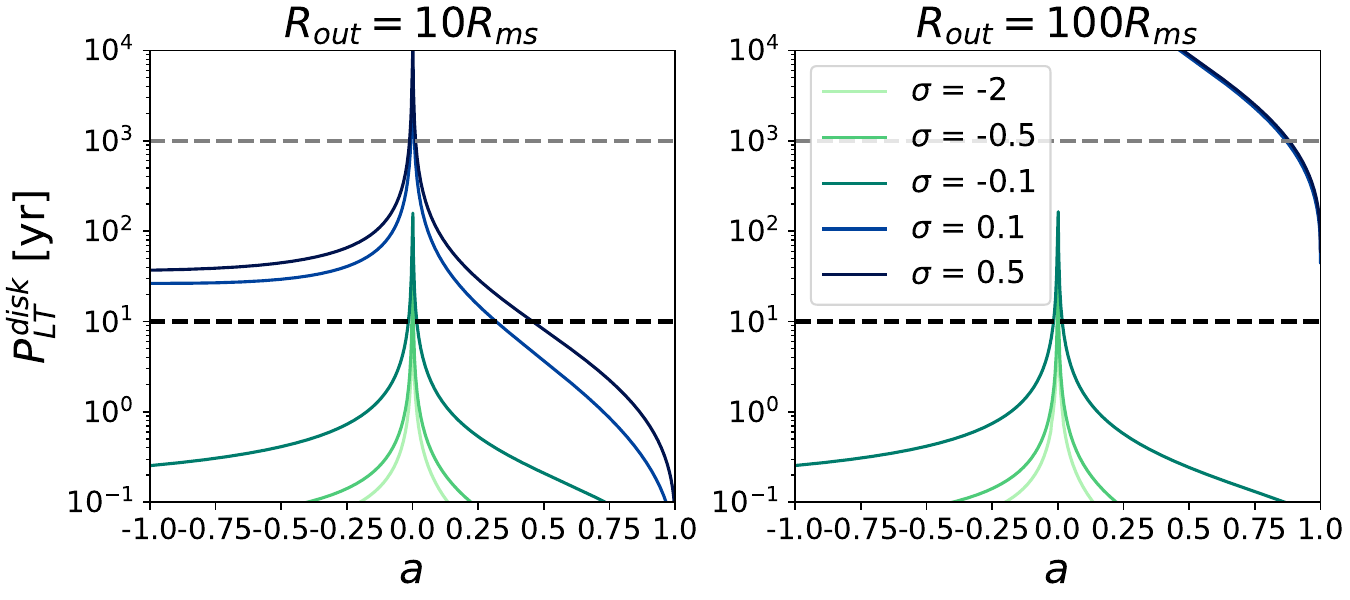}
      \caption{Period of LT disk precession as a function of the spin parameter $a$, outer radius $R_{out}$ and density power law slope $s$ (upper panel) or density exponential factor $\sigma$ (lower panel). The black dashed line corresponds to $P_\mathrm{LT}^\mathrm{disk}=10$ years and the gray one to $P_\mathrm{LT}^\mathrm{disk}=1000$ years.}
    \label{fig8}
\end{figure}

The precession period of the rigid disk is then
\begin{equation}
    P_\mathrm{LT}^\mathrm{disk} = \frac{2\pi GM}{c^3}\frac{\int_{\xi_\mathrm{ms}}^{\xi_\mathrm{out}}\Sigma_\mathrm{d}(\xi)[\Upsilon(\xi)]^{-1}\xi^3d\xi}{\int_{\xi_\mathrm{ms}}^{\xi_\mathrm{out}}\Sigma_\mathrm{d}(\xi)\Psi(\xi)[\Upsilon(\xi)]^{-2}\xi^3d\xi}
,\end{equation}
with $\Upsilon(\xi)=\xi^{3/2}+a$ and $\Psi(\xi)=1-(1-4q\xi^{-3/2}+3a^2\xi^{-2})^{1/2}$.
In the case of a power-law surface density, we vary the power-law slope from $s=-2$ to $s=1$, fixing $\Sigma_0=1$. The resulting period as a function of $a$ is given in Figure \ref{fig8} (upper panels). For the exponential surface density case, we present the results for $\sigma=-2$ to $\sigma=0.5$ (lower panels). We have explored the range from $R_\mathrm{out}=10R_\mathrm{ms}$ to $R_\mathrm{out}=1000R_\mathrm{ms}$, but only show the most relevant curves corresponding to the cases $R_\mathrm{out}=10R_\mathrm{ms}$ (left panels) and $R_\mathrm{out}=100R_\mathrm{ms}$ (right panels).

First, we consider the short-term PA oscillation of $\sim$11 years. If we assume a power-law surface density, the right timescale for the LT precession can be achieved in a disk where the precessing ring has a small outer radius ($R_\mathrm{out}=10R_\mathrm{ms}$) if the BH spin is $-0.2<a<0.15$ approximately, for an expected disk surface density power law slope lower than 1. If the surface density must decrease with radius, then for a disk with such a small precessing ring the central BH must have a spin $|a|<0.05$ approximately. This allowed spin range decreases quickly with $s$, so that in the case of a steep power-law $s=-2$, the BH spin must be almost null. Conversely, if we allow for a larger part of the disk to precess ($R_\mathrm{out}=100R_\mathrm{ms}$ or more) then an $\sim$11 year LT precession period can be obtained for a wide range of negative values of $s$ and non-null BH spin, which better agrees with our expectations for an accretion disk. If we assume an exponential function for the surface density, it seems less likely that the LT disk precession can account for the short-term PA periodicity. Indeed, considering $\sigma<0$ implies that the BH spin is practically null in any disk configuration, whether $R_\mathrm{out}$ is small or large. However, if the disk surface density satisfies an exponential increase with radius, then the right timescale can be achieved for $R_\mathrm{out}\approx10R_\mathrm{ms}$ and $a>0$. 

Second, we investigate the $\sim$1000 year period that we expect from the long-term PA re-orientation. For a power-law surface density, whether the slope $s$ is negative or positive, the precessing part of the disk must have an outer radius larger than approximately $30R_\mathrm{ms}$ in order to have a spinning BH with $|a|\geq0.1$. To have a decreasing surface density in the disk, for instance $s=-1$, $R_\mathrm{out}$ must reach the order of $100R_\mathrm{ms}$. The value of $R_\mathrm{out}$ must be even larger for a steeper power-law profile. In the case where the surface density follows an exponential function, even with $R_\mathrm{out}>100R_\mathrm{ms}$ the period of $\sim$1000 years can only be obtained for positive values of $\sigma$.

We have shown that LT precession of the disk can account for the observed PA oscillation on short or on long timescales. In addition, there is a wider acceptable parameter space to match the observed periods when assuming that the disk surface density is governed by a power-law function. We note, however, that the $\sim$11 and $\sim$1000 year oscillations can only be reproduced for distinct assumptions regarding the size of the precessing ring in the disk and the BH spin. Therefore, if 3C\,66A hosts a single SMBH, then LT precession of the disk can either be the cause of the fast PA oscillation or of its long-term re-orientation, but cannot account for both. 

\subsection{Disk-induced BH precession}
\label{discussion_LT_BH}
\begin{figure}
   \centering
   \includegraphics[width=\linewidth]{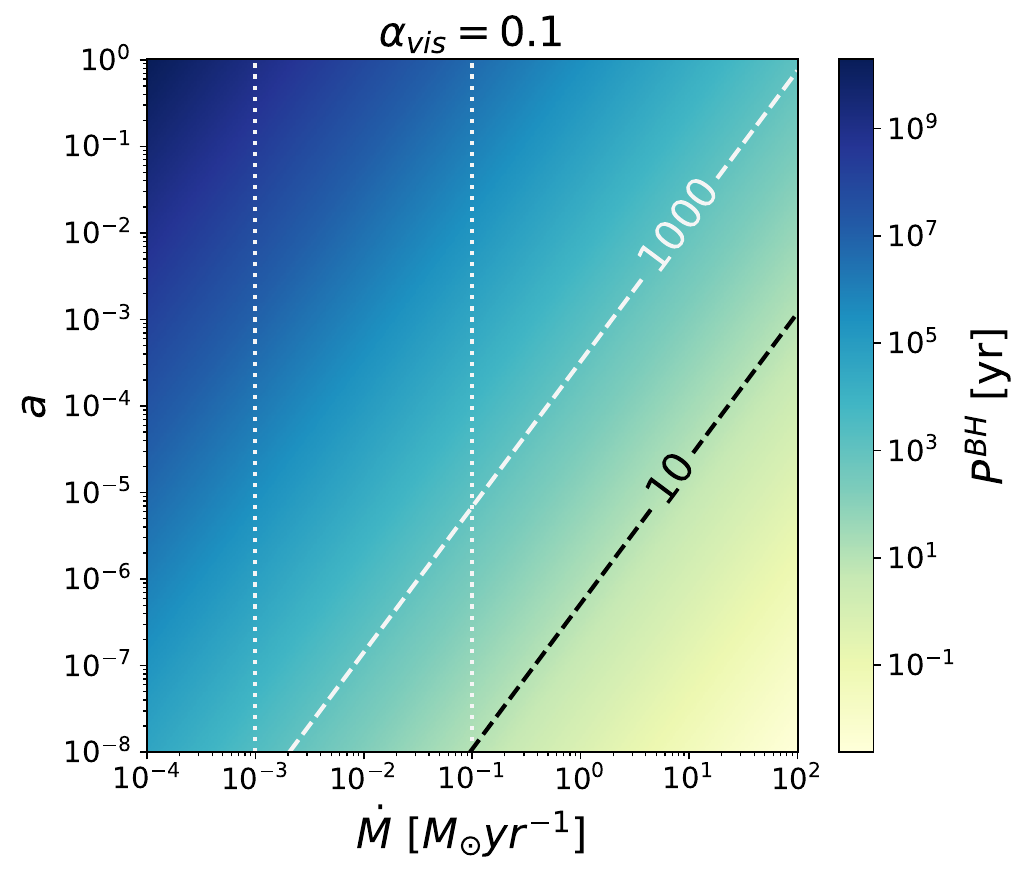}
      \caption{2D plot of the BH spin axis precession period as a function of the spin parameter $a$ and the accretion rate $\dot{M}$ for $\alpha_\mathrm{vis}=0.1$. The black dashed line corresponds to $P^\mathrm{BH}=10$ years and the gray one to $P^\mathrm{BH}=1000$ years. The area in between the vertical dotted white lines corresponds to the typical range of accretion rates for BL Lac objects.}
    \label{fig9}
\end{figure}

Next, we considered  how the disk axis to spin misalignment may induce precession of the BH spin axis. If a massive, misaligned outer accretion disk exerts sufficient torque, it can drive both the inner disk region and the BH spin axis to precess. Similar to the case of LT disk precession, this may lead the entire disk-jet system to precess and explain the observed variations in the jet orientation. The timescale for such a configuration is of the order of thousands of years or more. Given the dimensionless viscosity parameter, $\alpha_\mathrm{vis}$ \citep{Shakura1973A&A....24..337S}, and the mass accretion rate, $\dot{M}$, we can compute the expected precession period for 3C\,66A \citep{Lu1992}:
\begin{equation}
    P^\mathrm{BH} = 10^{9.25}a^{5/7}\alpha_\mathrm{vis}^{48/35}\left(\frac{M}{10^8M_{\odot}}\right)^{1/7}\left(\frac{\dot{M}}{10^{-2}M_{\odot}yr^{-1}}\right)^{-6/5} yr.
\end{equation}
Assuming $\alpha_\mathrm{vis}=0.1$ \citep{Armitage1998ApJ...501L.189A, King2007}, we show in Figure \ref{fig9} how the period changes with the accretion rate and spin. We first consider the Eddington accretion rate $\dot{M}_\mathrm{Edd}$ from the estimated Eddington luminosity $L_\mathrm{Edd}$ of 3C\,66A:
\begin{equation}
    \dot{M}_\mathrm{Edd} = \frac{L_\mathrm{Edd}}{\eta c^2}=\frac{4\pi GMm_\mathrm{P}}{\eta c\sigma_\mathrm{T}},
\end{equation}
where $m_\mathrm{P}$ is the proton mass, $\sigma_\mathrm{T}$ is the Thomson cross-section and $\eta$ translates the efficiency in converting gravitational into radiative energy. We ignore general relativistic effects and take the typical value $\eta=0.1$ expected for a moderately spinning BH \citep{Shakura1973A&A....24..337S}. We obtain $\dot{M}_\mathrm{Edd} \approx 3M_{\odot}/yr$. BL Lac objects such as 3C\,66A are expected to be powered by radiatively inefficient, hot, geometrically thin and optically thick disks, known as ADAF \citep[advection dominated accretion flow; ][]{Narayan1994ApJ...428L..13N}. Population studies have shown that BL Lac objects are highly sub-Eddington systems typically accreting with a rate $\dot{M}<0.01\dot{M}_\mathrm{Edd}$ \citep{Wang2002ApJ...579..554W, Ghisellini2010MNRAS.402..497G}. We therefore expect 3C\,66A to accrete with a rate of, at the very most, $\dot{M}\approx0.1M_{\odot}/yr$. In Figure \ref{fig9}, the expected range of $\dot{M} = 10^{-3}-10^{-1}M_{\odot}/yr$ for BL Lac objects corresponds to the area in between the vertical dotted white lines.

For the maximal value $\dot{M}=0.1M_{\odot}/yr$, we obtain a precession period of approximately $10^6$ years when $a=0.1$. This timescale is orders of magnitude above the $\sim$11 year period of the inner jet PA. The BH precession could only account for the short-term PA oscillation if the accretion rate were much larger than expected for 3C\,66A, or if the BH spin were of the order of $10^{-8}$. Even when decreasing the disk viscosity to a reasonable value of $10^{-2}$, the spin parameter would have to be smaller than $10^{-6}$. It is therefore highly unlikely that precession of the BH can explain the short PA periodicity. A similar conclusion is drawn by \cite{Caproni2004ApJ...602..625C} or \cite{2018MNRAS.478.3199B} regarding the observed PA oscillations of 3C 345 and OJ 287, respectively, that are also of the order of 10 years. Regarding the long-term PA change occurring over approximately 1000 years, the acceptable parameter range for $\alpha_\mathrm{vis}$, $a$ and $\dot{M}$ is more reasonable. In the best scenario, the spin parameter would be approximately $10^{-5}$ which remains very small. While this interpretation for the long-term PA periodicity is possible, we do not favor it due to the very low probability of a single SMBH at the center of 3C\,66A having such a negligible spin.

Therefore, the misalignment between the accretion disk and the BH spin axis can explain one of the observed PA periods in 3C\,66A only through LT precession of the disk, not through BH spin axis precession. In addition, LT effects alone cannot account for both the short-term jet precession and the long-term one. Additional mechanisms are needed to explain all of the observed features in 3C\,66A.

\subsection{Jet instabilities} \label{discussion_jet_instabilities}

A broad array of instability types are known to potentially occur in relativistic jets, thanks both to early analytical investigation and to the fast development over recent decades of (R)MHD and PIC (particle-in-cell) simulations \citep{Nishikawa2019Galax...7...29N, Ortuno2022ApJ...931..137O}. One of the most discussed instabilities to explain changes in AGN jets is the current-driven kink instability (CDI) \citep{McKinney2006MNRAS.368.1561M, Mizuno2009ApJ...700..684M, 2014ApJ...784..167M, Singh2016ApJ...824...48S}. In the presence of strong toroidal magnetic fields it leads to helical jet distortion. 
Recently, \cite{Jorstad2022Natur.609..265J} interpret a $\sim$13 hours quasi-periodic oscillation in the optical and $\gamma$-ray flux of BL Lacertae as CDI propagating near a recollimation shock, at approximately 5 pc from the BH core. CDI is excited by strong magnetic fields \citep{Hu2025A&A...693A.154H}, which is often an argument in favor of this instability to explain jet variability near the base. We verify whether CDI could explain the short-term PA periodic pattern in the jet of 3C\,66A. The growth timescale of CDI ($m=1$) kink-modes is governed by the crossing time $\tau_\mathrm{cross}=R_\mathrm{j}/v_\mathrm{A}$ where $R_\mathrm{j}$ is the jet radius and $v_\mathrm{A}$ the Alfvén speed. 
We take $R_\mathrm{j}\approx0.1$\,mas from the average FWHM of the fitted inner jet circular components and consider  $v_\mathrm{A}\approx 0.8c$ for the relativistic jet. The resulting crossing time is $\tau_\mathrm{cross}\approx2$ years. \cite{Mizuno2009ApJ...700..684M} show that CDI affects the jet structure over a characteristic timescale of $\sim$100$R_\mathrm{j}/v_\mathrm{A}$ and during this linear phase the instability grows. 
To understand whether CDI affects a relativistic jet, we can check whether it satisfies the condition
\begin{equation}
    \frac{\mathcal{D}}{\Gamma c}>100\frac{R_\mathrm{j}}{v_\mathrm{A}} \label{eq_CDI_condition}
,\end{equation}
with $\Gamma$ the Lorentz factor and $\mathcal{D}$ the jet distance over which we expect the instability to grow. We take $\Gamma\approx20$ from the study of 5 jet knots in 3C\,66A by \cite{Weaver2022}. For the length-scale $\mathcal{D}$, we choose the distance of 18 pc which corresponds to the average at which the PA of the inner jet component is measured. If CDI is responsible for the periodic pattern observed in the inner jet, then it should have been able to grow significantly by this distance. The left-hand side of eq. \ref{eq_CDI_condition} yields $\mathcal{D}/\Gamma c\approx1\times10^3$ years, while $100\tau_\mathrm{cross}\approx2\times10^2$ years. The proper propagation time is approximately one order of magnitude larger than 100 crossing times. 
It is therefore possible for the inner jet region to be significantly affected by CDI. 
We can moreover compare the CDI growing timescale to the travel time of the presumed wave in 3C\,66A, which is manifested by the PA oscillation shift with distance from the core. 

As done in Section \ref{modeling_precession_MHDwaves}, taking $\Delta\tau_\mathrm{shift}\approx1.5$ years between the inner and downstream jet regions, the wave appears to propagate with the apparent speed $\beta_\mathrm{wave}^\mathrm{app}\approx5$. Between 0.1 and 1\,mas from the core (limits of the inner and downstream jet regions), the travel time of the wave is therefore approximately 3 years, which is larger than the crossing time. Ours is an approximate estimate due to the fact that we can only identify a periodic pattern for two regions in the jet, and their distance from the core falls in a wide ring. Nonetheless, the properties of the presumed wave in the jet of 3C\,66A agree with the CDI scenario.

The second instability of interest is the Kelvin-Helmholtz instability (KHI), which arises at the interface between the jet and the external medium where the velocity shear can lead to turbulence and wave-like disturbances \citep{1987ApJ...318...78H, 2006ApJ...647..172S, 2010MNRAS.402...87A, 2012ApJ...749...55P, Nishikawa2019Galax...7...29N}. It typically occurs over timescales of years but is suppressed by high magnetic fields, typically ruling out the possibility of its development at the jet base. 
For 3C\,66A, the inner jet PA probes the orientation of the inner 10-25 pc, while the downstream jet PA probes the 25-60 pc approximately. At these distances from the central engine, it is possible that the magnetic field is low enough for the KHI to propagate. 
We draw a comparison with M87, for which \cite{Walker2018ApJ...855..128W} identify a quasi-period of $\sim$8-10 years in the jet offset with respect to a central line. While this does not correspond to the PA change, it equivalently describes the jet orientation evolution since the central line is fixed along a given PA. 
\cite{Walker2018ApJ...855..128W} find that the quasi-periodicity in the jet offset propagates down the jet, meaning that its phase is progressively shifted when measuring it at further distances from the core. The quasi-periodicity is observed between 2 and 8\,mas from the core, which goes up to approximately 2.5 pc in deprojected distance. They interpret this transverse jet position shift as KHI propagation, showing that a KH helical mode can explain the observed pattern motion for reasonable physical parameters of the jet. \cite{Matveyenko2015AstL...41..712M} also argue that a hydrodynamical instability in the flow ejection is responsible for precession in the jet of M87. However, an equivalent PA periodicity to that found by \cite{Walker2018ApJ...855..128W}, of approximately 11 years, was observed at the base of M87's jet using a larger dataset by~\cite{Cui2023}. These authors discarded the KHI hypothesis due to its suppression by strong magnetic fields and argue for LT precession instead. This manifests how the interpretation of periodic signals in jet orientation evolution is still much debated, even for extensively studied sources such as M87. Recently, \cite{Kino2025ApJ...986...49K} also considered the possibility of the presence of a SMBHB in M87, based on the $\sim$11 year precession period and a $\sim$0.9 year transverse period.

In the case of 3C\,66A, the $\sim$11 year period observed corresponds to an angular frequency $\omega_\mathrm{j}\approx2\times10^{-8}$ rad/s. \cite{Walker2018ApJ...855..128W} showed that a lower limit to the helical mode resonant frequency can be expressed as
\begin{equation}
    \omega_{*} \approx\frac{3\pi}{4}\frac{c/R_\mathrm{j}}{(c/v_\mathrm{w}-1)},
\end{equation}
where $v_\mathrm{w}$ represents the speed of sound if we consider the sonic limit (low magnetization) or the Alfvén speed if we consider the Alfvénic limit (high magnetization) for the jet. In order to obtain this expression, the authors assume that the external medium satisfies the no flow limit, implying that its speed $u_\mathrm{e}=0$. In addition, one must assume that the sound and Alfvén speeds are the same inside and outside of the jet, which can be motivated by the presence of a cocooning medium surrounding it. In this case we take $R_\mathrm{j}\approx0.4$\,mas, from the average FWHM of the fitted downstream jet circular components, since the condition $\omega_*>\omega$ must also be satisfied in the downstream jet if we interpret the PA periodicity of this region as a KHI signature.
Then, for the resonant frequency $\omega_*$ to exceed the observed angular frequency $\omega$, we must have $v_\mathrm{w}\gtrsim0.6c$. Below this value, we have $\omega>\omega_*$ and the instability growth rate is vanishingly small \citep{Hardee2007ApJ...664...26H}. The condition on $v_\mathrm{w}$ rules out the sonic limit interpretation since the maximum speed of sound is $c/\sqrt{3}$. The Alfvén speed can theoretically satisfy $v_\mathrm{A}\gtrsim0.6c$ in the jet, although it may not be this large in 3C\,66A for the jet regions considered, due to the distance of more than 10 pc from the core implying a moderate magnetization. If we assume that the jet is in the Alfvénic limit and the magnetic pressure is balanced between the jet and the surrounding medium, the KH helical mode can propagate near the resonant frequency. 
In this case, the helical mode wave speed is \citep{Hardee2007ApJ...664...26H}:
\begin{equation}
    v_\mathrm{h}^*=\frac{\Gamma_\mathrm{j}(\Gamma_\mathrm{we}v_\mathrm{we})u_\mathrm{j}+\Gamma_\mathrm{e}(\Gamma_\mathrm{wj}v_\mathrm{wj})u_\mathrm{e}}{\Gamma_\mathrm{j}(\Gamma_\mathrm{we}v_\mathrm{we})+\Gamma_\mathrm{e}(\Gamma_\mathrm{wj}v_\mathrm{wj})}, \label{eq_KH_speed}
\end{equation}
where the j and e subscripts indicate parameters of the jet and of the external medium, respectively, $u_\mathrm{j}$ is the jet flow speed and $\Gamma_\mathrm{w}=(1-v_\mathrm{w}^2/c^2)^{-1/2}$ is the sonic or Alfvénic Lorentz factor. Here, we let $u_\mathrm{e}$ be positive. The jet is in the Alfvénic limit and we assumed that the Alfvén speed is equal in the jet and external medium, leading to: $v_\mathrm{wj}=v_\mathrm{we}=v_\mathrm{A}$ and $\Gamma_\mathrm{wj}=\Gamma_\mathrm{we}=\Gamma_\mathrm{A}$. The KH helical mode speed from eq. \ref{eq_KH_speed} thus simplifies into:
\begin{equation}
    v_\mathrm{h}^*=\frac{\Gamma_\mathrm{j}u_\mathrm{j}+\Gamma_\mathrm{e}u_\mathrm{e}}{\Gamma_\mathrm{j}+\Gamma_\mathrm{e}}.
\end{equation}

We estimate a lower and upper limit for this speed using the Lorentz factor and apparent component speeds reported by \cite{Weaver2022}. For the jet parameters, we take the average values $\Gamma_\mathrm{j}\approx20$ and $\beta_\mathrm{j}^\mathrm{app}\approx18$ from the components studied. Assuming the viewing angle is $\Theta\approx5\degree$, the jet speed is $u_\mathrm{j}\approx0.999c$. The lower limit of $v_\mathrm{h}^*$ is obtained when assuming $u_\mathrm{e}=0$ and $\Gamma_\mathrm{e}=1$. The upper limit is obtained by considering that the external medium properties correspond to the smallest propagation speed observed in the jet and to the smallest estimated Lorentz factor: $u_\mathrm{e}\approx0.997c$ and $\Gamma_\mathrm{e}\approx14$.
The KH helical mode speed is then in the range $0.951c<v_\mathrm{h}^*<0.998c$.
We can compare this speed to that of the presumed wave in 3C\,66A. The latter is given by $\beta_\mathrm{wave}^\mathrm{app}\approx5$, equivalent to the propagation speed $v_\mathrm{wave}\approx0.987c$. KHI propagation can therefore explain the observed phase-shift in the PA periodic pattern between the inner and downstream jet. 
In addition, strong magnetic fields can lead to the damping of the KHI as it propagates along the jet \citep{Mizuno2009ApJ...700..684M}, which is in agreement with the observed amplitude reduction of the periodic pattern further down the jet. Once again, this would imply that the jet is strongly magnetized at tens of pc from the core. In addition, in the case of M87 \cite{Walker2018ApJ...855..128W} note that there is no sign of the quasi-period pattern at arcseconds scales until knot A. A similar property is observed for 3C\,66A, since we do not observe the periodic pattern starting at approximately 1.3\,mas from the core, equivalent to $\sim$80 pc deprojected. Therefore, the KHI can explain the observed PA oscillation in 3C\,66A. It requires a jet in the Alfvénic limit and an equivalent magnetization of the jet and of the immediate external medium. 

Another instability known to occur in jets is the Rayleigh-Taylor instability (RTI), that can arise at the interface between the accelerated jet and the ambient medium due to density gradients, growing into what is often dubbed finger-like structures. Recently, \cite{Hu2025A&A...693A.154H} showed through 3D and 2D RMHD simulations that this instability is stabilized by the presence of a magnetic field. For a helical magnetic field they observe the CDI to develop instead, after passing recollimation shocks, supporting the findings by \cite{Jorstad2022Natur.609..265J}. In the case of 3C\,66A, RTI may play a role in the complex jet morphology observed, but unlikely explains any of the periodicities observed. 
A similar conclusion can be drawn regarding the mushroom instability (MI) which is caused by electron-scale velocity shears and leads to mushroom-like structures in 2D simulations, perpendicular to the jet propagation direction. PIC simulations indicate that this instability grows stronger with stronger magnetic fields \citep{Nishikawa2019Galax...7...29N}. It has also been associated with particle acceleration and possible magnetic reconnection \citep{Kawashima2022ApJ...928...62K}. It is likely that MI acts in the 3C\,66A jet and causes the propagation of disturbances as well as flux enhancement, but not on periodic timescales.
Other internal instabilities that may occur in the jet, but on timescales too short ($\sim$ seconds to days) to explain our flux or PA oscillations, are the two-stream instability \citep{Horky2013RAA....13..687H}, the Weibel instability \citep{Nishikawa2019Galax...7...29N} and the pressure-driven instability \citep{Longaretti2008LNP...754..131L, Das2019MNRAS.482.2107D}.

Therefore, many internal instabilities can occur in relativistic jets, with the CDI and KHI able to explain one of the PA periods of 3C\,66A. We have shown that the CD kink instability or the KH helical mode can account for the the wave-like propagating PA pattern of $\sim$11 years. For the latter, this requires that the jet is in the Alfvénic limit and its magnetization is equal to that of the surrounding medium. The instability would propagate while being damped by the high magnetic field and disappear beyond $\sim$80 pc from the core. This finding leads us to consider an alternative interpretation to the SMBHB system, presented in Section \ref{modeling_SMBHB}. We can indeed imagine a combination of LT precession of the disk, explaining the PA monotonic decrease through secular jet precession, and of either CDI or KHI propagating at the jet base, leading to the PA oscillation on decade-scales.

\subsection{Disk instabilities and dynamics} \label{discussion_disk_instabilities}

On top of processes occurring in the jet, we investigate disk instabilities that may lead to flux variability and or to precession
\citep{Armitage2003MNRAS.341.1041A, King2004MNRAS.348..111K}.
For example, \cite{Ferreira2022A&A...660A..66F} explore instabilities in the disk to explain low frequency quasi-periodic oscillations in black hole X-ray binaries, which could also lead to jet wobbling in accreting sources.

Except for the LT effect discussed in Section \ref{discussion_LT_disk}, a relevant disk instability is radiation-induced warping of the accretion disk, which could lead to jet precession and hence explain both PA and flux variability \citep{Pringle1997MNRAS.292..136P, Lai2003ApJ...591L.119L}. An accretion disk irradiated by a central source (which might in fact be the central disk regions) is unstable to becoming nonplanar: if the disk is warped, light hits its surface unevenly and, in the optically thick case, it causes the re-emitted radiation to exert uneven torques on different parts of the disk. These lead to further warping over time, effectively acting as a precession. For example, such a warped disk has been considered to explain jet precession at the center of the Seyfert galaxy NGC 1275 \citep{Dunn2006MNRAS.366..758D}. Assuming an optically thick disk, \cite{Pringle1997MNRAS.292..136P} estimates the characteristic disk-warping timescale to be
\begin{equation}
    P_\mathrm{warp} \approx2\times10^{6}\alpha_\mathrm{vis}^{-1}\frac{M}{10^8M_{\odot}}\text{ yr}.
\end{equation}
Choosing again the typical value of $\alpha_\mathrm{vis}=0.1$, we obtain $P_\mathrm{warp}\approx10^7$ years, which is several orders of magnitude larger than both the short- and long-term periods observed for 3C\,66A. We therefore rule out disk warping as a mechanism explaining jet precession for our system.

Other well-known disk instabilities that could account for flux variability include thermal-viscous instabilities. These broadly include the ionization instability and the radiation pressure instability. They have been mostly studied in the context of cataclysmic variables or X-ray binaries, but some authors have shown interest in their potential impact on AGN disks \citep{Hameury2009A&A...496..413H, Janiuk2011MNRAS.414.2186J}. Ionization instability is caused by hydrogen transitions in the outer disk and may lead to quasi-periodic emission variability (which is well-established for X-ray binaries) by modulating the accretion rate. For a BH of mass $M\sim10^8M_{\odot}$, the ionization instability timescale is expected to be of the order of $\sim10^6$ years or more \citep{Janiuk2011MNRAS.414.2186J}. Radiation pressure instability, for which there is little observational background, occurs in the radiation pressure dominated inner disk and may also lead to accretion modulation, with a characteristic timescale of $\sim10^6$ years between outbursts for a $M\sim10^8M_{\odot}$ BH. This may be reduced to $\sim10^3-10^4$ years in the case of high accretion rates \citep{Czerny2009ApJ...698..840C, Janiuk2011MNRAS.414.2186J}, although we do not expect such rates for a BL Lac object like 3C\,66A \citep{Ghisellini2010MNRAS.402..497G}. The timescales for both of these instabilities remain far too large to explain the (possible) periodicity of 3C\,66A in the (radio) optical LCs.  While \cite{Godfrey2012ApJ...758L..27G} consider that radiation pressure instability may cause the observed periodic structure in the jet of PKS 0637$-$752 through accretion modulation, we disfavor this mechanism in 3C\,66A. We note that the presence of a SMBHB is another strong scenario that \cite{Godfrey2012ApJ...758L..27G} propose for PKS 0637$-$752.

To explain the observed flux variability in 3C\,66A, especially the optical period of $\sim$2 years, we can moreover consider a hot-spot rotating in the accretion disk at some radius, $r_\mathrm{hs}$ \citep{Gupta2009ApJ...690..216G, Prince2023A&A...678A.100P}. Assuming the hot-spot is the source of optical flux with a period of $P_\mathrm{obs}^\mathrm{optical}=2.38\pm0.19$ years, we can estimate the central BH mass as
\begin{equation}
    M = \frac{3.24\times10^4P_\mathrm{obs}^\mathrm{optical}}{(r_\mathrm{hs}^{3/2}+a)(1+z)}M_{\odot}.
\end{equation}
We consider the Schwarzschild BH scenario, in which $r_\mathrm{hs}=6$ (in units of $GM/c^2$) and $a=0$, and the maximal Kerr BH scenario, in which $r_\mathrm{hs}=1.2$ and $a=0.9982$. In the first case, the resulting BH mass is $M=1.23\times10^{11}M_{\odot}$ and in the second, it is $M=7.83\times10^{11}M_{\odot}$. Both mass values are too large by several orders of magnitude to be in agreement with the expected system in 3C\,66A. We therefore rule out the presence of a hot-spot in the disk to explain the optical periodicity. The same conclusion can be drawn for the radio flux variability. This hot-spot scenario is similarly discarded by \cite{Prince2023A&A...678A.100P} and \cite{Mao2024MNRAS.531.3927M} regarding the periodicity features of $\sim$100 days and $\sim$600 days found in the LCs of the blazars PKS 0346-27 and PKS J2156-0037, respectively. Hence, we find that none of the disk phenomena explored can be applied to 3C\,66A to explain the observed PA or flux periodicity.

\subsection{External factors} \label{discussion_external}

Alternative scenarios for the observed bending in jets, such as the twisted structure found for 3C\,66A, have been proposed.
For instance, bent jet structures might occur because of external causes.
\citet{2002ApJ...580..742H} argue that the bending structure of the quasar PKS 1510$-$089 would possibly be due to a density gradient between the host galaxy and intergalactic medium or a ram pressure from the intracluster medium.
However, this scenario is unlikely in the parsec-scale jet bending because it is generally related to much larger-scale jet bending \citep{1972Natur.237..269M, 1973A&A....26..423J, 1979ApJ...234..818J}.
Another possible explanation which would arise within the host galaxy is a colliding jet with surrounding clouds \citep{2000Sci...289.2317G, 2001MNRAS.328..719G, 2003ApJ...589L...9H}. 
\citet{2003ApJ...589L...9H} find a bent trajectory with increasing flux density and decreasing size for the C4 component in the quasar 3C 279, indicating that C4 has been deflected by the interstellar medium. 
It is less likely responsible for the whole twisted structure of 3C\,66A because the deflecting jet model requires a number of well-distributed clouds near the jet for the multiple bendings observed.

\section{Conclusions}
\label{conclusions}

In this paper, we present three major features of the 3C\,66A jet based on KaVA observations in 2014 and VLBA observations over 29 years.
We find a twisted jet structure on the KaVA maps and, from the VLBA archival data, for the first time, we show  that the inner jet PA varies periodically. We observe a periodic pattern of $\sim$11 years in both the inner jet and the downstream jet, with a $\sim$1.5 year phase-shift and an amplitude reduction further from the core. The PA oscillation is accompanied by a monotonic decrease of about $-0.8\degree$/year.
Moreover, we  find strong variability in the radio core flux with possible periods of $\sim$10 years, $\sim$7 years, and $\sim$4 years. In addition, we detect a long-term decrease of approximately $-9.4$\,mJy/year in the baseline. The core flux presents a possible correlation with the periodic inner jet PA swing. 

These findings, together with the well-known $\sim$2 year period of the optical LC in 3C\,66A and with multi-wavelength baseline drops coincident with the jet's PA decrease, suggest a geometric scenario for both the PA oscillation and its monotonic decrease. We interpret these observations as the signature of a precessing jet over a decadal timescale superimposed on a longer-term secular drift. The short-term nozzle precession may excite MHD waves such as Alfvén transverse waves that propagate down the jet, possibly explaining the downstream phase-shift and amplitude attenuation of the oscillation pattern. 

The origin of the precession timescales  is a  topic subject to some debate on its own. We propose that a SMBHB system is the driver behind both precession timescales, with orbital motion and disk-orbit precession accounting for the short- and long-term trends respectively. Such a system can naturally explain all of the observables reported in this paper. The putative binary system orbits with a rest-frame period of approximately 17 years and the separation between the BHs is $1.65\times10^{-2}$ pc. The SMBHB is expected to merge in $\sim$934 Myr, currently emitting GWs below the sensitivity of PTA experiments. 

Recently, many SMBHB candidates have been reported and some of them  have been deemed controversial. We adopt this scenario as a parsimonious explanation, providing a simple interpretation for the large number of complex properties of 3C\,66A, in particular, its different periodic timescales regarding flux and orientation. However, we have investigated a wide range of other possible interpretations. We show that LT precession of the accretion disk can reproduce either the short- or long-term precession timescale, especially for a power-law disk density profile which allows a wider range of spin parameters. 
We also evaluated the potential of assuming jet instabilities as an origin of the observed variability. The CDI kink mode and the KH helical mode are both possible interpretations for the $\sim$11 year periodic pattern, assuming that it corresponds to a propagating wave. We further explored disk-related mechanisms, such as radiation-induced disk warping, thermal-viscous instabilities, and a rotating hot-spot in the disk. None of these scenarios can account for  the PA or the flux variability observed in 3C\,66A. Therefore, excluding the SMBHB model, only one alternative appears viable: LT precession of the accretion disk drives long-term jet precession, while the observed decadal PA oscillation arises from CDI or KHI propagating in a magnetized jet. In the case of KH helical modes, the jet must be remain in the Alfvénic limit out to $\sim$100 pc and we must assume a magnetized external medium with comparable properties to the jet. In the future, additional kinematic, polarimetric, and astrometric analyses for 3C\,66A will allow further investigation of the complex origins of its variability.

\begin{acknowledgements} 
The authors thank Prof. Eduardo Ros, the MPIfR internal referee, for reading the manuscript and providing constructive comments that have improved the paper.
The authors also express their gratitude to Dr. Hyunwook Ro, Dr. Andrei Lobanov and Dr. Wu Jiang for insightful discussions.
This research was supported by funding from Korea AeroSpace
Administration (KASA)) (grant numbers RS-2024-00509838 and R25TA0065942000).
P.T. acknowledges support from France 2030 through the project named Académie Spatiale d'Île-de-France (\url{https://academiespatiale.fr/}) managed by the National Research Agency under the reference ANR-23-CMAS-0041.
J. K. and B.W.S. are grateful for the support of the Korea Astronomy and Space Science Institute under the R$\&$D program (Project No. 2017-1-840-90) supervised by the Ministry of Science and ICT.
G.-Y.Z acknowledges support from the project M2FINDERS that is funded from the European Research Council (ERC) under the European Union's Horizon 2020 research and innovation programme (grant agreement No 101018682).
S.-J.Y. acknowledges ($a$) support from the Mid-career Researcher Program (RS-2024-00344283) through Korea's NRF funded by the Ministry of Science and ICT, and ($b$) support from the Basic Science Research Program (RS-2022-NR070872) through Korea's NRF funded by the Ministry of Education.
This study makes use of VLBA data from the VLBA-BU Blazar Monitoring Program (BEAM-ME and VLBA-BU-BLAZAR; http://www.bu.edu/blazars/BEAM-ME.html), funded by NASA through the Fermi Guest Investigator Program. The VLBA is an instrument of the National Radio Astronomy Observatory. The National Radio Astronomy Observatory is a facility of the National Science Foundation operated by Associated Universities, Inc.
This study makes use of the following open-source packages: \textit{NumPy} \citep{harris2020array}, \textit{Matplotlib} \citep{Hunter:2007}, \textit{SciPy} \citep{2020SciPy-NMeth}, \textit{Astropy} \citep{Astropy2022ApJ...935..167A}, \textit{emcee} \citep{emcee2013PASP..125..306F}, \textit{libwwz}, and \textit{psresp} \citep{psresp2002MNRAS.332..231U}.

\end{acknowledgements}

\bibliographystyle{aa}
\bibliography{bibliography.bib}
\begin{appendix}

\section{PA evolution of three jet components}
\label{app:PA_components}

\begin{figure}
   \centering
   \includegraphics[trim={10mm, 00mm, 10mm, 10mm},clip=true,width=\linewidth]{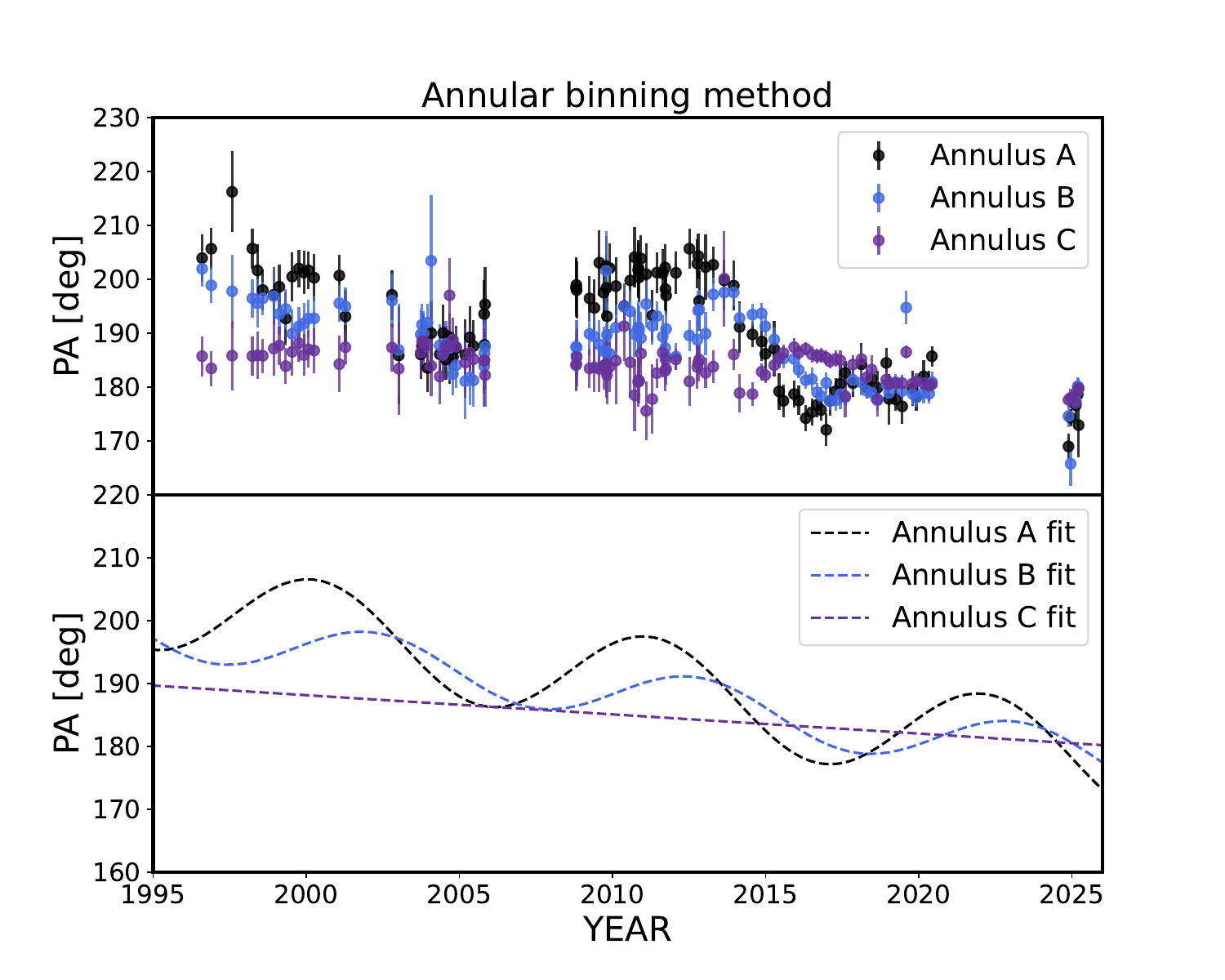}
      \caption{Evolution of the PA for the annuli A (inner jet:\ black), B (downstream jet:\ blue), and C (purple) at 43\,GHz, obtained with the annular bin method. The data is shown in the upper panel and the analytical fits in the lower panel: sinusoidal+linear fits for the annuli A and B and linear fit for the annulus C.}
    \label{PA_three_components}
\end{figure}

\begin{figure}
   \centering
   \includegraphics[trim={10mm, 00mm, 10mm, 10mm},clip=true,width=\linewidth]{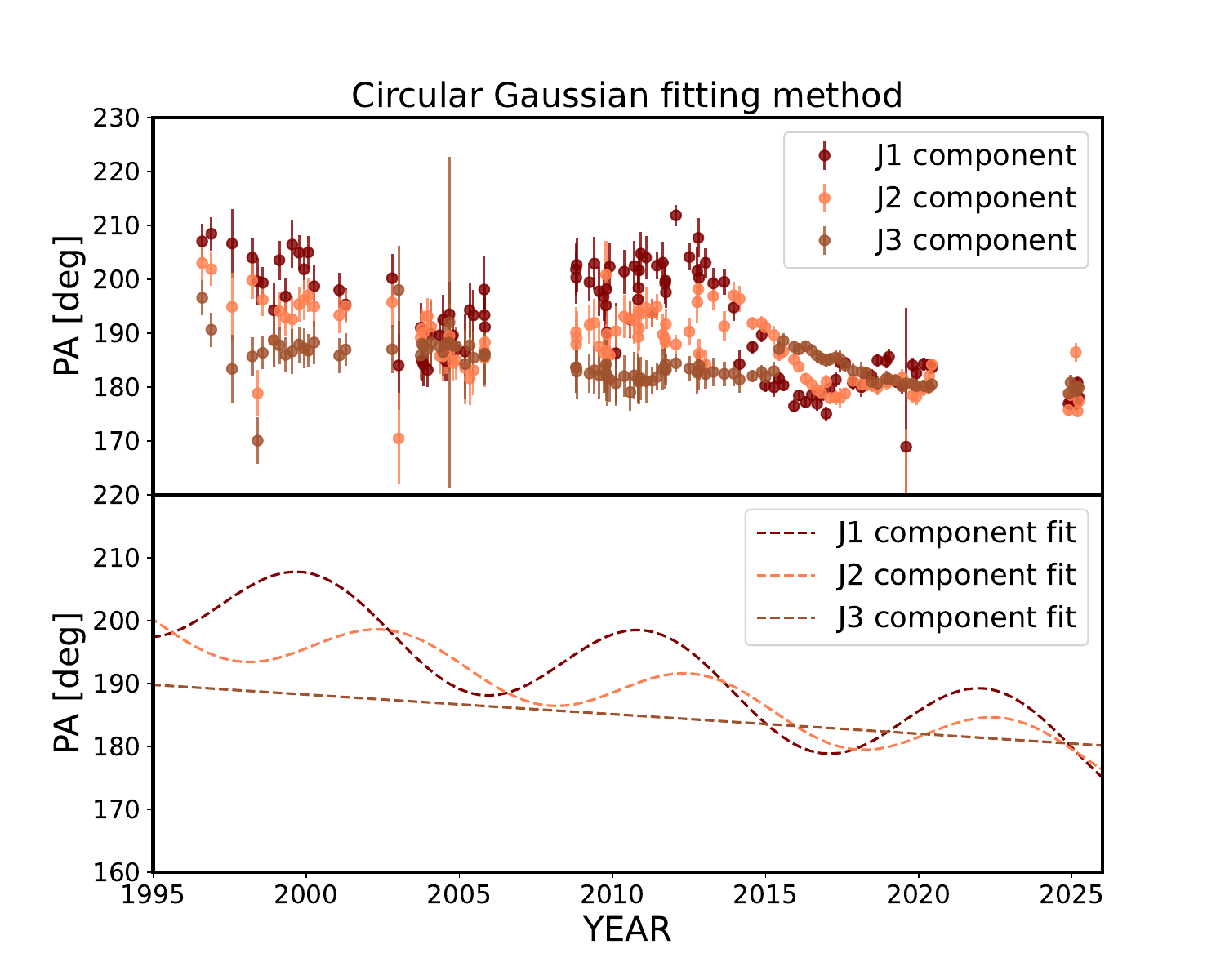}
      \caption{Evolution of the PA for the Gaussian components J1 (inner jet - maroon), J2 (downstream jet - orange) and J3 (brown) at 43\,GHz, obtained with the circular Gaussian fitting method. The data is shown in the upper panel and the analytical fits in the lower panel: sinusoidal+linear fits for the J1 and J2 components and linear fit for the J3 component.}
    \label{PA_three_components_Gaussian}
\end{figure}

\begin{figure*}
    \centering
\includegraphics[width=\textwidth]{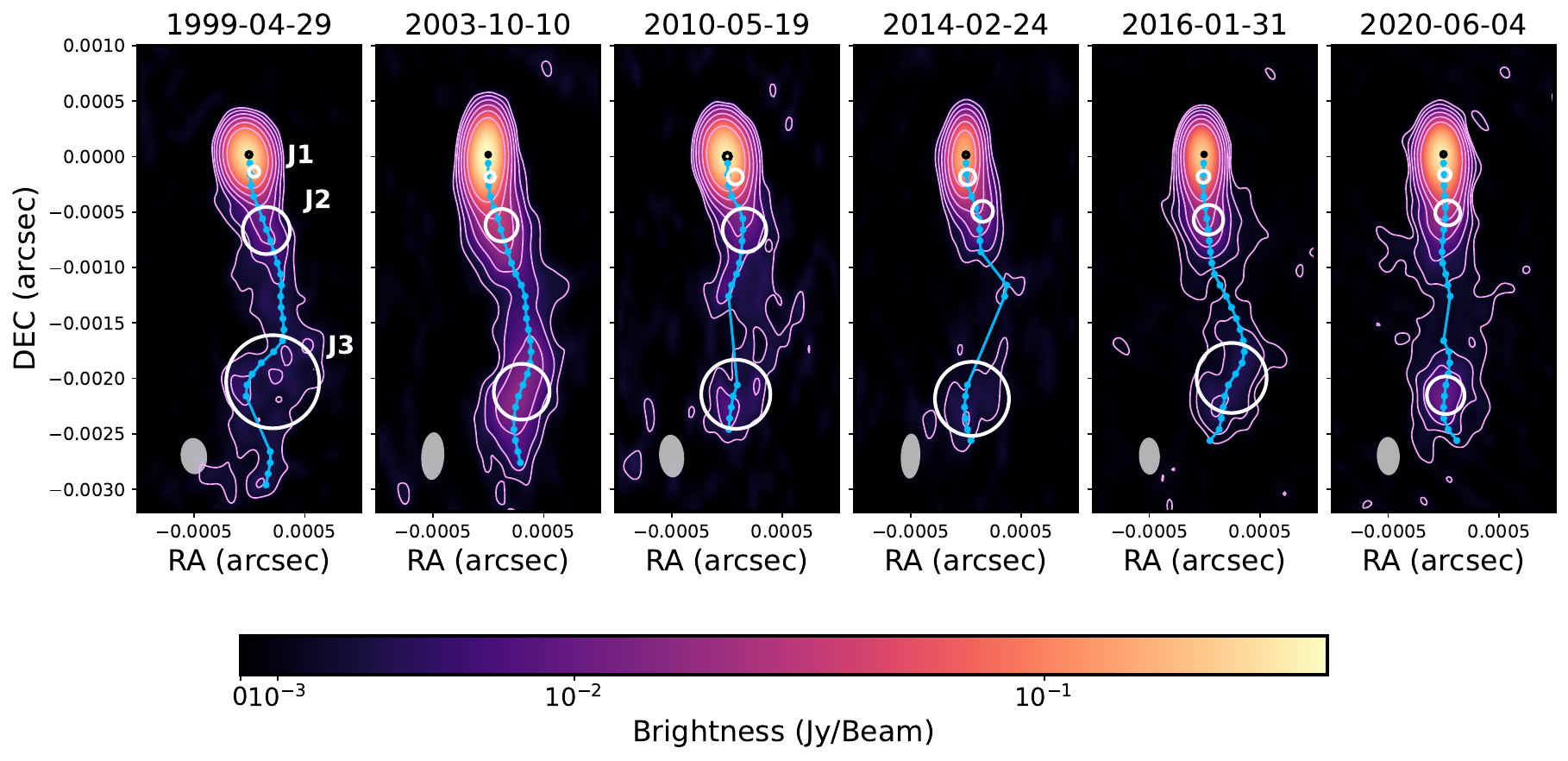}
    \caption{VLBA intensity maps at 43\,GHz overplotted with three Gaussian fitted jet components (white) and core (black) at six epochs. In the leftmost panel, the labels J1, J2 and J3 denote the three jet components. The blue connected dots indicate the jet ridgeline. The pink contour levels start at three times the rms value, scaling twice. The gray ellipse at the bottom of each panel corresponds to the restoring beam.}
    \label{components}
\end{figure*}

The evolution of the PA in the jet of 3C\,66A is studied for three jet components with two distinct methods (annular bins and circular Gaussian fits). These jet components are found progressively further away from the core. The first two are denominated inner jet and downstream jet. Using the binning method to compute their PA, they correspond to the annuli $(0.15<r<0.45)$\,mas (annulus A), $(0.45<r<1.00)$\,mas (annulus B) and $(1.50<r<3.00)$\,mas (annulus C) respectively, where $r$ is the radius from the center $(0,0)$. Their PA is also evaluated using circular and elliptical Gaussians from \textit{modelfit}. In Figure \ref{components}, we show the white circular Gaussian fits to the three jet components (denoted J1, J2 and J3) at six different epochs between 1999--2020, highlighting their different evolution. From Figure \ref{components}, as well as Figures \ref{PA_three_components} and \ref{PA_three_components_Gaussian}, it is clear, for instance, that the inner jet and the downstream jet's PAs do not oscillate in phase. 

In Figure \ref{PA_three_components} we show the PA evolution for the annuli A, B and C. The data is visible on the top panel, while the corresponding fits are shown in the lower panel for visibility. Figure \ref{PA_three_components_Gaussian} presents equivalent results but obtained with the circular Gaussian fitting approach (components J1, J2 and J3). The fit results of Eq.\ref{eq_fit} for both methods are given in Table \ref{tab1}. The figures, as well as the fitted values, indicate an almost identical evolution of the PA for each considered component with the two methods.

\section{Periodicity search methods} 
\label{app:periodicity_methods}

We describe the three approaches used to search for periodicity in the inner jet PA (from the annular bin results) and the core flux density. We detail the results obtained in each case and summarize the major periodicities found in Table \ref{tab_periods}. All of the reported periods' standard deviations are obtained from the FWHM of the periodicity peaks.

\subsection{Jurkevic analysis}
We first implement the Jurkevich method \citep{Jurkevich1971}: the data is divided into $m$ bins based on trial periods and, for each period, the total variance of all bins $V_\mathrm{m}^2$ is computed. A minimum in the variance indicates a possible periodicity. This method allows to detect a periodicity even if it is not sinusoidal, but it doesn't provide statistical confidence levels. In order to determine the significance of a minimum in the $V_\mathrm{m}^2$ plot, we rely instead on the F-test suggested by \cite{Kidger1992}:
  \begin{equation}
      F = \frac{1-V_\mathrm{m}^2}{V_\mathrm{m}^2}.
  \end{equation}
If $F\geq0.5$, the period can be considered strong, while it is considered weak if $F<0.25$.

\begin{figure}
    \centering\includegraphics[width=\linewidth]{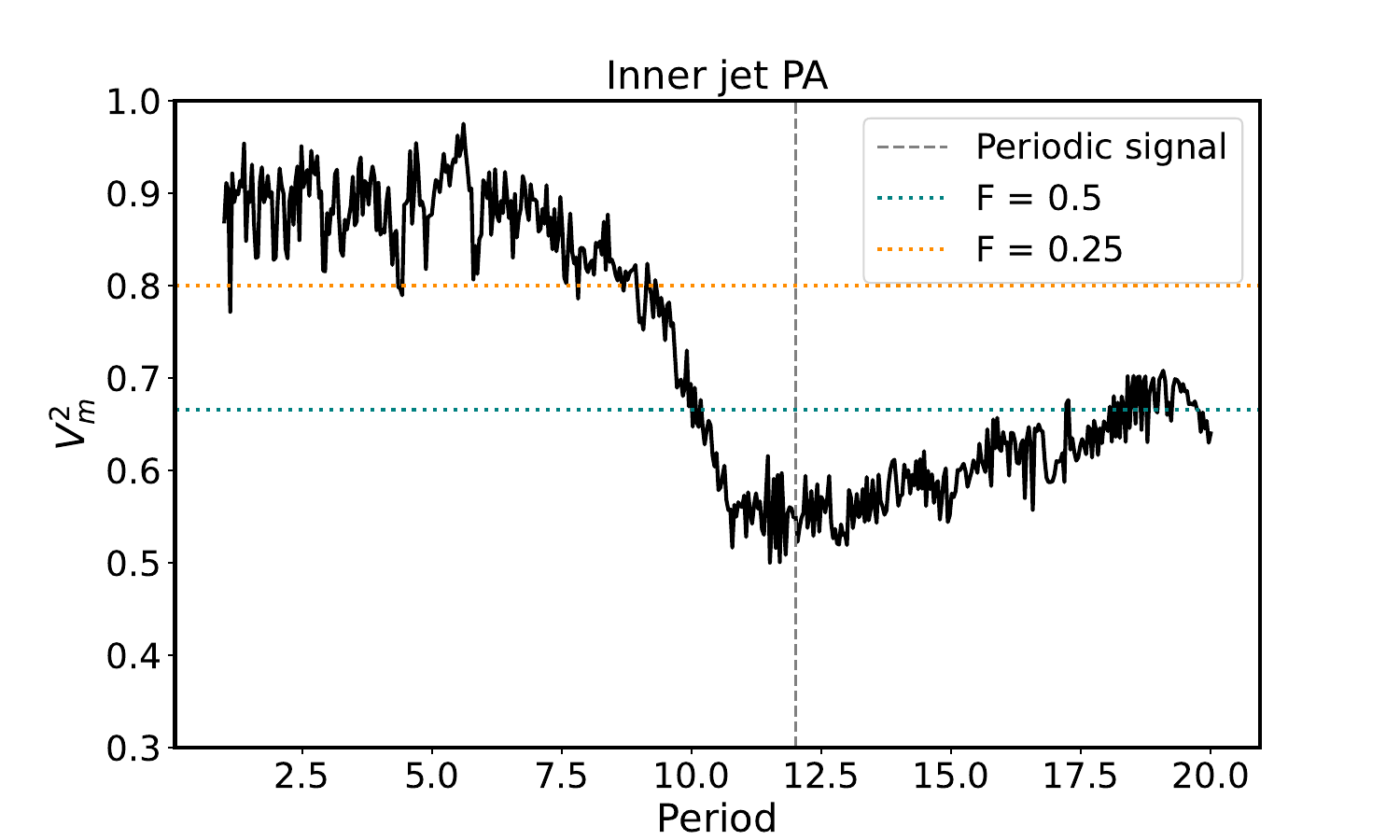}
      \caption{Jurkevic test plot for the inner jet PA. The black line corresponds to the $V_\mathrm{m}^2$ values whose minima indicate possible periodicities and the dotted blue (orange) line to the level $F=0.5$ ($F=0.25$). The vertical gray dashed line indicates the period at the minimum.}
    \label{Jurkevic_PA}
\end{figure}

We set the bins number to 15 and the number of trial periods to 500. The plot of $V_\mathrm{m}^2$ as a function of period is shown in Figure \ref{Jurkevic_PA} for the PA, where we explore the period range between 1 and 20 years. There is a single significant drop in the $V_\mathrm{m}^2$ values that reaches 
a minimum around 12 years, and gaussian fitting of the variance data leads to a period of $14.4\pm5.1$ years. While this is a very wide drop, it reaches clearly below the $F=0.5$ threshold used to determine significance.
Figure \ref{Jurkevic_PA} shows the results for the core flux density, in the period range between 1 and 15 years. This Jurkevic test appears less conclusive: there are a number of peaks, more or less well defined, that fall below $F=0.25$ and two that reach $F=0.5$. We report three peaks of interest from this analysis, considering the strength of their minima and the clarity of their peak shape. The corresponding periods are: $3.5\pm0.3$ years, $6.9\pm0.2$ years and $10.5\pm0.6$, all reaching the $F=0.5$ level. These periods are also found with additional methods.

\begin{figure}
    \centering\includegraphics[width=\linewidth]{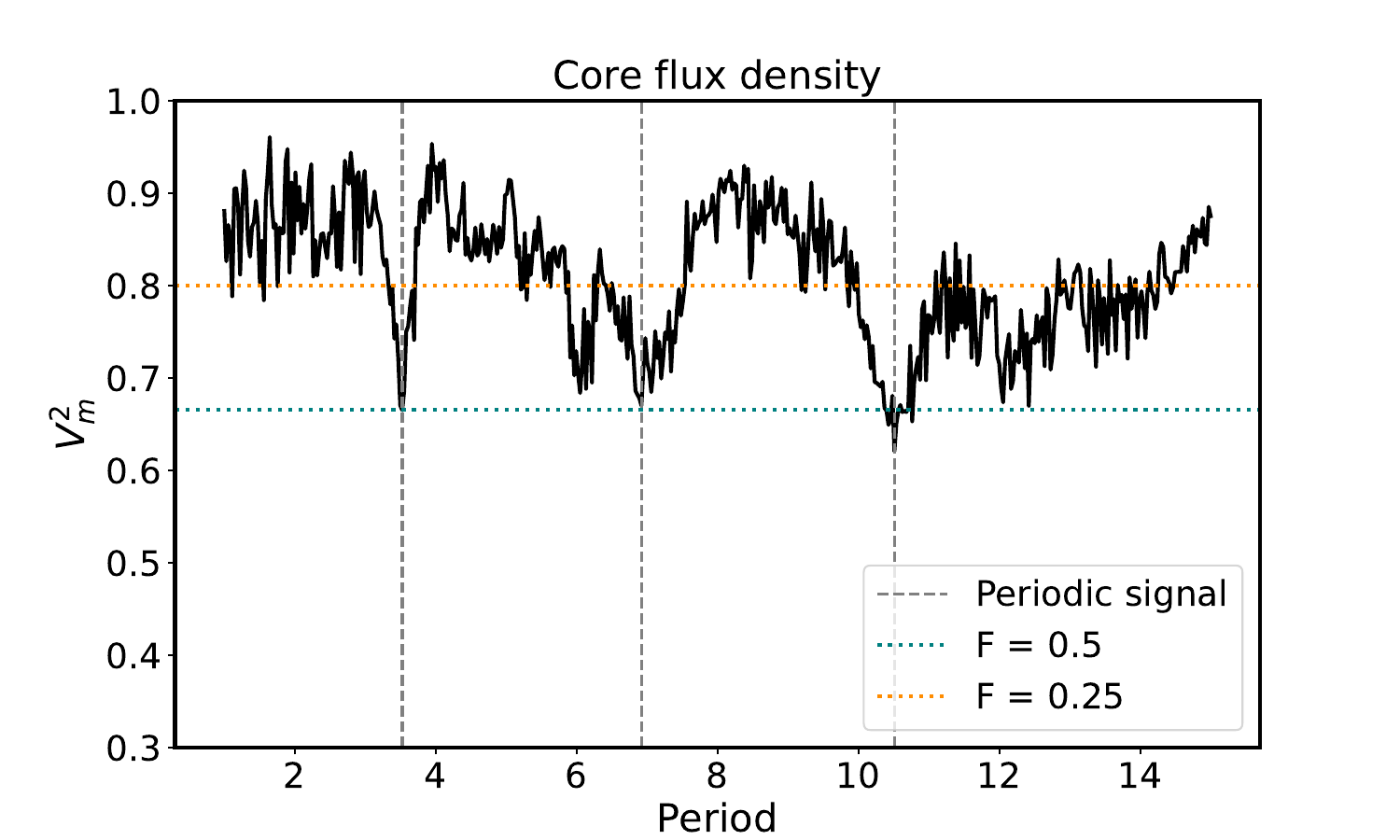}
      \caption{Jurkevic test plot for the core flux density. The black line corresponds to the $V_\mathrm{m}^2$ values whose minima indicate possible periodicities and the dotted blue (orange) line to the level $F=0.5$ ($F=0.25$). The vertical gray dashed lines indicate the periods at the minima of interest.}
    \label{Jurkevic_CFD}
\end{figure}

\subsection{Lomb-Scargle periodogram}

The LS periodogram \citep{1976Ap&SS..39..447L, 1982ApJ...263..835S} is a method based on sinusoidal least-squares fitting. This approach can only detect sinusoidal periodicities, but it is more robust to noise than the Jurkevich method. We use the LS periodogram from the \textit{AstroPy} package. The significance levels account for red-noise, using the method described in Section \ref{appendix_period_levels}.

\begin{figure}
    \centering
    \includegraphics[trim={00mm, 12mm, 00mm, 0mm},clip=true,width=\linewidth]{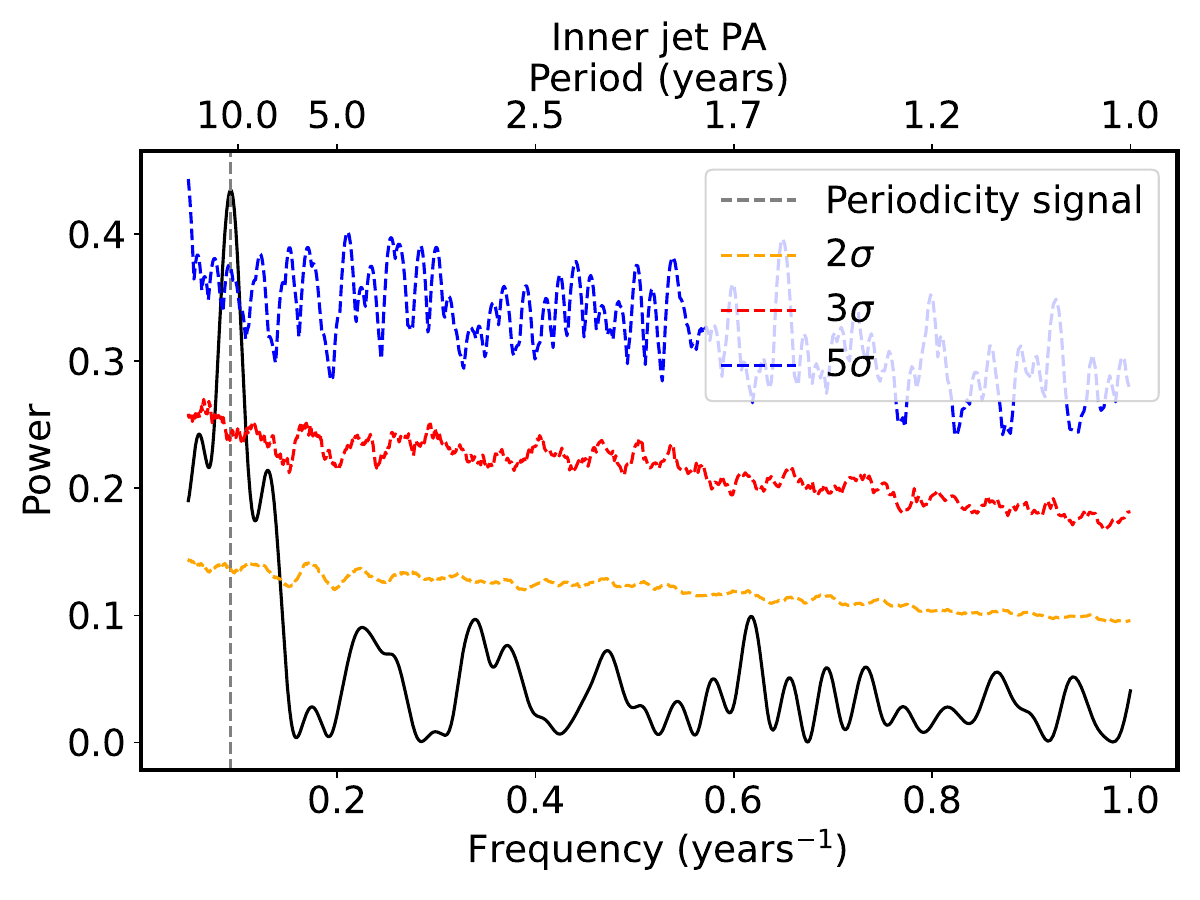}
    \includegraphics[trim={2mm, 5mm, 2mm, 3mm},clip=true,width=\linewidth]{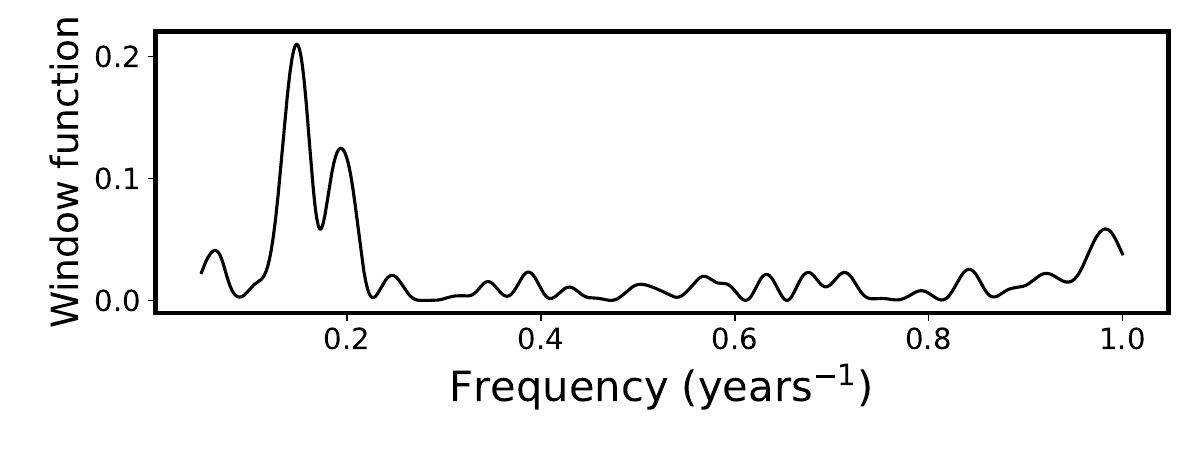}
      \caption{LS periodogram for the PA (upper panel) and window function (lower panel). The significance levels are indicated by the (yellow, red, blue) dashed lines ($2\sigma$, $3\sigma$, $5\sigma$). The gray dashed line marks the peak at 10.8 years.}
    \label{LS_PA}
\end{figure}

In the PA search we explore the frequency range corresponding to periods between 1--20 years and we use 1000 frequencies for the logarithmic grid. The top panel of Figure \ref{LS_PA} shows the LS power at a given period for the inner jet PA.
We find the highest peak ($power \sim 0.4$) at $10.8\pm2.3$ years, denoted by the gray dashed line in the frequency domain. In order to verify the significance of the peak, we firstly test the effect of uneven-sampling on the LS periodogram by plotting the window function (bottom panel of Figure \ref{LS_PA}). While there is a small peak around 7 years, the value of the window functinon is below 0.05 at the PA maximum of $\sim$11 years, so that uneven sampling does not lead to a falsely detected peak here. In addition, the LS power at this period is larger than the $5\sigma$ significance level computed from red noise, indicating that this is a strong detection.

\begin{figure}
    \centering
    \includegraphics[trim={00mm, 12mm, 00mm, 0mm},clip=true,width=\linewidth]{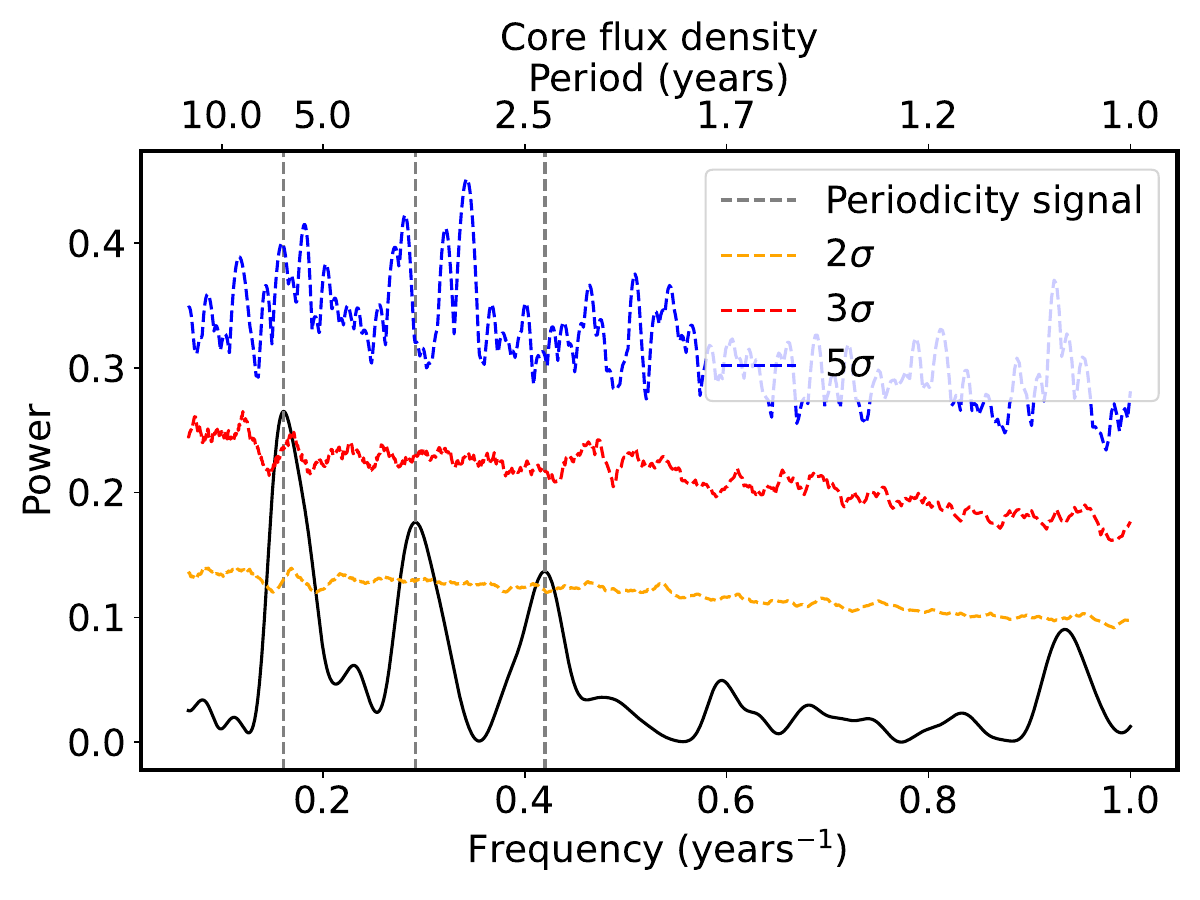}
    \includegraphics[trim={2mm, 5mm, 2mm, 3mm},clip=true,width=\linewidth]{LS_windowfunc.pdf}
      \caption{LS periodogram for the core flux density (upper panel) and window function (lower panel). The gray dashed lines mark the peaks at 6.2, 3.4 and 2.4 years.}
    \label{LS_CFD}
\end{figure}

The top panel of Figure \ref{LS_CFD} shows the LS result for the core flux density, where the explored periods are 1--15 years with 1000 frequencies. We find three peaks above the $2\sigma$ significance level: at $6.2\pm0.9$ years, $3.4\pm0.3$ years and $2.4\pm0.2$ years. We note that the $\sim$6 years and $\sim$3 years peaks coincide with the two first significant peaks obtained from the Jurkevic analysis. However, while the peak at $\sim$6 years attains the $3\sigma$ significance level, it also corresponds to a strong peak in the window function of power $\sim$0.20, which indicates that there are likely non-negligible effects from uneven sampling.

\subsection{Weighted-wavelet Z transform}

Finally, we perform the WWZ transform. The WWZ method relies on wavelet analysis to search for periodicity in both time and frequency, and is able to detect signals with varying shape and amplitude. In addition, it is well-adapted to analyze nonuniformly sampled data. We use the \textit{libwwz} publicly available package\footnote{\url{https://pypi.org/project/libwwz/}}, based on the derivations in \cite{Foster1996}, to compute the WWZ power.

For a function of time $x(t)$, which describes in our case either the flux density or the PA, the wavelet transform is:
\begin{equation}
    W(\upsilon, \tau; x(t)) = \upsilon^{1/2}\int x(t)f^*(\upsilon(t-\tau))dt,
\end{equation}
where $\upsilon$ is the scale factor or frequency, with $\upsilon^{-1}$ typically called dilation, $\tau$ is the time shift, $f$ is the wavelet kernel and $f^*$ is its complex conjugate.
We adopt the abbreviated Morlet wavelet \citep{Grossman1984},
\begin{equation}
    f(\upsilon(t-\tau))=e^{i\upsilon(t-\tau)-c\upsilon^2(t-\tau)^2},
\end{equation}
which is acceptable because we set $c$ to be small enough for the additional constant term $e^{-1/4c}$ in the traditional Morlet wavelet definition to be negligible. The Morlet wavelet describes a Gaussian with harmonic modulation and the abbreviated form is very similar to a windowed Fourier transform with frequency-dependent window $e^{-c\upsilon^2(t-\tau)^2}$. 

Following the \textit{libwwz} approach, we perform the WWZ transform analysis by splitting the time domain into $n_{\tau}=100$. The linear frequency range spans $1-24$ years for the PA and $1-15$ years for the core flux density, with a frequency step of 0.001. We set $c=1/32\pi^2=3.17\times10^{-3}<0.02$ for the decay constant. As for the LS periodogram, we include significance level estimations that account for red-noise (see Section \ref{appendix_period_levels}).

\begin{figure*}
    \includegraphics[width=0.5\linewidth]{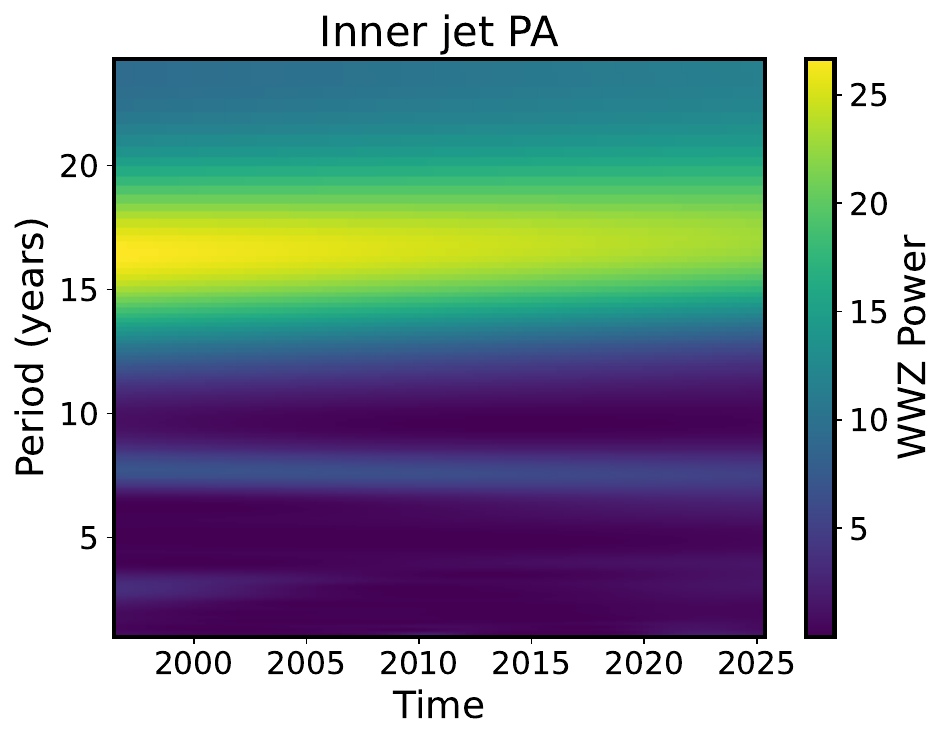}
    \raisebox{4mm}{\includegraphics[trim={0mm, 0mm, 0mm, 10mm},clip=true,width=0.5\linewidth]{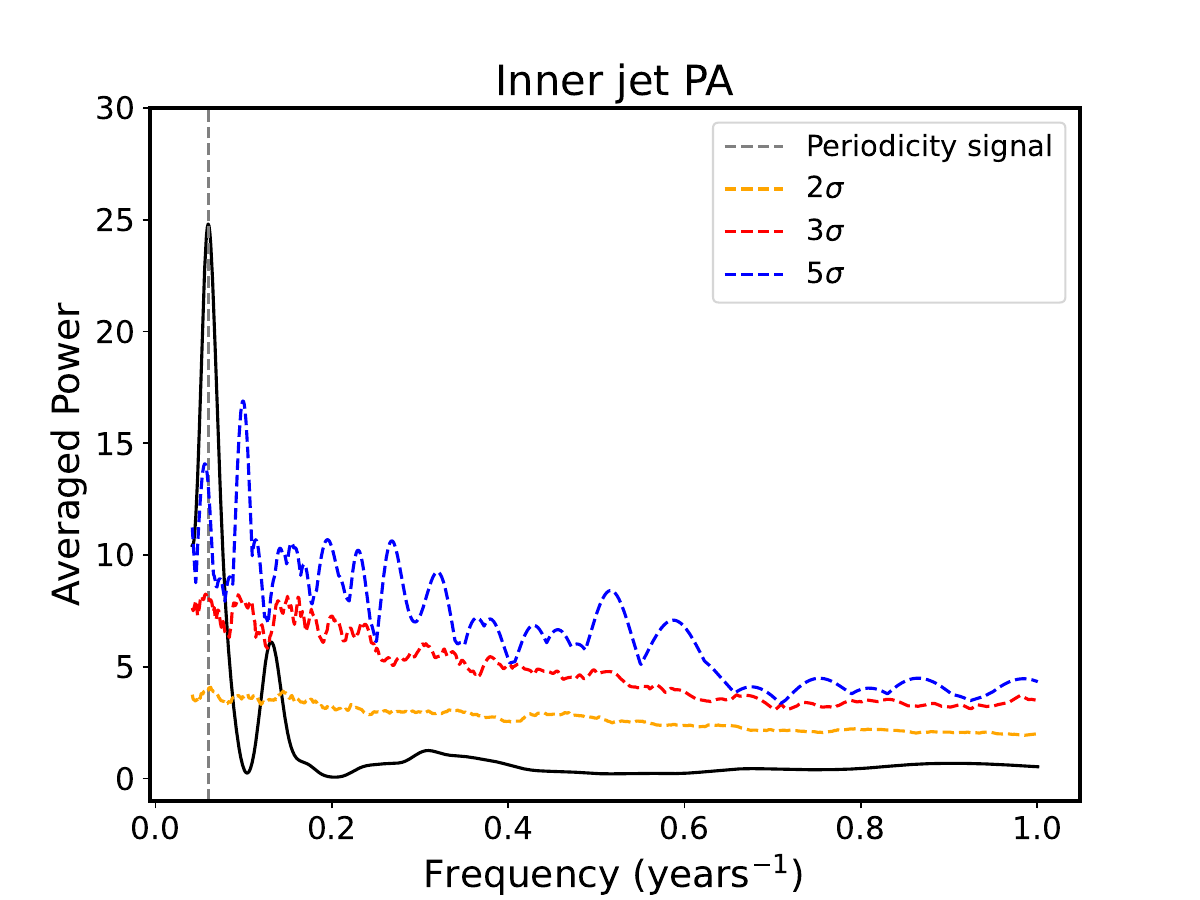}}
    \includegraphics[width=0.5\linewidth]{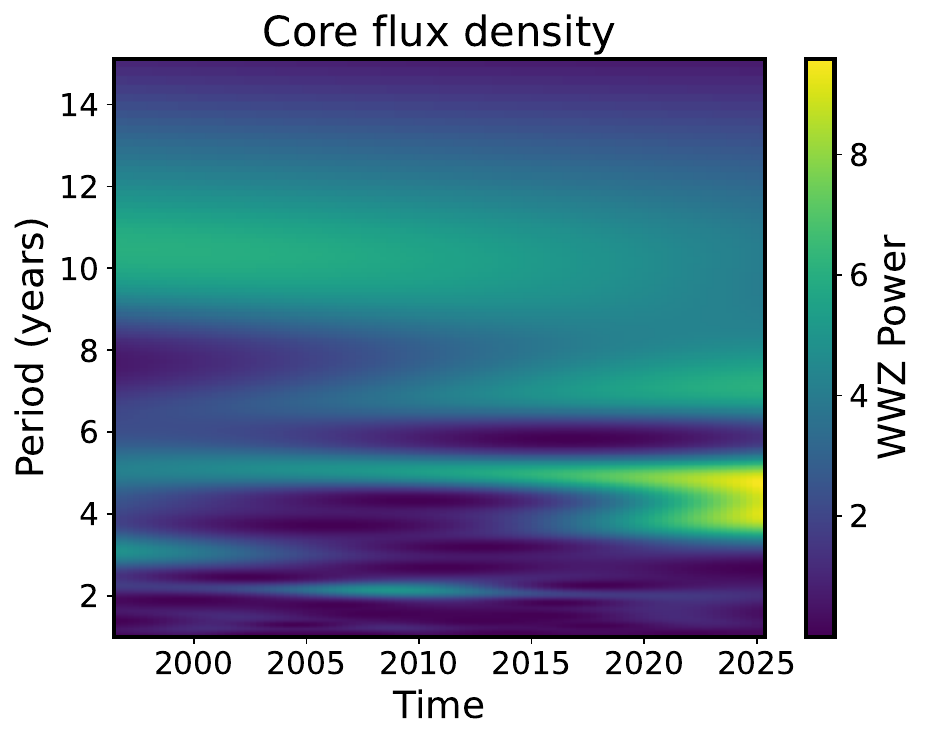}
    \raisebox{4mm}{\includegraphics[trim={0mm, 0mm, 0mm, 10mm},clip=true,width=0.5\linewidth]{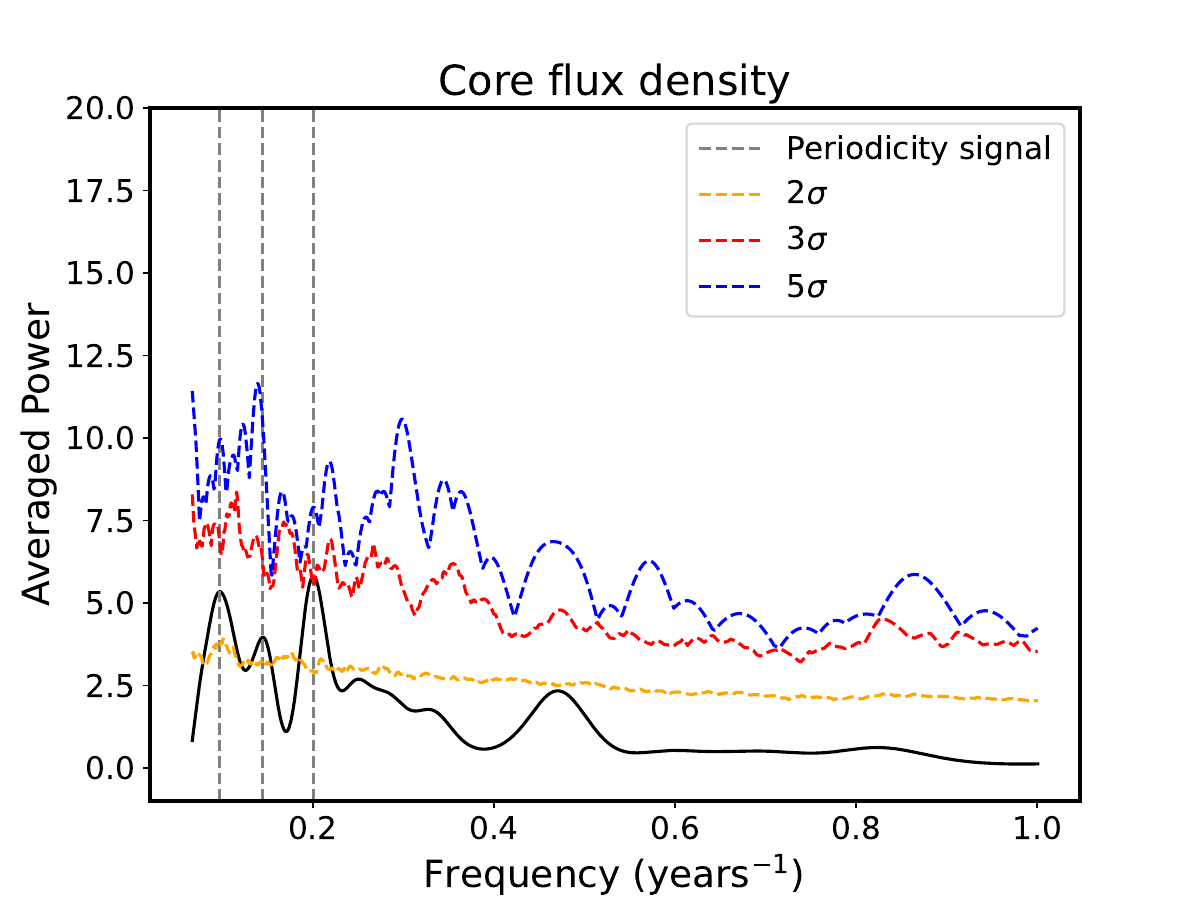}}
      \caption{Left: WWZ power as a function of frequency and time for the inner jet PA (top) and the core flux density (bottom). Right: $\tau$-averaged power and ($2\sigma$, $3\sigma$, $5\sigma$) significance levels in the (yellow, red, blue) dashed curves.}
    \label{WWZ}
\end{figure*}

The results for the PA are given in Figure \ref{WWZ} (bottom panel). The left panel shows the 2D WWZ power as a function of both time and frequency and the right panel shows the $\tau$-averaged power with frequency, together with the significance levels. We find a clear peak at $16.8\pm4.3 $ years, once again above the $5\sigma$ significance level. Looking at the 2D map, this is a persistent periodicity signal over time. Given the strong signals obtained with the three analyses, we conclude that a PA period of the order of a decade is firmly detected.
In addition, we notice from the WWZ transform a new period at $7.7\pm0.8$ years, above the $2\sigma$ significance level. The strength of this signal being stronger at earlier times, it is not revealed by the Jurkevic or LS methods.

Figure \ref{WWZ} shows the WWZ transform results for the core flux density (upper panel). We find a signal reaching the 3$\sigma$ significance level, which peaks at $5.0\pm0.4$ years. From the left panel of Figure \ref{WWZ}, this $\sim$5 years signal is persistent over time but merges around 2020 with a period of approximately 3.5 years becoming stronger with time. This explains the larger period detected with the WWZ transform compared to the two previous methods. 
We also find two other peaks above the $2\sigma$ significance levels.
The strongest one is detected at $10.3\pm3.5$ years, which is very wide and is the continuation of the weaker peak at $6.9\pm0.1$, which barely reaches the $2\sigma$ threshold.
While the significance of the results is lower than for the PA, we do detect a $\sim$7 years and $\sim$4 years signal with all of our periodicity search methods, warranting consideration. The $\sim$10 year period is also detected with two of the three methods. Additional data and better sampling would allow us to confirm whether these periodicities are meaningful.

\subsection{Significance levels estimation}
\label{appendix_period_levels}

We expect the 43\,GHz data of 3C\,66A to be affected by red noise. It is especially known that this noise impacts blazar LCs and must be taken into account when looking for periodicity signals. Red noise is the term used to denote stochastic processes, whose Fourier power spectrum $\mathcal{P}$ can be modeled simply as a power-law following $\mathcal{P}(\nu)=\nu^{-\alpha}$, where $\nu$ is the frequency and $\alpha$ the slope. In order to estimate significance levels in the LS periodogram and WWZ transform search methods that account for red noise, we apply the well-established \textit{PSRESP}\footnote{\url{https://github.com/wegenmat-privat/psresp/blob/master/README.rst}} (power spectral response) approach \cite{psresp2002MNRAS.332..231U}.
It generates artificial LCs following the method by \citep{Timmer1995A&A...300..707T} that have a power-law PSD and the same sampling as the observed data, for a range of $\alpha$ values and a range of periodogram and LC bins. The periodograms of the simulated LCs and of the real data are then compared. A success fraction based on chi-squared statistics is obtained. Although the PSRESP model was developed for LCs and red-noise estimation is generally applied to flux density data, we choose to perform the same red-noise estimation also for the PA. Indeed, such red noise may be present in the data and results in larger significance levels compared to white noise. We therefore model it in order to ensure the significance of the periodicity found for the inner jet PA.

We obtained $\alpha_\mathrm{PA}=1.7\pm0.7$ for the inner jet PA and $\alpha_\mathrm{CFD}= 1.5\pm0.3$ for the core flux density. We use these mean values in order to simulate artificial LCs that reproduce the level of red noise of our data when performing the LS periodogram and the WWZ transform. In the first case, we produce 10 000 artificial LCs and measure the LS power to obtain the 1--5$\sigma$ significance levels. However, we use 1000 LCs in the case of the WWZ method because this analysis is computationally expensive.

\begin{table}
    \caption{Periodicity search results in the inner jet PA and core flux density (CFD) of 3C\,66A.}
         \centering
         \begin{tabular}{cccc}
            \hline
            \noalign{\smallskip}
            Data & Method & Period $\pm$ error [yr] & Significance\\
        \noalign{\smallskip}
     \hline
     \noalign{\smallskip}
     PA & Jurkevich & 14.4 $\pm$ 5.1 & $F>0.5$ \\\noalign{\smallskip}
      & L-S & 10.8 $\pm$ 2.3 & $\sigma > 5$\\\noalign{\smallskip}
      & WWZ & 7.7 $\pm$ 0.8 & $\sigma > 2$\\
      \noalign{\smallskip}
      & & 16.8 $\pm$ 4.3 & $\sigma > 5$\\
      \noalign{\smallskip}
     CFD & Jurkevich & 3.5 $\pm$ 0.3 & $F>0.5$ \\
     \noalign{\smallskip}
      &  & 6.9 $\pm$ 0.2 & $F>0.5$ \\
      \noalign{\smallskip}
      &  & 10.5 $\pm$ 0.6 & $F>0.5$ \\
      \noalign{\smallskip}
      & L-S & 2.4 $\pm$ 0.2 & $\sigma > 2$\\
      \noalign{\smallskip}
      & & 3.4 $\pm$ 0.3 & $\sigma > 2$\\
      \noalign{\smallskip}
      & & 6.2 $\pm$ 0.9 & $\sigma > 3$\\\noalign{\smallskip}
      & WWZ & 5.0 $\pm$ 0.4 & $\sigma > 3$\\
      \noalign{\smallskip}
      & & 6.9 $\pm$ 0.1 & $\sigma > 2$\\
      \noalign{\smallskip}
      & & 10.3 $\pm$ 3.5 & $\sigma > 2$\\
      \noalign{\smallskip}
     \hline
    \end{tabular}
    \vspace{3ex}
    \label{tab_periods}
   \end{table}

\section{Precession model fitting}
\label{app:precession_fitting}

\begin{figure}
    \centering\includegraphics[trim={10mm, 40mm, 10mm, 15mm},clip=tru,width=\linewidth]{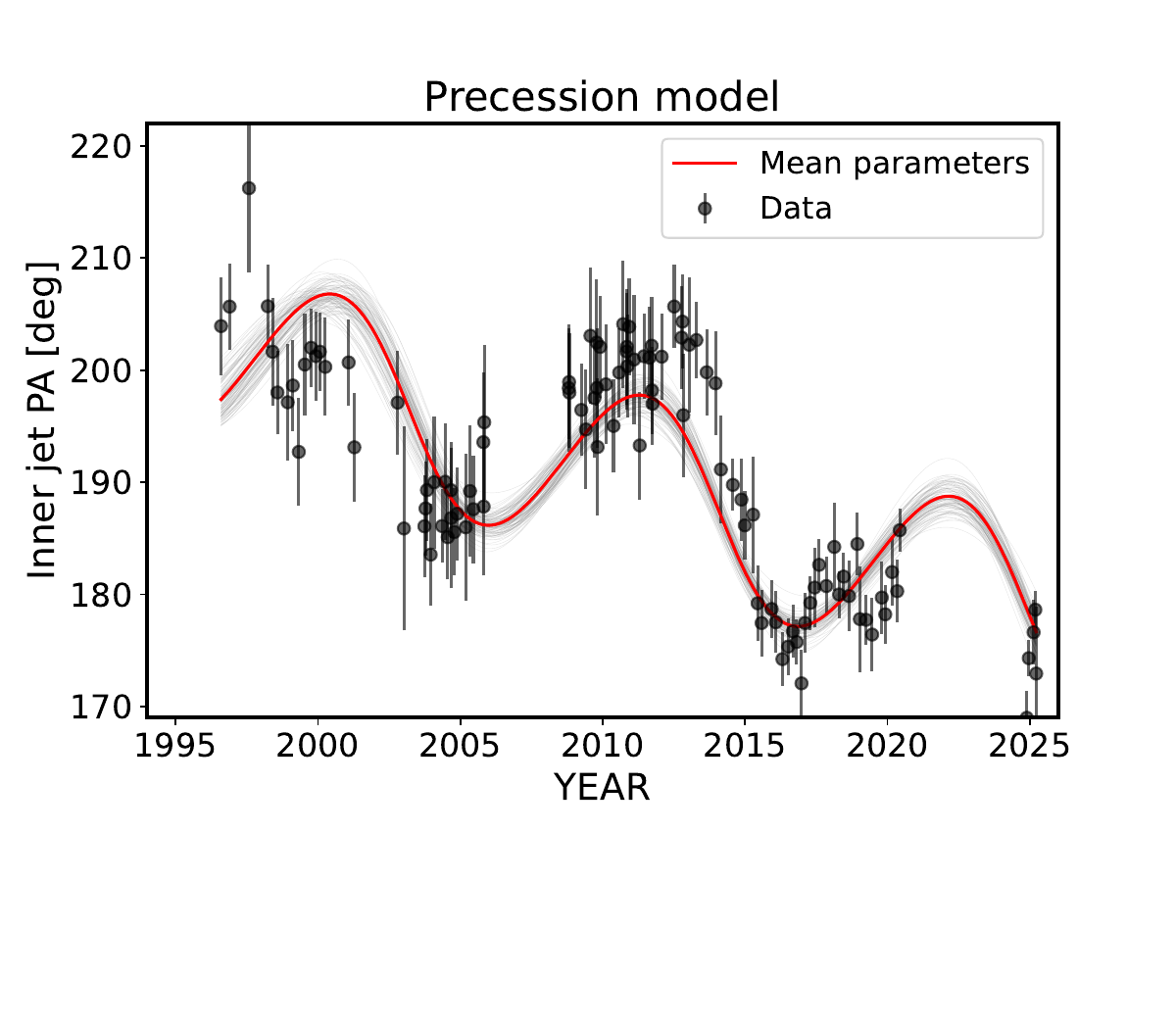}
      \caption{Precession fit to the inner jet PA for 100 random MCMC samples (thin gray lines) and the mean parameters (red line).}
    \label{Precession_PA}
\end{figure}

We perform a likelihood analysis to model the inner jet PA with a precessing cone. We consider the set of five free parameters $\vartheta = (\Omega,P_\mathrm{p},t_0,\phi_0,\theta_0)$ and a Gaussian log-likelihood function for the PA:
\begin{equation}
    \ln\mathcal{L}(\vartheta) = -\frac{1}{2}\sum_i\frac{(y_i-g(t_i,\vartheta))^2}{\sigma_i^2}
\end{equation}
with $y_i$ the PA data at time $t_i$, $\sigma_i$ the corresponding uncertainty from the observations and $g(t_i,\vartheta)$ the model prediction for the PA at time $t_i$ and for the parameters $\vartheta$. The precession model $g$ is described by Eq. \ref{eq_PA_precession}, topped with the linear decrease of $-0.83\degree$/yr obtained in Section \ref{orientation}. This ensures that the precession model considered here is applied only to the short-term oscillation of the inner jet PA.

We assume a uniform prior on the five parameters searched. For the base case: $\Omega\in[0,10]\degree$, $P_\mathrm{p}\in[6,20]$ yr, $t_0\in[2014,2020]$ yr, $\phi_0\in[0,90]\degree$ and $\theta_0\in[170,200]\degree$. In the relaxed constraint case $\phi_0\in[0,15]\degree$ and in the tight constraint case $\phi_0\in[0,5]\degree$. The second case is motivated by the fact that 3C\,66A is a BL Lac object, and the third is motivated by reported viewing angle values of 3C\,66A ranging between 1 and 5 degrees approximately \citep{2017ApJ...846...98J, Weaver2022}. Due to the uncertainty on the viewing angle, we choose not to set a lower limit but only an upper limit on the prior even for the tight constraint case. The viewing angle $\Theta$ and the angle between the precession cone axis and the line of sight $\phi_0$ are related by Eq. \ref{eq_viewing_angle}. We have checked that, for the parameter ranges considered in this analysis, setting $\phi_0<15\degree$ is equivalent to $\Theta<18\degree$ and setting $\phi_0<5\degree$ is equivalent to $\Theta<6\degree$. The log-prior reads,
\begin{equation}
   \ln P(\vartheta) =
\begin{cases}
    0, & \text{if parameters are within bounds} \\
    -\infty, & \text{otherwise.}
\end{cases} 
\end{equation}
Finally, the posterior probability is given by Bayes' theorem:
\begin{equation}
    P(\vartheta|y)=\frac{P(y|\vartheta)P(\vartheta)}{P(y)}.
\end{equation}
We use the \textit{emcee} MCMC module to sample from the posterior distribution. We set up 30 walkers and iterate over 20 000 steps, discarding the first 2 000 as burn-in to remove transient behavior. We obtain a robust mean acceptance fraction of 0.5. The corner plot results are shown in Figure \ref{fig7_MCMC_corner_plots} for the three cases studied. We take the mean and standard deviation of the samples in the tight constraint case $\phi_0<5\degree$ for our resulting precession model values (see Table \ref{tab_MCMC_parameters}). 
The PA evolution with 100 random MCMC samples is shown in figure \ref{Precession_PA}, highlighting the successful fit of the model to the data. 

\section{BH mass estimate}
\label{app:mass_estimate}

We estimate the primary BH mass $M_1$ in 3C\,66A based on optical variability timescale measurements, with the corrected redshift value of $z=0.340$ compared to the value of 0.444 originally used in the literature \citep{Kaur2017, Fan2018}.
We assume that the minimum variability time-scale $\Delta t_\mathrm{min}$ observed for 3C\,66A corresponds to the orbital time-scale of the innermost stable orbit around a central BH. This gives an upper limit that we consider to be the primary BH mass $M_1$ \citep[e.g.,][]{Xie2002MNRAS.334..459X, Kaur2017}:
\begin{equation}
    M_1 = 1.62\times10^4\frac{\delta}{1+z}\Delta t_\mathrm{min} M_{\odot},
\end{equation}
where $\delta$ is the Doppler factor of the jet. For this parameter, a large number of values have been reported in the literature on 3C\,66A. We take the weighted mean and standard deviation of the following list of factors: $\delta=1.99$ \citep{Lahteenmaki1999ApJ...521..493L}, $\delta=6.0$ \citep{Bottcher2005ApJ...631..169B}, $\delta=2.6$ \citep{Hovatta2009A&A...494..527H}, $\delta=5.69$ \citep{Fan2014RAA....14.1135F}, $\delta=4.3$ \citep{Liodakis2017MNRAS.466.4625L}, $\delta=16.5$ \citep{2017ApJ...846...98J}, $\delta=11.4$ \citep{Weaver2022}, where the last two values are in turn the average Doppler factor measured from 2 and 5 radio jet knots. For the values without reported error, we use an informed estimation from the measurements for which we have a standard deviation. We obtain $\delta=4.46\pm0.19$.
We note that for all the Doppler factors in the literature the authors assumed a redshift $z=0.444$. While it is in disagreement with the value used in this paper, the average Doppler factor remains valid for our study. Indeed, in the method used by the above authors to measure $\delta$, the choice between $z=0.340$ and $z=0.444$ has a negligible impact compared to the uncertainties on the other parameters and ultimately on the Doppler factor value that we use.

We take $\Delta t_\mathrm{min}=37$ min, which is the shortest among the optical intra-day variability timescales found by \cite{Kaur2017}, to measure the primary BH mass. This leads to $M_1=(1.20\pm0.05)\times10^8M_{\odot}$. We consider the typical range of mass ratios 30:1-3:1 derived by \cite{Gergely2009ApJ...697.1621G} for SMBH mergers. Taking $q=0.18\pm0.15$, we obtain the central BH mass $M=(1.42\pm0.19)\times10^8M_{\odot}$.

\section{Tables} \label{app:tables}
In Tables \ref{tab1} and \ref{tab2} are respectively given the parameters of the KaVA maps and of the Gaussian components of the images at each epoch, both at 22\,GHz and 43\,GHz, shown in Figure \ref{fig1}. 

   \begin{table*}
      \centering
      \caption[]{KaVA map parameters shown in Figure 1.}
         \label{tab1}
         \begin{tabular}{ccccccc}
            \hline
                 & & & \multicolumn{3}{c}{Restoring Beam} & \\\cline{4-6}
                $\nu$ & Epoch & $S_{\mathrm{peak}}$ & Major & Minor & P.A. & Lowest contour level \\
                (GHz) & (YYYY/MM/DD) & (Jy/beam) & (mas) & (mas) & (deg) & (mJy/beam) \\
                (1) & (2) & (3) & (4) & (5) & (6) & (7) \\
            \hline
                22\,GHz & 2014/04/16 & 0.325 & 1.23 & 1.16 & -58.9 & $0.73$ \\
                 & 2014/09/30 & 0.339 & 1.20 & 1.06 & -58.5 & $1.82$ \\
                 & 2014/12/15 & 0.350 & 1.49 & 1.13 & 31.9 & $2.41$\\ 
                43\,GHz & 2014/04/17 & 0.283 & 0.62 & 0.57 & -66.0 & $1.40$\\
                 & 2014/10/01 & 0.283 & 0.62 & 0.57 & -24.0 & $1.89$\\
                 & 2014/12/16 & 0.253 & 0.80 & 0.65 & 62.3 & $1.52$\\
            \hline
         \end{tabular}
         \tablefoot{(1) Observing frequency; (2) observing epoch; (3) peak flux density; (4), and (5) FWHM of the major and minor axes of the restoring Gaussian beam, respectively; (6) PA of the major axis of the beam measured from the north to east; and (7) lowest contour level of the map, corresponding to three times the rms noise in the map. The higher contour levels are obtained scaling twice.}
   \end{table*}

   \begin{table*}
      \caption[]{Parameters of Gaussian components of the KaVA image shown in Figure 1.}
         \label{tab2}
         \centering
         \begin{tabular}{ccccccc}
            \hline
                $\nu$ & Comp. ID & S [mJy] & R [mas] & $\theta$ [deg] & Major [mas] & $T_{\mathrm{b}}$ [K]\\
                (1) & (2) & (3) & (4) & (5) & (6) & (7)\\
            \noalign{\smallskip}
            \hline
                \multicolumn{7}{c}{First epoch} \\
                \hline
                22\,GHz & K & 288 $\pm$ 19 & 0.0 & 0.0 & 0.038 $\pm$ 0.002 & $6.7\times10^{11}$\\
                 & C8 & 48 $\pm$ 6 & 0.591 $\pm$ 0.007 & 194.0 $\pm$ 0.66 & 0.166 $\pm$ 0.014 & $5.8\times10^{9}$ \\
                 & C7 & 13 $\pm$ 3 & 1.725 $\pm$ 0.006 & 183.5 $\pm$ 0.22 & 0.09 $\pm$ 0.013 & $5.5\times10^{9}$ \\
                 & C6 & 22 $\pm$ 3 & 2.439 $\pm$ 0.002 & 180.3 $\pm$ 0.06 & 0.058 $\pm$ 0.005 & $2.1\times10^{10}$ \\
                 & C4 & 7 $\pm$ 1 & 3.8 $\pm$ 0.12 & 182.9 $\pm$ 1.81 & 1.348 $\pm$ 0.24 & $1.3\times10^{7}$ \\
                 & C3 & 2 $\pm$ 1 & 5.497 $\pm$ 0.223 & 179.3 $\pm$ 2.32 & 1.363 $\pm$ 0.446 & $4.2\times10^{6}$  \\
                 & C2 & 1 $\pm$ 1 & 7.653 $\pm$ 0.093 & 182.5 $\pm$ 0.7 & 0.515 $\pm$ 0.186 & $1.5\times10^{7}$ \\
                 & C1 & 17 $\pm$ 5 & 14.78 $\pm$ 0.343 & 166.4 $\pm$ 1.33 & 2.384 $\pm$ 0.686 & $1.0\times10^{7}$ \\
                43\,GHz & K & 240 $\pm$ 9 & 0.0 & 0.0 & 0.064 $\pm$ 0.002 & $5.2\times10^{10}$ \\ 
                 & C9 & 40 $\pm$ 4 & 0.371 $\pm$ 0.006 & 195.8 $\pm$ 0.9 & 0.166 $\pm$ 0.012 & $1.3\times10^{9}$ \\
                 & C8 & 11 $\pm$ 4 & 0.823 $\pm$ 0.015 & 191 $\pm$ 1.01 & 0.12 $\pm$ 0.029 & $6.7\times10^{8}$ \\
                 & C7 & 6 $\pm$ 1 & 1.459 $\pm$ 0.035 & 187.9 $\pm$ 1.36 & 0.421 $\pm$ 0.069 & $3.2\times10^{7}$ \\
                 & C6 & 16 $\pm$ 3 & 2.321 $\pm$ 0.017 & 180.7 $\pm$ 0.43 & 0.231 $\pm$ 0.035 & $2.7\times10^{8}$ \\
            \noalign{\smallskip}
            \hline
            \noalign{\smallskip}
                \multicolumn{7}{c}{Second epoch} \\
            \hline
                22\,GHz & K & 309 $\pm$ 17 & 0.0 & 0.0 & 0.07 $\pm$ 0.003 & $2.1\times10^{11}$ \\
                & C8 & 53 $\pm$ 8 & 0.658 $\pm$ 0.016 & 191.5 $\pm$ 1.36 & 0.284 $\pm$ 0.031 & $2.2\times10^{9}$ \\
                & C6 & 38 $\pm$ 7 & 2.27 $\pm$ 0.041 & 182 $\pm$ 1.03 & 0.565 $\pm$ 0.081 & $3.9\times10^{8}$ \\
                & C5 & 2 $\pm$ 2 & 3.137 $\pm$ 0.049 & 177.5 $\pm$ 0.89 & 0.243 $\pm$ 0.098 & $1.3\times10^{8}$ \\
                & C4 & 7 $\pm$ 3 & 5.03 $\pm$ 0.16 & 179.1 $\pm$ 1.82 & 0.851 $\pm$ 0.32 & $3.2\times10^{7}$ \\
                & C2 & 2 $\pm$ 1 & 8.489 $\pm$ 0.064 & 181.8 $\pm$ 0.43 & 0.305 $\pm$ 0.127 & $7.6\times10^{7}$ \\
                & C1 & 16 $\pm$ 8 & 15.21 $\pm$ 0.716 & 165 $\pm$ 2.7 & 2.993 $\pm$ 1.433 & $5.8\times10^{6}$\\
                43\,GHz & K & 250 $\pm$ 17 & 0.0 & 0.0 & 0.091 $\pm$ 0.004 & $2.7\times10^{10}$ \\
                & C9 & 44 $\pm$ 6 & 0.458 $\pm$ 0.005 & 192.6 $\pm$ 0.59 & 0.095 $\pm$ 0.009 & $4.3\times10^{9}$ \\
                & C8 & 11 $\pm$ 5 & 0.869 $\pm$ 0.022 & 193.7 $\pm$ 1.44 & 0.145 $\pm$ 0.044 & $4.8\times10^{8}$ \\
                & C7 & 7 $\pm$ 3 & 1.579 $\pm$ 0.019 & 190.2 $\pm$ 0.7 & 0.136 $\pm$ 0.038 & $3.2\times10^{8}$ \\
                & C6 & 24 $\pm$ 4 & 2.344 $\pm$ 0.027 & 181.8 $\pm$ 0.67 & 0.438 $\pm$ 0.055 & $1.1\times10^{8}$ \\
            \noalign{\smallskip}
            \hline
                \multicolumn{7}{c}{Third epoch} \\
            \hline
                22\,GHz & K & 311 $\pm$ 16 & 0.0 & 0.0 & 0.05 $\pm$ 0.002 & $4.1\times10^{11}$ \\
                & C8 & 33 $\pm$ 4 & 1.099 $\pm$ 0.01 & 187.5 $\pm$ 0.5 & 0.236 $\pm$ 0.019 & $2.0\times10^{9}$ \\
                & C6 & 27 $\pm$ 4 & 2.476 $\pm$ 0.021 & 180.7 $\pm$ 0.48 & 0.394 $\pm$ 0.042 & $5.8\times10^{8}$ \\
                & C4 & 4 $\pm$ 2 & 5.507 $\pm$ 0.03 & 178.5 $\pm$ 0.31 & 0.187 $\pm$ 0.059 & $3.5\times10^{8}$ \\
                & C1 & 14 $\pm$ 7 & 15.19 $\pm$ 0.595 & 166.8 $\pm$ 2.24 & 2.653 $\pm$ 1.191 & $6.5\times10^{6}$ \\
                43\,GHz & K & 226 $\pm$ 17 & 0.0 & 0.0 & 0.061 $\pm$ 0.003 & $5.4\times10^{10}$ \\
                & C9 & 34 $\pm$ 7 & 0.554 $\pm$ 0.007 & 190.2 $\pm$ 0.68 & 0.099 $\pm$ 0.013 & $3.1\times10^{9}$ \\
                & C7 & 7 $\pm$ 3 & 1.477 $\pm$ 0.07 & 193.9 $\pm$ 2.72 & 0.455 $\pm$ 0.14 & $2.9\times10^{7}$ \\
                & C6 & 22 $\pm$ 5 & 2.303 $\pm$ 0.054 & 182.6 $\pm$ 1.34 & 0.54 $\pm$ 0.108 & $6.7\times10^{7}$ \\
            \noalign{\smallskip}
            \hline
         \end{tabular}
         \tablefoot{(1) observing frequency; 
      (2) component identification;
      (3) flux density; 
      (4) separation from the core; 
      (5) PA with respect to the core; 
      (6) FWHM; 
      (7) brightness temperature}
   \end{table*}

\onecolumn
In Table \ref{tab3} are given the PA, core flux and total flux at each epoch for the 43\,GHz VLBA archival data, shown in Figure \ref{fig3}.

\begin{longtable}{cccc}
\caption{Epoch, PA, and flux measurements from the 43 GHz VLBA data (Figure \ref{fig3}).}\\
\toprule
Epoch & PA [deg] & Core flux [mJy] & Total flux [mJy] \\
\midrule
\endfirsthead

\caption{continued.}\\
\toprule
Epoch & PA [deg] & Core flux [mJy] & Total flux [mJy] \\
\midrule
\endhead

\bottomrule
\multicolumn{4}{r}{\textit{Continued on next page}} \\
\endfoot

\bottomrule
\endlastfoot

1996.6 & 203.9 $\pm$ 4.4 & 507 $\pm$ 51 & 687 $\pm$ 69 \\
1996.9 & 205.7 $\pm$ 3.8 & 489 $\pm$ 49 & 640 $\pm$ 64 \\
1997.58 & 216.3 $\pm$ 7.5 & 537 $\pm$ 54 & 734 $\pm$ 73 \\
1998.24 & 205.7 $\pm$ 3.7 & 423 $\pm$ 42 & 537 $\pm$ 54 \\
1998.41 & 201.7 $\pm$ 4.8 & 400 $\pm$ 40 & 587 $\pm$ 59 \\
1998.58 & 198.0 $\pm$ 3.8 & 333 $\pm$ 33 & 514 $\pm$ 51 \\
1998.94 & 197.1 $\pm$ 5.2 & 309 $\pm$ 31 & 440 $\pm$ 44 \\
1999.12 & 198.7 $\pm$ 4.0 & 363 $\pm$ 36 & 475 $\pm$ 47 \\
1999.33 & 192.7 $\pm$ 4.8 & 352 $\pm$ 35 & 438 $\pm$ 44 \\
1999.54 & 200.5 $\pm$ 4.5 & 485 $\pm$ 48 & 602 $\pm$ 60 \\
1999.76 & 202.0 $\pm$ 3.5 & 558 $\pm$ 56 & 701 $\pm$ 70 \\
1999.93 & 201.3 $\pm$ 4.0 & 631 $\pm$ 63 & 754 $\pm$ 75 \\
2000.07 & 201.6 $\pm$ 3.5 & 567 $\pm$ 57 & 724 $\pm$ 72 \\
2000.26 & 200.3 $\pm$ 4.4 & 526 $\pm$ 53 & 687 $\pm$ 69 \\
2001.08 & 200.7 $\pm$ 3.8 & 485 $\pm$ 49 & 632 $\pm$ 63 \\
2001.28 & 193.2 $\pm$ 4.8 & 751 $\pm$ 75 & 1072 $\pm$ 107 \\
2002.8 & 197.1 $\pm$ 4.6 & 362 $\pm$ 36 & 499 $\pm$ 50 \\
2003.02 & 185.9 $\pm$ 9.1 & 346 $\pm$ 35 & 452 $\pm$ 45 \\
2003.74 & 186.1 $\pm$ 4.6 & 368 $\pm$ 37 & 524 $\pm$ 52 \\
2003.78 & 187.6 $\pm$ 4.2 & 430 $\pm$ 43 & 587 $\pm$ 59 \\
2003.83 & 189.3 $\pm$ 4.5 & 378 $\pm$ 38 & 512 $\pm$ 51 \\
2003.96 & 183.5 $\pm$ 4.5 & 510 $\pm$ 51 & 680 $\pm$ 68 \\
2004.08 & 189.9 $\pm$ 5.9 & 302 $\pm$ 30 & 364 $\pm$ 36 \\
2004.36 & 186.1 $\pm$ 3.3 & 356 $\pm$ 36 & 442 $\pm$ 44 \\
2004.47 & 190.1 $\pm$ 5.2 & 345 $\pm$ 35 & 452 $\pm$ 45 \\
2004.56 & 185.1 $\pm$ 3.8 & 443 $\pm$ 44 & 577 $\pm$ 58 \\
2004.67 & 189.2 $\pm$ 4.3 & 477 $\pm$ 48 & 615 $\pm$ 61 \\
2004.68 & 186.9 $\pm$ 6.2 & 368 $\pm$ 37 & 447 $\pm$ 45 \\
2004.79 & 185.6 $\pm$ 3.9 & 497 $\pm$ 50 & 609 $\pm$ 61 \\
2004.89 & 187.2 $\pm$ 4.2 & 388 $\pm$ 39 & 492 $\pm$ 49 \\
2005.19 & 186.0 $\pm$ 6.6 & 297 $\pm$ 30 & 368 $\pm$ 37 \\
2005.34 & 189.2 $\pm$ 5.8 & 400 $\pm$ 40 & 508 $\pm$ 51 \\
2005.46 & 187.6 $\pm$ 4.8 & 358 $\pm$ 36 & 458 $\pm$ 46 \\
2005.81 & 193.6 $\pm$ 6.2 & 453 $\pm$ 45 & 589 $\pm$ 59 \\
2005.82 & 187.9 $\pm$ 6.2 & 412 $\pm$ 41 & 561 $\pm$ 56 \\
2005.84 & 195.3 $\pm$ 6.9 & 458 $\pm$ 46 & 614 $\pm$ 61 \\
2008.81 & 198.4 $\pm$ 5.3 & 394 $\pm$ 39 & 521 $\pm$ 52 \\
2008.82 & 199.0 $\pm$ 5.1 & 375 $\pm$ 37 & 509 $\pm$ 51 \\
2008.83 & 198.0 $\pm$ 5.3 & 344 $\pm$ 34 & 462 $\pm$ 46 \\
2009.25 & 196.5 $\pm$ 4.1 & 172 $\pm$ 17 & 285 $\pm$ 28 \\
2009.41 & 194.7 $\pm$ 5.3 & 148 $\pm$ 15 & 250 $\pm$ 25 \\
2009.57 & 203.1 $\pm$ 6.2 & 196 $\pm$ 20 & 292 $\pm$ 29 \\
2009.71 & 197.5 $\pm$ 5.3 & 310 $\pm$ 31 & 409 $\pm$ 41 \\
2009.79 & 202.5 $\pm$ 5.6 & 317 $\pm$ 32 & 405 $\pm$ 40 \\
2009.8 & 198.4 $\pm$ 5.3 & 326 $\pm$ 33 & 432 $\pm$ 43 \\
2009.82 & 193.1 $\pm$ 6.2 & 323 $\pm$ 32 & 438 $\pm$ 44 \\
2009.91 & 202.0 $\pm$ 4.5 & 336 $\pm$ 34 & 439 $\pm$ 44 \\
2010.11 & 198.7 $\pm$ 5.3 & 313 $\pm$ 31 & 369 $\pm$ 37 \\
2010.38 & 195.0 $\pm$ 4.2 & 366 $\pm$ 37 & 458 $\pm$ 46 \\
2010.58 & 199.8 $\pm$ 4.0 & 306 $\pm$ 31 & 375 $\pm$ 37 \\
2010.72 & 204.1 $\pm$ 5.7 & 330 $\pm$ 33 & 361 $\pm$ 36 \\
2010.84 & 201.7 $\pm$ 5.2 & 575 $\pm$ 58 & 742 $\pm$ 74 \\
2010.85 & 202.0 $\pm$ 5.1 & 510 $\pm$ 51 & 637 $\pm$ 64 \\
2010.87 & 200.4 $\pm$ 4.6 & 469 $\pm$ 47 & 578 $\pm$ 58 \\
2010.93 & 203.9 $\pm$ 4.3 & 295 $\pm$ 30 & 381 $\pm$ 38 \\
2011.1 & 201.0 $\pm$ 5.7 & 426 $\pm$ 43 & 495 $\pm$ 50 \\
2011.3 & 193.3 $\pm$ 4.8 & 264 $\pm$ 26 & 391 $\pm$ 39 \\
2011.45 & 201.2 $\pm$ 3.8 & 350 $\pm$ 35 & 444 $\pm$ 44 \\
2011.64 & 201.1 $\pm$ 4.5 & 358 $\pm$ 36 & 466 $\pm$ 47 \\
2011.72 & 202.1 $\pm$ 4.4 & 410 $\pm$ 41 & 562 $\pm$ 56 \\
2011.73 & 198.1 $\pm$ 3.9 & 351 $\pm$ 35 & 483 $\pm$ 48 \\
2011.75 & 197.0 $\pm$ 3.7 & 383 $\pm$ 38 & 550 $\pm$ 55 \\
2012.07 & 201.2 $\pm$ 3.8 & 223 $\pm$ 22 & 294 $\pm$ 29 \\
2012.51 & 205.7 $\pm$ 3.7 & 175 $\pm$ 17 & 243 $\pm$ 24 \\
2012.77 & 202.9 $\pm$ 4.6 & 198 $\pm$ 20 & 288 $\pm$ 29 \\
2012.8 & 204.3 $\pm$ 4.2 & 200 $\pm$ 20 & 298 $\pm$ 30 \\
2012.83 & 196.0 $\pm$ 5.5 & 231 $\pm$ 23 & 364 $\pm$ 36 \\
2013.04 & 202.4 $\pm$ 5.9 & 195 $\pm$ 20 & 304 $\pm$ 30 \\
2013.29 & 202.7 $\pm$ 3.4 & 235 $\pm$ 23 & 318 $\pm$ 32 \\
2013.65 & 199.8 $\pm$ 3.8 & 178 $\pm$ 18 & 231 $\pm$ 23 \\
2013.96 & 198.9 $\pm$ 4.7 & 228 $\pm$ 23 & 305 $\pm$ 30 \\
2014.15 & 191.1 $\pm$ 4.8 & 185 $\pm$ 19 & 247 $\pm$ 25 \\
2014.57 & 189.8 $\pm$ 2.3 & 165 $\pm$ 17 & 233 $\pm$ 23 \\
2014.87 & 188.4 $\pm$ 3.7 & 165 $\pm$ 17 & 245 $\pm$ 25 \\
2014.99 & 186.2 $\pm$ 3.0 & 199 $\pm$ 20 & 288 $\pm$ 29 \\
2015.28 & 187.1 $\pm$ 5.3 & 242 $\pm$ 24 & 318 $\pm$ 32 \\
2015.44 & 179.2 $\pm$ 3.3 & 215 $\pm$ 22 & 328 $\pm$ 33 \\
2015.58 & 177.5 $\pm$ 3.0 & 218 $\pm$ 22 & 314 $\pm$ 31 \\
2015.93 & 178.7 $\pm$ 2.5 & 241 $\pm$ 24 & 339 $\pm$ 34 \\
2016.08 & 177.5 $\pm$ 2.7 & 244 $\pm$ 24 & 341 $\pm$ 34 \\
2016.31 & 174.2 $\pm$ 2.4 & 225 $\pm$ 23 & 316 $\pm$ 32 \\
2016.51 & 175.3 $\pm$ 2.5 & 206 $\pm$ 21 & 301 $\pm$ 30 \\
2016.68 & 176.7 $\pm$ 2.3 & 218 $\pm$ 22 & 310 $\pm$ 31 \\
2016.81 & 175.7 $\pm$ 2.0 & 234 $\pm$ 23 & 335 $\pm$ 34 \\
2016.98 & 172.1 $\pm$ 3.0 & 240 $\pm$ 24 & 355 $\pm$ 35 \\
2017.1 & 177.4 $\pm$ 2.7 & 322 $\pm$ 32 & 484 $\pm$ 48 \\
2017.29 & 179.2 $\pm$ 2.4 & 220 $\pm$ 22 & 316 $\pm$ 32 \\
2017.44 & 180.5 $\pm$ 3.6 & 387 $\pm$ 39 & 509 $\pm$ 51 \\
2017.6 & 182.6 $\pm$ 2.3 & 368 $\pm$ 37 & 464 $\pm$ 46 \\
2017.85 & 180.7 $\pm$ 2.7 & 385 $\pm$ 38 & 511 $\pm$ 51 \\
2018.13 & 184.2 $\pm$ 4.0 & 459 $\pm$ 46 & 613 $\pm$ 61 \\
2018.3 & 180.0 $\pm$ 2.1 & 544 $\pm$ 54 & 756 $\pm$ 76 \\
2018.46 & 181.6 $\pm$ 2.1 & 441 $\pm$ 44 & 590 $\pm$ 59 \\
2018.65 & 179.8 $\pm$ 3.1 & 229 $\pm$ 23 & 349 $\pm$ 35 \\
2018.94 & 184.4 $\pm$ 2.8 & 161 $\pm$ 16 & 289 $\pm$ 29 \\
2019.03 & 177.9 $\pm$ 4.7 & 147 $\pm$ 15 & 243 $\pm$ 24 \\
2019.25 & 177.7 $\pm$ 2.2 & 118 $\pm$ 12 & 192 $\pm$ 19 \\
2019.46 & 176.5 $\pm$ 3.3 & 345 $\pm$ 34 & 516 $\pm$ 52 \\
2019.59 & 245.6 $\pm$ 11.6 & 163 $\pm$ 16 & 456 $\pm$ 46 \\
2019.8 & 179.7 $\pm$ 3.3 & 175 $\pm$ 17 & 259 $\pm$ 26 \\
2019.92 & 178.3 $\pm$ 2.7 & 144 $\pm$ 14 & 211 $\pm$ 21 \\
2020.16 & 182.1 $\pm$ 3.0 & 232 $\pm$ 23 & 324 $\pm$ 32 \\
2020.34 & 180.3 $\pm$ 2.8 & 338 $\pm$ 34 & 444 $\pm$ 44 \\
2020.43 & 185.7 $\pm$ 1.9 & 354 $\pm$ 35 & 473 $\pm$ 47 \\
2024.89 & 169.0 $\pm$ 2.4 & 375 $\pm$ 38 & 625 $\pm$ 62 \\
2024.96 & 174.3 $\pm$ 1.6 & 364 $\pm$ 36 & 550 $\pm$ 55 \\
2025.13 & 176.5 $\pm$ 2.8 & 535 $\pm$ 53 & 829 $\pm$ 83 \\
2025.19 & 178.6 $\pm$ 1.7 & 324 $\pm$ 32 & 510 $\pm$ 51 \\
2025.22 & 173.0 $\pm$ 5.9 & 395 $\pm$ 40 & 600 $\pm$ 60 \\
\label{tab3}
\end{longtable}
\twocolumn

\end{appendix}

\end{document}